\documentclass[twocolumn]{aa}
\usepackage{graphicx}
\usepackage{txfonts}
\usepackage{txfonts}
\usepackage{amsmath,mathtools}
\usepackage{natbib}
\bibpunct{(}{)}{;}{a}{}{,} 
\usepackage{here}
\usepackage[usenames,dvipsnames]{xcolor}
\usepackage{booktabs}
\usepackage{amssymb}
\usepackage{amsmath}
\usepackage{rotating}
\usepackage[normalem]{ulem}

\newcommand{\Msun}{\ensuremath{\,\mathrm{M_\odot}}\xspace}
\newcommand{\Lsun}{\ensuremath{\,\mathrm{L_\odot}}\xspace}

\graphicspath{{figures/}}

\definecolor{blue700}{HTML}{536DFE}

\definecolor{deeppurple}{HTML}{6200EA}
\usepackage[breaklinks=true,colorlinks,linkcolor=blue,citecolor=deeppurple]{hyperref} 
\begin{document} 

   \title{
	It's written in the massive stars: The role of stellar physics in the formation of black holes
	}

   \author{E. Laplace\inst{1}\thanks{eva.laplace@h-its.org}
   \and
   	F. R. N. Schneider\inst{1,2}
   \and
   	Ph. Podsiadlowski\inst{3, 1}
   }
  
  \institute{
  	Heidelberger Institut f\"ur Theoretische Studien, Schloss-Wolfsbrunnenweg 35, 69118 Heidelberg, Germany
  	\and
  	Astronomisches Rechen-Institut, Zentrum f\"ur Astronomie der Universit\"at Heidelberg, M\"onchhofstr. 12-14, 69120 Heidelberg, Germany
  	\and
  	University of Oxford, St Edmund Hall, Oxford, OX1 4AR, United Kingdom 
}
  
  \date{Received ...; accepted ...}

 
  \abstract
   {
   	In the age of gravitational-wave (GW) sources and newly discovered local black holes (BH) and neutron stars (NS), understanding the fate of stars is a key question. Not every massive star is expected to successfully explode as a supernova (SN) and leave behind a NS; some stars form BHs. The remnant left after core collapse depends on explosion physics but also on the final core structure, often summarized by the compactness parameter or iron core mass, where high values have been linked to BH formation. Several independent groups have reported similar patterns in these parameters as a function of mass, characterized by a prominent ``compactness peak'' followed by another peak at higher masses, pointing to a common underlying physical mechanism. Here, we investigate its origin by computing detailed single star models from 17 to 50 solar masses with MESA. We show that the timing and energetics of the last nuclear burning phases determine whether stars will reach a high final compactness and iron-core mass and likely form BHs. 
   	The first and second compactness increases originate from core carbon and neon burning, respectively, becoming neutrino dominated, which enhances the core contraction and ultimately increases the iron-core mass and compactness. An early core neon ignition during carbon burning, and an early silicon ignition during oxygen burning both help counter the core contraction and decrease the final iron core mass and compactness. Shell mergers between C/Ne and O-burning shells further decrease the compactness and we show that they are due to an enhanced entropy production in these layers. 
   	We find that the final structure of massive stars is not random but already ``written'' in their cores at core helium exhaustion when it is characterized by the central carbon mass fraction $X_{\mathrm{C}}$ and the CO core mass.
   	The same mechanisms determine the final structure of any star in this core mass range, including binary products, though binary interactions induce a systematical shift in the range of expected BH formation due to changes in $X_{\mathrm{C}}$. Finally, we discuss the role of stellar physics uncertainties and how to apply these findings to studies of GW sources.
   }


   \maketitle
%

\section{Introduction} 
Massive stars with masses larger than 10\Msun are rare compared to their low-mass counterparts but their contribution to the evolution and chemical enrichment of galaxies is disproportionately important \citep{hopkins_astrophysics_2014}. Their strong winds and supernova explosions create mechanical and energetic feedback and enrich their surroundings with heavier elements, determining the properties and evolution of galaxies and of the next generations of stars \citep{geen_bringing_2023}. However, not every massive star explodes. A fraction is expected to collapse directly and form black holes. Core-collapse supernovae are not always successful and failed explosions are another common path for black hole formation \citep[see, e.g., the recent review by][]{heger_black_2023}. Even successful explosion can be associated with BH formation through fallback accretion \citep{burrows_black-hole_2023}. Understanding and predicting the fate of stars remains one of the main unsolved problems in astrophysics.

The fate of stars is determined by the explosion mechanism, fallback dynamics, and their final core structure. In this work, we focus on the latter. Progress in stellar and supernova physics in recent years has shown that there is no simple initial mass threshold for a star to form a BH. In fact, almost 30 years ago, \citet{timmes_neutron_1996} pointed out that the final structure of a star, characterized for example by the final iron core mass, is not monotonic with mass. This picture was confirmed by multiple studies \citep[e.g.,][]{brown_formation_2001, oconnor_black_2011,sukhbold_compactness_2014,pejcha_landscape_2015,ertl_two-parameter_2016,sukhbold_high-resolution_2018, limongi_presupernova_2018,schneider_pre-supernova_2021,temaj_convective-core_2024}. Moreover, it is now established that most massive stars live in close binary or multiple systems \citep[e.g.,][]{sana_binary_2012}, further complicating this picture. Binary interactions have been shown to affect the pre-collapse core structures of stars, both based on studies of pure He star models approximating stripped stars in binaries \citep{brown_formation_2001, woosley_evolution_2019, aguilera-dena_stripped-envelope_2022, aguilera-dena_stripped-envelope_2023} and binary evolution models of stripped stars, accretors, and mergers \citep{laplace_different_2021,schneider_pre-supernova_2021,schneider_bimodal_2023,schneider_pre-supernova_2024}. In turn, this affects their explodability \citep{muller_three-dimensional_2019, vartanyan_binary-stripped_2021,woosley_birth_2020,antoniadis_explodability_2022}, nucleosynthesis \citep{farmer_cosmic_2021,farmer_nucleosynthesis_2023}, significantly reduces the parameter space for the formation of compact object mergers observable with gravitational-wave (GW) observations \citep{schneider_pre-supernova_2021,schneider_bimodal_2023} and leads to features in the chirp-mass distribution of binary BH mergers \citep{schneider_bimodal_2023}. However, to study the properties of BH populations and the formation of GW sources, current state-of-the-art studies necessarily have to make simplifying assumptions regarding the formation of BHs, often using analytical prescriptions solely based on the core mass of single-star progenitors \citep[e.g.,][]{fryer_compact_2012}. These can lead to substantially different outcomes compared to models that take the structure of stars into account \citep{patton_comparing_2022}. 

Observationally, only few direct hints of the link between the pre-supernova structure of massive stellar progenitors and the formation of BHs exist. A red supergiant (RSG), N6946-BH1, that was observed to suddenly vanish in the optical after a short outburst \citep{gerke_search_2015,adams_search_2017,sukhbold_missing_2020,basinger_search_2021} could be the first direct progenitor of a BH ever observed. This event is compatible with model predictions for a single star with high compactness undergoing a failed supenova explosion and eventually forming a BH \citep{lovegrove_very_2013,sukhbold_missing_2020,temaj_convective-core_2024}. Very recent observations with JWST potentially challenge this interpretation by identifying an infrared source at the location of this object, which may correspond to a surviving star enshrouded by dust \citep{beasor_jwst_2024} or to the emission from a failed supernova \citep{kochanek_search_2024}. Future observations are needed to better understand if this event was indeed a RSG forming a black hole after a failed explosion.\\
An additional, indirect observational clue on the link between RSGs and BH formation comes from observations of hydrogen-rich (type II) core-collapse SNe. Archival data searches have unambiguously identified several RSG progenitors at the location of these SNe. These observed RSG SN progenitors tend to have low luminosities of $\log L/ \Lsun \leq 5.1$ \citep{smartt_progenitors_2009}. This is in tension with the observed maximum luminosity of about $\log L/ \Lsun \approxeq 5.5$ found for RSGs in the galaxy and in the Magellanic Clouds \citep{davies_red_2020}, and is known as the missing RSG problem. A possible interpretation for these ``missing'' luminous RSG SN progenitors is that these are high-mass stars that ``quietly'' form black holes instead of exploding. The exact value of this maximum luminosity remains to be determined because of systematic uncertainties associated with photometric data of these objects \citep{davies_luminosities_2018,davies_red_2020}. Nonetheless, these observations offer important insights into the structures of stars that may be the progenitors of BHs. In principle, the bolometric pre-SN luminosity of a RSG can be directly linked to its final core mass, independently of the uncertainties in convective boundary mixing or rotation \citep{temaj_convective-core_2024}\footnote{In practice, limited photometry, dust production and uncertainties in the distance can affect the inferred observational luminosity \citep[e.g.,][]{davies_luminosities_2018}}. However, the initial mass of these progenitors is very uncertain. This is because, even assuming single-star evolution, variations in internal mixing, mass loss history, and rotation all affect the relation between the initial and final core mass of stars \citep{farrell_uncertain_2020}. Taking binary evolution into account further complicates this relation \citep{zapartas_diverse_2019,zapartas_how_2021}. Generally, the connection between the final core mass and final observable properties of stars is much better constrained \citep{temaj_convective-core_2024}. 
Finally, the lack of stars with high core masses exploding as supernovae is also supported by studies of their late-time supernova spectra \citep[e.g.,][]{jerkstrand_progenitor_2012} and by age-dating of supernova remnant environments \citep[e.g, ][]{jennings_supernova_2014}.\\

Based on detailed stellar models, several summarizing quantities have been defined to evaluate the fate of massive stars. The compactness parameter $\xi_{m}$ \citep{oconnor_black_2011} is commonly used in recent literature and defined as
\begin{equation}
\centering
\xi_{m} = \frac{m/\Msun}{R(m) / 1000 \rm{km}},
\label{eq:compactness}
\end{equation}
where $m$ is the mass coordinate at which the compactness is evaluated, typically at a chosen value of $2.5\Msun$, and $R$ is the radius at this mass coordinate. Essentially, it is a measure of the density (or mass-radius relation \citealt{chieffi_presupernova_2020}) outside the iron-rich core. This quantity, though arguably arbitrary in its definition, is known to correlate with other key properties, such as the iron-core mass, and the binding energy above the iron-rich core \citep[e.g.,][see also Fig.~\ref{fig:xi_overview}]{sukhbold_compactness_2014, schneider_pre-supernova_2021, temaj_convective-core_2024}. Stars with large iron core masses tend to have a high binding energy outside this core, and are thus difficult to explode by any explosion mechanism and tend to ultimately from BHs \citep{brown_formation_2001,sukhbold_compactness_2014,heger_black_2023,temaj_convective-core_2024}. In recent literature, the compactness parameter has been used as a predictor for the final remnant expected after core collapse, with high values indicating BH- and low values NS formation  \citep{oconnor_black_2011,ugliano_progenitor-explosion_2012,sukhbold_compactness_2014,limongi_presupernova_2018,schneider_pre-supernova_2021, schneider_bimodal_2023, schneider_pre-supernova_2024,heger_black_2023}. Multiple studies have pointed out that more sophisticated metrics are needed to accurately capture the explosion physics and understand the conditions for shock revival \citep{pejcha_landscape_2015,ertl_two-parameter_2016,muller_simple_2016,sukhbold_core-collapse_2016, vartanyan_binary-stripped_2021,burrows_core-collapse_2021}.
The explodability of stars is a subject of active discussion in the community and several explodability criteria have been proposed and explored in recent years. These include the two-parameter criterion of \citet{ertl_two-parameter_2016}, the presence of steep density profile with a density discontinuity around the Si/O interface \citep{vartanyan_binary-stripped_2021,tsang_applications_2022, wang_essential_2022, boccioli_explosion_2023}, or a forced explosion condition \citep{murphy_integral_2017,gogilashvili_force_2023}.

Independently of the discussion about explodability criteria, several unrelated groups making different assumptions regarding the microphysics and using different methods have reported remarkably similar patterns in the final core structure of stars (often summarized by the compactness parameter) as a function of their core or initial mass \citep{oconnor_black_2011,sukhbold_compactness_2014,sukhbold_high-resolution_2018,limongi_presupernova_2018,chieffi_presupernova_2020,chieffi_impact_2021,schneider_pre-supernova_2021, patton_towards_2020,takahashi_monotonicity_2023, temaj_convective-core_2024}. Typically, it consists of two prominent peaks in the compactness parameter, separated by $\approx15\Msun$ in initial mass and $\approx 7\Msun$ in CO core mass. The robustness of this pattern points to a common underlying physical process determining the final core structure of stars, and with it, their fate.

Pioneering work by \citet{sukhbold_compactness_2014} analyzed the pattern in the compactness parameter for the first time in great detail and linked it to the different nuclear burning conditions in the last evolutionary stages of massive stars post helium burning. \citet{patton_towards_2020} demonstrated that variations in the final compactness are linked to the initial conditions for core carbon burning, i.e. the mass of the CO core and the initial central carbon abundance at core helium exhaustion. \citet{sukhbold_high-resolution_2018} found that the final compactness of a star is influenced by small variations in  physical assumptions and resolution and interpreted this as a sign of intrinsic randomness in the core structure of stars. However,  \citet{chieffi_presupernova_2020} argued that these apparently random variations can be traced back to their assumptions regarding the core helium-burning evolution, where in particular semi-convection can result in late ingestion of helium in the core, generating ``breathing pulses'' which change the central carbon abundance and lead to different initial conditions for core carbon burning and ultimately, to a different final core structure. In our work, which includes convective boundary mixing above the helium-burning core, we do not find these signs of intrinsic stochasticity either \citep{schneider_bimodal_2023,temaj_convective-core_2024}.\\
Independent studies connected variations in the final core structure of stars to the number and size of carbon-burning shells and to the transition from convective to radiative carbon burning \citep{brown_formation_2001,sukhbold_compactness_2014,sukhbold_missing_2020,chieffi_presupernova_2020}. However, this explanation appears incomplete. Even after the transition from convective to radiative core carbon burning, models of stars that undergo radiative core carbon burning can result in a low compactness (see, e.g., Fig. 2 of \citealt{sukhbold_missing_2020}). In addition, the cause of the prominent drop in compactness after the first peak remains unclear, though \citet{sukhbold_compactness_2014} identified a link between the base of the carbon shell exceeding the effective Chandrasekhar mass and a smaller oxygen-burning core.

\citet{schneider_pre-supernova_2021} identified a connection between models with high compactness and the mass range for which carbon burning and neon burning become neutrino-dominated. Following these findings, in the present study, we investigate the origin of the observed patterns in the final structures of massive stars. We compute detailed simulations of massive single, non-rotating stars at solar metallicity that are the common progenitors of core-collapse events (17 -- 50\Msun) and focus on the evolution of the innermost 6\Msun.

We present our computational setup in Sect.~\ref{sec:methods} and the overall properties of our models in Sect.~\ref{sec:results}. In Sect.~\ref{sec:results:experiment}, we conduct a simplified experiment to identify and understand the general physical mechanisms responsible for the observed trends in final core structure. These insights are then applied to our fiducial set of stellar models in Sect.~\ref{sec:results:stars}. We summarize the physical mechanisms identified as being responsible for determining the main pattern in the final core structure of stars in Sect.~\ref{sec:results:global_picture}. We discuss the implications and uncertainties of our results in Sect.~\ref{sec:discussion} and present our conclusions in Sect.~\ref{sec:conclusion}.

\section{Methods}\label{sec:methods}

We compute the interior structure of massive single stars with initial masses between 17 and 50\Msun with the MESA stellar evolution code \citep[version 10398,][]{paxton_modules_2011,paxton_modules_2013,paxton_modules_2015,paxton_modules_2018}.
Our models build upon \citet{schneider_pre-supernova_2021, schneider_bimodal_2023}, with similar assumptions. 
Specifically, our models are computed at solar metallicity Z=0.0142 \citep{asplund_chemical_2009} and are non-rotating. We compute convective mixing with an mixing-length theory \citep{bohm-vitense_uber_1958} parameter of $\alpha_{\rm{MLT}}= 1.8$ and assume step overshooting of 0.2 pressure-scale height, which is only applied over the H and He-burning convective cores. We adopt the Ledoux criterion for convection and
assume a semi-convection efficiency of $\alpha_{\rm{sc}}=1.0 $ \citep{schootemeijer_constraining_2019}. We adopt the re-scaled ``Dutch'' wind mass loss rates of \citet{schneider_pre-supernova_2021} and enable the MLT++ method of MESA that boosts the local energy transport for outer layers of our massive star models that locally exceed the Eddington limit. The models are computed with the MESA \texttt{approx21\_cr60\_plus\_co56.net} nuclear network until the onset of core collapse, which is defined as the moment when the infall velocity of the iron-rich core exceeds $900 \,\mathrm{km}\,\mathrm{s}^{-1}$. This network effectively sets $Y_{\rm{e}}$ in the entire iron-rich core by making the approximation that deleptonizations in this core only occur through electron-captures onto $^{56}\rm{Fe}$. However, variations in $Y_{\rm{e}}$ between different stellar models are generally small \citep{woosley_evolution_2002} and do not affect the formation of the main qualitative patterns in the final core structure that are the subject of this work, which already appear at the end of core Ne burning (see also Sec. \ref{sec:results:pattern}). Using this nuclear reaction network is sufficient for our purpose, but we caution against employing these models as input for predictions of three-dimensional supernova simulations or nucleosynthesis yields, which require a larger nuclear network \citep{farmer_variations_2016,renzo_progenitor_2024}. To further verify that the final structure pattern is reproduced in models with larger networks, we compute three additional models with initial masses of $21\Msun$, $22\Msun$, and $23\Msun$, for which we employ a nuclear reaction network of 128 isotopes \citep[as recommended by ][]{farmer_variations_2016}. The $21\Msun$ and $22\Msun$ models encounter numerical difficulties after the iron-core infall velocity exceeds $250\,\mathrm{km}\,\mathrm{s}^{-1}$. By this point the iron core mass and the central entropy change only slightly (by less than 2\%), so we consider these to be comparable to our default collapse models. The reaction rates in our models are based on the JINA REACLIB database version 2.2 \citep{cyburt_jina_2010}.

We ensure a high spatial resolution in our models, in particular in zones of high temperature and density \citep{farmer_variations_2016}. More specifically, we adopt a minimum of 2000 grid points and a grid spacing option \texttt{mesh\_delta\_coeff = 0.6}, which results in an average of 5000--6000 grid points for each model. We also ensure a high temporal resolution throughout the evolution with a maximum time step of $10^{-4}$ years that is further limited based on changes in composition. 
Our final core collapse models and further information are available online\footnote{\href{https://zenodo.org/doi/10.5281/zenodo.13645155}{https://zenodo.org/doi/10.5281/zenodo.13645155}}.

To better disentangle the effects responsible for determining the final structure of massive stars, we perform a controlled experiment (see Sec. \ref{sec:results:experiment}). These models have the same core masses but a different central carbon mass fraction at the moment of core helium depletion (when the central helium mass fraction is lower than $10^{-4}$). This is achieved by computing additional models using the MESA \texttt{relax\_initial\_composition} method in which we artificially modify the central $^{12}\mathrm{C}$ abundance and the central $^{16}\mathrm{O}$ abundance while keeping the total mass fractions constant. Our base model for this experiment is our fiducial model with an initial mass of 22\Msun, which corresponds to the first compactness peak, at the moment of core helium depletion.

We examine the effect of changing the  $^{12}\mathrm{C}(\alpha,\gamma)^{16}\mathrm{O}$ reaction rate and discuss these in Appendix~\ref{sec:appendix:cag}. This notoriously uncertain nuclear reaction plays a crucial role for the evolution of stars, including their fate \citep{weaver_nucleosynthesis_1993,austin_effective_2014,sukhbold_missing_2020,farmer_constraints_2020}, as it determines the final central abundances at the end core helium depletion and the mass of the CO core (see also Section \ref{sec:results:pattern}). Our fiducial set of models adopts the default MESA rate from \citet{xu_nacre_2013}. The other two sets we compute adopt the approximately 15\% lower rate from \cite{kunz_astrophysical_2002} which is often used in the literature, and a rate that is 10\% higher than that of our default model. We compute additional sets of models for which we vary the semi-convection efficiency, discussed in detail in Appendix~\ref{sec:appendix:sc}. Finally, we perform a resolution test in Appendix~\ref{sec:appendix:res_test} to explore the effect of numerical uncertainties on the occurrence of shell mergers.

To evaluate the final fate of our massive star models, we compute the expected explosion outcome of our models using the semi-analytical parametric neutrino-driven supernova explosion model of \citet{muller_simple_2016}, with the same assumptions as in \citet{schneider_pre-supernova_2021,schneider_bimodal_2023,schneider_pre-supernova_2024} and \cite{temaj_convective-core_2024}. For simplicity, we do not consider black holes formed by supernova fallback in this work.

For comparison, we also employ the two-parameter explodability criterion by \citet{ertl_two-parameter_2016}. This criterion depends on the mass $M_{4}$, which is mass coordinate m where the entropy reaches a value of $s / (N_{A}k_{\rm{B}}) = 4 $ and on $\mu_4 = \frac{\Delta m/\Msun}{\Delta r/1000\rm{km}}\bigg\rvert_{s=4}$, the radial mass gradient at $M_4$. Physically, this location $M_4$ typically corresponds to the Si/O interface (i.e. the transition point between the Si-rich and O-rich layers), which has been found to be a good predictor for successful multi-dimensional neutrino-driven supernova explosions \citep{ertl_two-parameter_2016,muller_simple_2016,ertl_explosion_2020}. We use the \texttt{s19.8} calibration of \citet{ertl_two-parameter_2016} to models by \citet{woosley_evolution_2002} to distinguish successful and failed explosions.

\section{Final stellar properties}\label{sec:results}

\subsection{Final stellar structure}\label{sec:results:final_structure}
\begin{figure}
	\centering
	\includegraphics[width=0.5\textwidth]{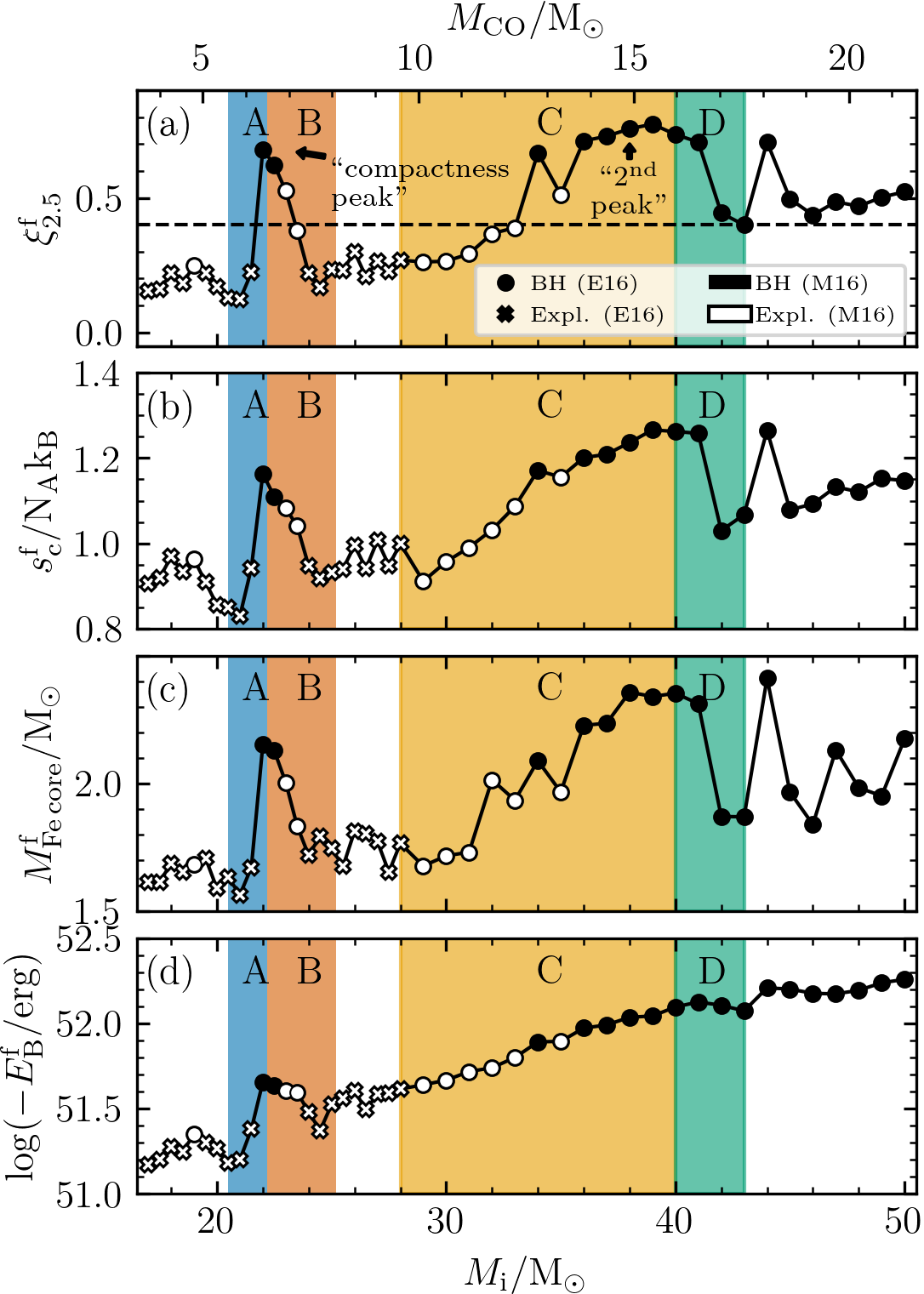}
	\caption{Final (a) compactness, (b) specific central entropy, (c) iron core mass, and (d) binding energy above $M_{4}$ at the onset of core collapse as a function of the initial mass. The top axis shows the CO core mass at core helium exhaustion. Circles and crosses  represent black hole formation and  explosions, respectively according to the \citet{ertl_two-parameter_2016} criterion, while black and white colors indicate BH formation and explosions based on the \citet{muller_simple_2016} supernova model.}
	\label{fig:xi_overview}
\end{figure}

We characterize the final stellar structure of our models by several quantities, including the compactness parameter, final specific central entropy, iron-core mass, and final binding energy (Fig.~\ref{fig:xi_overview}). Here, the binding energy $E_{B}$ of a star of mass $M$ above $M_{4}$ is defined as
\begin{equation}
E_{B} = - \int_{M_{4}}^{M} \frac{\rm{G} m}{r} dm.
\label{eq:binding_en}
\end{equation}
We identify specific mass ranges, labeled A, B, C, and D in Fig.~\ref{fig:xi_overview}, during which the compactness parameter, final specific central entropy, iron-core mass, and final binding energy follow specific trends (significant increase or decrease), whose physical origin we investigate in Sections \ref{sec:results:stars:regA}, \ref{sec:results:stars:regB}, \ref{sec:results:stars:regC}, and \ref{sec:results:stars:regD}, respectively. These mass ranges are indicated as a function of initial mass (bottom axis of Fig.~\ref{fig:xi_overview}) and of the CO core mass, based on the linear relation we derive between the two for our default assumptions (see also Fig.~\ref{fig:core_mass_sc}): 
\begin{equation}\label{eq:MCO_Mi}
M_{\rm{CO}} = 0.532\,M_{i} - 5.31.
\end{equation}

Fig.~\ref{fig:xi_overview} demonstrates that all quantities summarizing the final stellar structure follow a very similar, non-linear trend as a function of mass, confirming earlier findings \citep[e.g.,][]{,timmes_neutron_1996,brown_formation_2001,sukhbold_compactness_2014, chieffi_presupernova_2020,schneider_pre-supernova_2021,schneider_bimodal_2023,schneider_pre-supernova_2024,takahashi_monotonicity_2023,temaj_convective-core_2024}. For low masses, the values are approximately constant, with only small variations. A characteristic increase (region A) begins at initial (CO core) masses of about 19 (5.8)\Msun and reaches a maximum at CO core (initial) masses of 22 (6.5)\Msun, commonly referred to as the ``compactness peak''. It is followed by a decrease (region B) until a minimum is reached at about 25 (8)\Msun. After a mass range of about 3\Msun with small variations, all quantities then experience a significant second increase from 30 (11)\Msun (region C) until a second peak is reached and the values generally decrease from masses of 40 (16)\Msun (region D). As shown in the lowest panel, at this point all models have a high binding energy above $M_4$. Even if a supernova shock were successfully propagating after core collapse in these models, potentially leading to an observable supernova, the high binding energy means that the formation of a black hole is likely for these mass ranges, independently of the explosion mechanism \citep{sukhbold_compactness_2014,heger_black_2023}.\\

Our models show a lower intrinsic variability in the final structure than models by \citet{sukhbold_compactness_2014} and \citet{sukhbold_core-collapse_2016,sukhbold_high-resolution_2018}, just like the studies by \citet{chieffi_presupernova_2020,chieffi_impact_2021}, and \citet{takahashi_monotonicity_2023}. This is likely due to our choices of convective boundary mixing that prevent the occurrence of ``breathing pulses'' during core helium burning \citep{chieffi_presupernova_2020} that can change the relation between the CO core mass and the core carbon mass fraction $X_{\rm{C}}$ at core helium exhaustion. Secondary peaks in the quantities shown in Fig.~\ref{fig:xi_overview} are not observed for our default assumptions and mass sampling but can appear for different choices of physics. We argue that they are probably real (see Appendices~\ref{sec:appendix:cag} and ~\ref{sec:appendix:sc} for more details). 

All quantities shown in Fig.~\ref{fig:xi_overview} follow similar patterns because they are intrinsically linked \citep{fryer_remnant_2014,sukhbold_compactness_2014, takahashi_monotonicity_2023, schneider_pre-supernova_2021, schneider_pre-supernova_2024, temaj_convective-core_2024}. For completeness, we repeat these arguments below. The connection between the final central
entropy (Fig.~\ref{fig:xi_overview}b) and the final iron-core mass (Fig.~\ref{fig:xi_overview}c) can be understood through the effective Chandrasekhar mass  \cite[e.g.,][]{timmes_neutron_1996, woosley_evolution_2002,sukhbold_compactness_2014, schneider_pre-supernova_2024}. In massive stars, core collapse is triggered once the iron-rich core has reached a critical mass. However, in contrast to the cores of low-mass stars, the degenerate iron-rich cores of massive stars are hot, i.e. they have a finite temperature (entropy) that needs to be taken into account. This thermal structure leads to additional corrections compared to the classical Chandrasekhar mass $M_{\rm{Ch,0}} = 1.457 \left(\frac{Y_e}{0.5}\right)^{2}$ \citep[e.g.,][and references therein]{timmes_neutron_1996}, and lead to the approximate effective Chandrasekhar mass (ignoring special and relativistic corrections, and Coulomb corrections)
\begin{equation}
\label{eq:Mch}
\centering
M_{\rm{Ch}}\approx M_{\rm{Ch,0}} \left[\left(\frac{s_e}{\pi Y_e \rm{k_{\rm{B}}}\rm{N_{A}}}\right)^{2} + 1\right]\rm{,}
\end{equation}
where $s_e$ is the average electronic entropy, which is roughly a third of the average central entropy \citep{baron_effect_1990,timmes_neutron_1996}. This critical mass is mainly sensitive to changes in entropy because the electron fraction $Y_{e}$ tends to be very similar between different progenitors\footnote{In our models, $Y_{e}$ is pre-determined at the end of core oxygen burning and similar by construction (values between 0.4607 and 0.4629), see Sect.~\ref{sec:methods}.} \citep{sukhbold_compactness_2014}. Thus, the final iron-rich core mass that can form in a massive star before it exceeds $M_{\rm{Ch}}$ and its core begins to dynamically collapse is directly linked to the central entropy.

As shown in Fig.~\ref{fig:xi_overview}, the compactness parameter also correlates well with the final entropy and iron-core mass, which may appear surprising at first. This connection can be understood by considering that the compactness is essentially a measure of the mass-radius relation outside the iron core \citep{chieffi_presupernova_2020}. By taking into account that the final iron-rich core is adiabatic, this mass radius relation is directly connected to entropy through polytrope relations \citep{schneider_pre-supernova_2021}. Thus the final compactness can be understood as being equivalent to central entropy \citep[see also][]{fryer_remnant_2014}, itself linked to the final iron-core mass (and $M_{\rm{Ch}}$) as described above. The mass-radius relation and entropy both enter into the definition of binding energy above $M_4$ (see Eq. \ref{eq:binding_en}), which explains the observed correlation (Fig.~\ref{fig:xi_overview}d).

\subsection{Link to black hole formation}\label{sec:results:BH_formation}

To estimate the range of stars for which BH formation is expected, beyond considering the binding energy, we apply the \citet{ertl_two-parameter_2016} explodability criterion to our models. This is shown by the different markers (crosses for explosions and circles for black holes) in Fig.~\ref{fig:xi_overview}. Based on this criterion, stars with initial (core) masses below 30 (12) \Msun are expected to preferentially explode. Models in the first compactness peak and beyond the second compactness increase are predicted to form black holes. In addition, we compute the expected explosion outcome of our models using the semi-analytical neutrino-driven supernova explosion model of \citet{muller_simple_2016}. Black hole formation is indicated with black and successful explosions with white markers in Fig.~\ref{fig:xi_overview}. We find that the \citet{muller_simple_2016} model also predicts black hole formation for stars with a high compactness. The mass range for which successful explosions are expected is more extended compared to the outcome of the \citet{ertl_explosion_2020} criterion, reaching up to models with masses of 34 (12.5)\Msun.\\
Overall, we find that based on these explosion criteria, BH formation is consistently expected for models with high compactness and high central entropy, including the compactness-peak models at a mass of 22--23 (6.5--7)\Msun and models beyond a mass of 34 (14)\Msun. We confirm that using a compactness threshold for models in which BH formation is expected (e.g., $\xi^{f}_{2.5} > 0.4$, see dashed line in Fig.~\ref{fig:xi_overview}a) is too simplified to fully reproduce the predictions by more sophisticated measures of explodability. Considering a larger set of models \citep{schneider_pre-supernova_2021,temaj_convective-core_2024,schneider_pre-supernova_2024} we note that there is no clear compactness threshold separating successful and unsuccessful explosions (Maltsev et al. in prep.). For $\xi^{f}_{2.5}\lesssim 0.3$ successful explosions are found, and for $\xi^{f}_{2.5}\gtrsim 0.45$ explosions are unsuccessful, similarly to previous works  \citep{muller_simple_2016,takahashi_monotonicity_2023,zha_light_2023}. A compactness threshold can thus only be regarded as a rough first approximation of the core-collapse outcome. Approaches based on the entire final stellar structure, such as the \citet{muller_simple_2016} model, can be applied more generally.

\subsection{Emergence of the final structure pattern}\label{sec:results:pattern}

\begin{figure}
	\centering
	\includegraphics[width=0.5\textwidth]{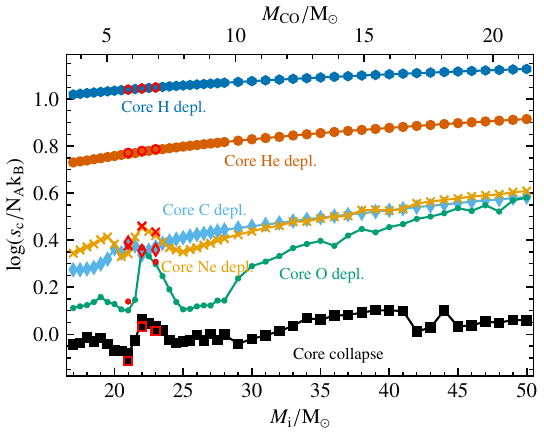}
	\caption{Evolution of the central specific entropy as a function of the initial mass for different evolutionary stages from core hydrogen depletion until core-collapse. Key difference in the final entropy start to appear at the end of core carbon burning. Red markers indicate three models computed with a nuclear network comprising 128 isotopes. Taking into account more nuclear reactions generally leads to a slightly lower final central entropy, though the general qualitative pattern of the compactness peak is unchanged. The characteristic final landscape is already mostly determined at the end of core Ne burning and remains until core collapse.}
	\label{fig:ev_entropy_diff_times}
\end{figure}

To understand at which point the observed pattern in the final stellar structure emerges, we trace the value of the central entropy at key evolutionary stages, shown in Fig.~\ref{fig:ev_entropy_diff_times}. At every stage, there is a general underlying trend of an increasing entropy as function of initial (or CO core) mass, which reflects the increasing core mass of stars as a function of their mass. As a function of time, the central entropy generally decreases. This can be understood as the effect of a changing mass-radius relation of the core, as it becomes increasingly compact due to core contraction \citep{chieffi_presupernova_2020,schneider_pre-supernova_2021}. Central nuclear burning episodes temporarily increase the central entropy, while thermal neutrino losses, which become important after core carbon burning, decrease it. At the end of core C burning (blue diamonds in Fig.~\ref{fig:ev_entropy_diff_times}), a first peak feature in the central entropy appears around 20 (5)\Msun initial (CO core) masses. It can be understood as a difference in heat content between models before and after 21.5 (6)\Msun. Models below this mass experience convective core carbon burning while models above this mass burn carbon radiatively as the burning becomes more neutrino-dominated (see also the Kippenhahn diagrams in the appendix Fig.~\ref{fig:eps_grav_overview_17_20.5} and \ref{fig:eps_grav_overview_21_25.5}). From the end of core neon burning on (orange crosses in Fig.~\ref{fig:ev_entropy_diff_times}), the characteristic feature corresponding to the compactness peak emerges around 22 (6.4)\Msun and remains until core collapse (black squares in Fig.~\ref{fig:ev_entropy_diff_times}). As we discuss later, this is because central entropy reflects the degree of contraction the core experiences in these models. 
Between the end of core O burning and core collapse, an additional feature emerges in the central entropy landscape for models above 40 (15)\Msun. The central entropy drops to lower values, though the general trend of increasing entropy as a function of mass remains. The key processes that determine the final entropy or compactness landscape thus already occur close to the time of core neon burning. For the models above 40 (15)\Msun additional processes play a role around the time of core silicon burning.

The emergence of the final structure landscape at the moment of core neon burning demonstrates that using a small nuclear reaction network is sufficient for characterizing the final structure landscape of stars. However, such models are not well-suited as input for multi-dimensional simulations of core collapse, which require larger reaction networks \citep{farmer_variations_2016,renzo_progenitor_2024}. We compare our default models to three additional models with initial masses of 21\Msun, 22\Msun, and 23\Msun, for which we employ a nuclear reaction network of 128 isotopes \citep{farmer_variations_2016}, shown with red markers in Fig~\ref{fig:ev_entropy_diff_times}. We find the same pattern in the formation of a compactness peak as in models computed with a smaller nuclear reaction network, though the final central entropy is 5\% smaller. In addition, the patterns in final central entropy shown here are very similar to the ones found by \citet{takahashi_monotonicity_2023}, who used a network comprising 300 isotopes and a finer grid (see their Fig. 11). The exact quantitative details of the final structure landscape, especially at the higher mass end, are thus still affected by changes in $Y_{e}$ induced by taking into account more nuclear reactions, which ultimately change $M_{\rm{Ch, eff}}$ (see Eq.~\ref{eq:Mch}). This can slightly shift the mass range for which a high final central entropy is reached, but does not significantly affect the main qualitative trends described here. 

\section{A controlled experiment for understanding the origin of the final structure}
\label{sec:results:experiment}
\subsection{Post core helium burning evolution}
\label{sec:results:experiment:he_burn}
\begin{figure}
	\centering
	\includegraphics[width=0.5\textwidth]{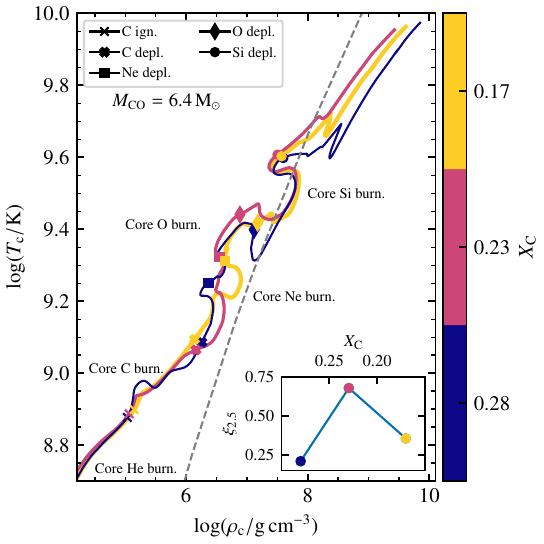}%
	\caption{ Central temperature as a function of the central density for models with the same core mass and different carbon mass fractions $X_{\rm{C}}$ after core helium burning (indicated in the color bar). The previous evolution of the core (not shown here for clarity) is indistinguishable for all models, while the post core helium burning evolution is greatly affected by the differences in central carbon abundance. Symbols indicate key evolutionary phases. To the right of the gray dashed line electron degeneracy dominates the pressure. The inset axis indicates the final compactness of the models as a function of decreasing $X_{\rm{C}}$.}
	\label{fig:Tc_rhoc_3ag}
\end{figure}

Beyond core helium burning, the evolution of massive stars proceeds differently compared to earlier stages, as thermal neutrino losses become important. Because of their minuscule interaction cross sections, the vast majority of neutrinos escape the star, causing a tremendous energy loss and greatly accelerating the evolution \citep[e.g.,][]{woosley_evolution_2002}. At the end of core helium burning, the core contracts. For temperatures around $\log (T/ \mathrm{K}) \approx 8.9$ and densities around $10^{5}\,\rm{g}\,\rm{cm}^{-3}$, carbon burning ignites in the center (see Fig. \ref{fig:Tc_rhoc_3ag}). The nuclear energy generation rate $\epsilon_{\rm{nuc}}$ can be expressed as $\epsilon_{\rm{nuc}}\sim \rho T^{23}X_{\mathrm{C}}$ during carbon burning, where $\rho$ and $T$ are the density and temperature conditions close to the core, and $X_{\mathrm{C}}$ the initial carbon abundance \citep[][]{woosley_evolution_2002}.
For high enough temperatures, $\log (T/ \mathrm{K}) \gtrsim 9$, pair production, followed by electron-positron pair annihilation, can occur. In rare cases, a neutrino-antineutrino pair is produced, which escapes the star. This process is significant enough to be the dominant energy sink in our models with an energy loss rate  $\epsilon_{\nu}\sim  \rho^{-1}T^{12}$ \citep[][]{sukhbold_missing_2020}. Photo-neutrinos from electron scattering processes also play a role, but the energy loss they cause is an order of magnitude lower than from pair-annihilation neutrinos. The balance between $\epsilon_{\nu}$ and $\epsilon_{\rm{nuc}}$ determines the final evolution of the core. As shown by the expressions for $\epsilon_{\rm{nuc}}$ and $\epsilon_{\nu}$, it mainly depends on $\rho$ and $T$, which are set by the core mass, and the initial central carbon abundance $X_{\mathrm{C}}$.

To disentangle the effects of the core mass and of $X_{\mathrm{C}}$ on the final structure \citep{patton_towards_2020} and understand the formation of the compactness peak identified between regions A and B in Fig.~\ref{fig:xi_overview}, we perform a controlled experiment. For a fixed CO core mass (6.62\Msun, corresponding to an initial mass of $M_{\rm{i}}=22$\Msun for our assumption), we compute models with a modified central carbon abundance. Our fiducial model ($X_{\mathrm{C}}= 0.23$) corresponds to the compactness peak model, while the models with carbon abundances of 0.28 and 0.17 have final compactness values before and after the peak, respectively. As shown in Fig.~\ref{fig:Tc_rhoc_3ag}, small differences in their central abundances cause large deviations in the post core helium evolution. Overall, the $X_{\mathrm{C}}= 0.23$ model (red line in Fig.~\ref{fig:Tc_rhoc_3ag}) reaches higher central temperatures and lower densities after core Ne burning compared to the other models at similar evolutionary stages.

\subsection{Carbon burning and beyond}
\label{sec:results:experiment:c_burn}
\begin{figure*}
    \centering
    \includegraphics[width=\textwidth]{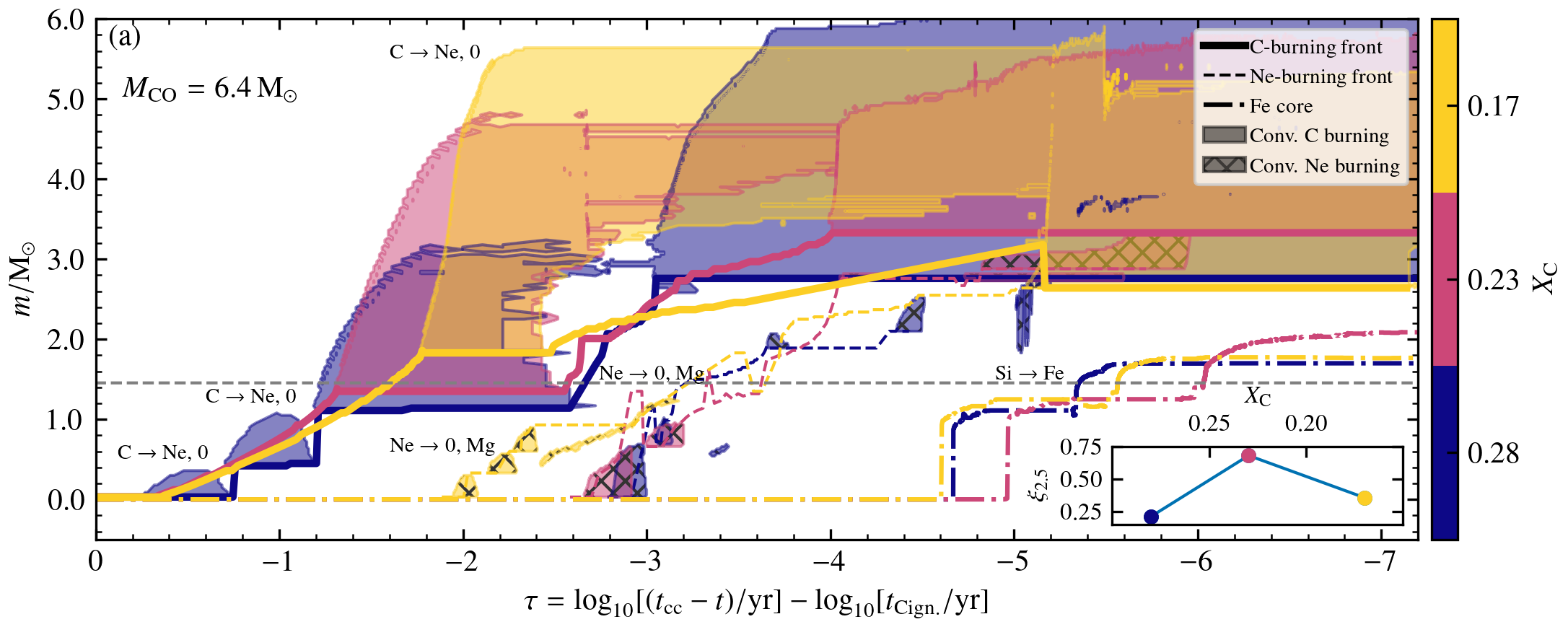}
    \includegraphics[width=\textwidth]{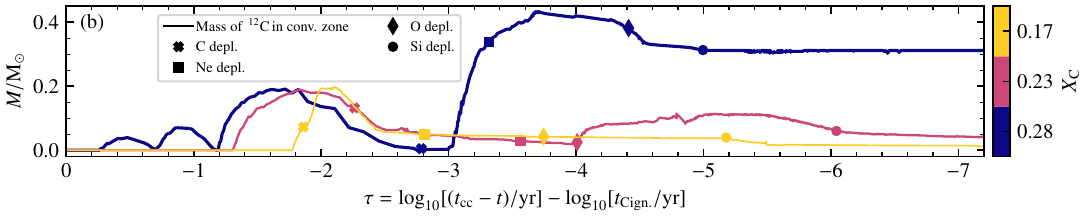}

    \caption{(a) Evolution of key elements of the inner 6\Msun core structure as a function of time from carbon ignition to core collapse for models with the same core mass and a different central carbon abundances $X_{C}$. Full and dashed lines indicate the carbon and neon-burning fronts, respectively. Shaded (hatched) regions show convective carbon (neon) burning regions in the core. The iron-core masses are indicated with dash-dotted lines. The gray horizontal dashed line indicates the classical Chandrasekhar limit. The inset figure shows the final compactness of these models as a function of decreasing $X_{C}$. (b) Total mass of $^{12}\rm{C}$ in the convective carbon-burning shells for each of the models as a function of time.}
    \label{fig:summary_kipp_3_cag}
\end{figure*}

In Fig.~\ref{fig:summary_kipp_3_cag}, we summarize the final evolution of the stellar structure for our three models with the same core mass and a different initial core carbon mass fraction $X_{\mathrm{C}}$. As indicated in the inset figure in Fig.~\ref{fig:summary_kipp_3_cag}a, these models vary in final compactness and represent models before ($X_{\mathrm{C}} = 0.27$), in ($X_{\mathrm{C}} = 0.23$), and after ($X_{\mathrm{C}} = 0.18$) the compactness peak, where the highest final central entropy and iron core mass are reached. Because they have the same core mass, their central temperature and density conditions at the end of core helium burning are the same (see also Fig.~\ref{fig:Tc_rhoc_3ag}) and therefore the neutrino loss rate $\epsilon_{\nu}$ is the same. However, at lower $X_{\mathrm{C}}$, neutrinos increasingly dominate the energetic balance during core carbon burning because the nuclear energy generation rate $\epsilon_{\rm{nuc}}$ decreases.

In Fig.~\ref{fig:summary_kipp_3_cag}a, we show the progression of the C- and Ne-burning fronts in these models in a Kippenhahn-like diagram as a function of the time from core carbon ignition to core collapse. The location of these nuclear burning fronts are defined as the mass coordinate at which the maximum energy generation rate for a particular burning process (e.g., C or Ne), is reached. In other words, a burning front traces the location at which the maximum burning occurs. These burning fronts are important for the final core structure because their location determines the maximum potential growth of the underlying core mass. A model with carbon burning front that reaches a higher mass coordinate forms a larger carbon-free core, and can eventually form a larger iron-rich core (dashed-dotted lines in Fig.~\ref{fig:summary_kipp_3_cag}).

After core carbon burning ignites in the stellar cores, the C-burning front moves outward in mass for all models as it burns through the available carbon fuel. In the $X_{\mathrm{C}} = 0.28$ model (dark blue lines in Fig.~\ref{fig:summary_kipp_3_cag}), the high core carbon abundance leads to a nuclear energy generation rate that is large enough to overcome the neutrino energy loss rate, which triggers the development of a relatively small (< 0.2\Msun) convective zone \citep{sukhbold_missing_2020}. It brings in 0.1\Msun of fresh $^{12}\mathrm{C}$ (see Fig.~\ref{fig:summary_kipp_3_cag} b) and keeps the carbon-burning front in the center until all the carbon has been burned. Subsequently, the core contracts until the temperature in the carbon-rich layers above the former convective zone reach high enough temperatures for carbon burning to take place and the carbon-burning front moves outward in mass. The location of the burning front thus reflects the amount of core contraction that occurred below it. In this model ($X_{\mathrm{C}} = 0.28$), this first convective episode is followed by two more successive convective regions that become more extended in mass as the temperature at the burning front increases. At the end of core carbon burning (blue cross in Fig.~\ref{fig:summary_kipp_3_cag}b), the core contracts even further, and core neon burning ignites in the center. The C-burning front moves further out in mass (though its progression is slowed by core neon burning, see also Fig.~\ref{fig:eps_ratio_burning_fronts}a) until the conditions for convection to occur are reached once again. At this point the carbon-burning front has a high temperature and reaches a region with unburned carbon, which means it reaches a high energy generation rate that highly exceeds the neutrino losses, triggering the formation of a large convective zone with over 0.4\Msun of carbon (see Fig.~\ref{fig:summary_kipp_3_cag}b). Because of the large amount of fuel remaining, the C-burning front keeps producing energy at a high rate and stays at this mass coordinate of 3\Msun until the end of the evolution. As a result, the carbon-free core below can only reach a relatively low mass. Hence, after silicon burning, a relatively low-mass iron core of 1.7\Msun forms.

For the other models, which have lower central carbon abundances (red and yellow lines in Fig.~\ref{fig:summary_kipp_3_cag}), central carbon-burning proceeds radiatively. Even though these models have the same mass, and therefore initially the same temperature and density conditions as the $X_{\mathrm{C}} = 0.28$ model, the lower core carbon abundance means that less energy is generated from carbon burning. Neutrino losses dominate and during the initial central carbon burning, energy is transported solely through radiation (see also Fig.~\ref{fig:eps_ratio_burning_fronts}). The available carbon in the center is depleted quicker, as no fresh carbon is brought in, and thus the core contracts and the C-burning front moves further out in mass. 
 
The contraction also increases the temperature at the burning front. Once the energy generation rate at the carbon-burning front reaches high enough values to exceed the neutrino losses (see also Fig.~\ref{fig:eps_ratio_burning_fronts}) two effects occur. First, a convective region forms and the carbon-burning front stays at a constant mass coordinate. Second, this convective carbon-burning shell acts like a mirror between the layers above and the core below. As the core below the shell contracts, the layers above greatly expand (as seen in the summary of the gravothermal energy in the stellar structure shown in Fig.~\ref{fig:eps_grav_overview_21_25.5}). The convective region increases in mass, bringing in more fuel for the carbon-burning shell (see Fig. ~\ref{fig:summary_kipp_3_cag}b) and prolonging the duration of this burning episode.

For $X_{\mathrm{C}} = 0.23$ and  $X_{\mathrm{C}} = 0.17$ (red and yellow lines in Fig.~\ref{fig:summary_kipp_3_cag}), the carbon-burning front reaches further out in mass during core carbon burning (time coordinate $t\approx -2$) than the $X_{\mathrm{C}} = 0.28$ model, signifying a larger core contraction. At this point, the mass coordinate of the C-burning front of the $X_{\mathrm{C}} = 0.23$ model, which corresponds to the compactness peak, is $m_{\mathrm{C}} = 1.35 \Msun$, close to the classical Chandrasekhar mass $M_{\rm{Ch, 0}}=1.455\Msun$, while that of the $X_{\mathrm{C}} = 0.17$ model exceeds it ($m_{\mathrm{C}} = 1.83\Msun$). As we discuss in Sect.~\ref{sec:results:experiment:degeneracy}, this is an important difference because electron degeneracy pressure plays a role in these models, slowing down the core contraction during core carbon burning in the $X_{\mathrm{C}} = 0.23$ and instead speeding it up in the $X_{\mathrm{C}} = 0.17$ model. 

The end of core carbon burning is the critical moment when the behavior of the compactness peak model ($X_{\mathrm{C}} = 0.23$) and the model beyond the compactness peak start to differ significantly. In the $X_{\mathrm{C}} = 0.17$ model, core neon burning already ignites while the C-burning front is still experiencing the first convective episode. In contrast, in the compactness-peak model ($X_{\mathrm{C}} = 0.23$) by the time neon ignites, convective carbon burning has mostly ended ($\tau =-2.6$). Because the core contraction is slower the model can burn a large fraction of the carbon in the convective region before core Ne burning ignites (see also Fig.~\ref{fig:summary_kipp_3_cag}b). Later on ($\tau \approx -3$), it quickly burns through the remaining carbon in this region, growing a large C-free core. The C-burning front in the $X_{\mathrm{C}} = 0.17$ model grows more slowly and less far ($m/\Msun = 3.18$ at $\tau = -5.15$) than in the compactness-peak model ($m/\Msun = 3.33$ at $\tau =-4$). As we show in Sect.~\ref{sec:results:experiment:neon_burning}, this is a direct consequence of the earlier core neon and oxygen ignition in the $X_{\mathrm{C}} = 0.17$ model, which suppresses carbon burning and slows down the progression of the C-burning front.

Moreover, towards the end of the evolution ($\tau =-5.2$) the C-burning front of the $X_{\mathrm{C}} = 0.17$ model suddenly drops in mass and reaches the same level as the Ne-burning front. As we discuss in more detail in Sect.~\ref{sec:results:experiment:shell_mergers}, this is caused by a shell merger between the C- and Ne-burning front. Eventually, a lower-mass iron core forms in this $X_{\mathrm{C}} = 0.17$ model compared to the compactness-peak model.

From this experiment, we have demonstrated that the final core structure of stars in this mass range is already determined at core helium depletion, where the initial conditions for core carbon burning are set. The same effect of a change in the final structure pattern due to a change in $X_{\mathrm{C}}$ can be observed for models with different initial or core masses. This is shown more generally in Appendix~\ref{sec:appendix:cag}, where we investigate changes induced by the $^{12}\mathrm{C}(\alpha,\gamma)^{16}\mathrm{O}$ rate, which is largely responsible for determining $X_{\mathrm{C}}$ at the end of core helium burning.

We have seen that the transition from a final stellar structure with low compactness and iron core mass to high compactness and iron core mass can be understood as the effect of increasingly neutrino-dominated burning. It results in an overall larger core contraction and the formation of a larger carbon-free core, and eventually, to the formation of a larger iron-rich core. 
At the compactness peak, the neutrino-dominated carbon-burning conditions are such that most of the available carbon is burned before the end of core neon burning. Afterward, the carbon-burning front quickly moves out in mass and grows a particularly large core C-free core, which eventually forms a particularly large iron-rich core.
For the transition between the compactness peak model and the model beyond, we identify three mechanisms that are responsible for the observed drop in compactness: (1) the effect of an earlier central neon and oxygen ignition with decreasing $X_{\mathrm{C}}$, (2) an increased electron degeneracy, and (3) shell mergers.

\subsection{Effect of an early core neon ignition}
\label{sec:results:experiment:neon_burning}

\begin{figure*}
	\centering
	\includegraphics[width=\textwidth]{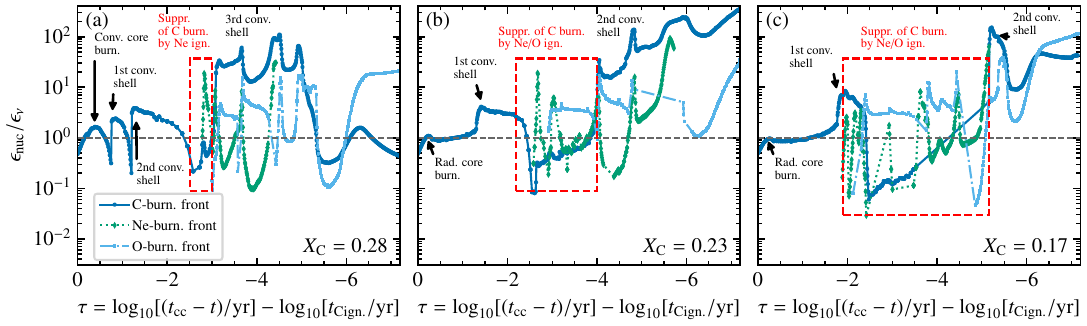}
	\caption{Time evolution of the ratio between the specific nuclear energy generation rate $\epsilon_{\rm{nuc}}$ and the neutrino loss rate $\epsilon_{\nu}$ at the location of the C, Ne, and O-burning fronts for models with the same core mass and different central carbon abundances. The dashed horizontal line indicates where $\epsilon_{\rm{nuc}}$ = $\epsilon_{\nu}$. When the energy ratio significantly exceeds this line, convection occurs \citep{sukhbold_missing_2020}. With a decreasing core carbon abundance, the core becomes more neutrino dominated and neon burning, followed by oxygen burning, occur earlier. As highlighted in the red boxes, this temporarily suppresses carbon burning.}
	\label{fig:eps_ratio_burning_fronts}
\end{figure*}

We have observed that the timing of core neon ignition varies significantly between models in our experiment with the same core mass and a changing initial core carbon abundance $X_{\mathrm{C}}$ (Fig.~\ref{fig:summary_kipp_3_cag}). The consequence of this early neon ignition is shown in Fig.~\ref{fig:eps_ratio_burning_fronts}, where the evolution of the ratio between the nuclear energy generation $\epsilon_{\rm{nuc}}$ and the neutrino loss rate $\epsilon_{\nu}$ at the location of the C, Ne, and O-burning fronts is plotted as a function of time until core collapse. Phases when $\epsilon_{\rm{nuc}}$ significantly exceeds $\epsilon_{\nu}$ correspond to convective episodes \citep{barkat_non-monotonic_1990,barkat_late_1994,sukhbold_missing_2020}.
In the $X_{\mathrm{C}} = 0.28$ model (Fig.~\ref{fig:eps_ratio_burning_fronts}a), core neon burning ignites immediately after core carbon depletion, when a core contraction occurs. Neon burning temporarily slows down the progression of the carbon-burning front and dominates the energy generation rate. However, it does not take long before the slowed down carbon-burning front reaches regions of unburned carbon, quickly becoming the main energy source again and forming a large convective region that remains at the same mass coordinate until the end of the evolution (see also Fig.~\ref{fig:summary_kipp_3_cag}b).

In the compactness-peak model ($X_{\mathrm{C}} = 0.23$, Fig.~\ref{fig:eps_ratio_burning_fronts}b), core neon and oxygen burning ignite in rapid succession after the convective carbon-burning region has exhausted its fuel. They dominate the energy generation, slowing down the progression of the carbon burning front until the end of core oxygen burning is reached. The ensuing core contraction helps the carbon-burning front reach layers where the nuclear energy generation rate dominates compared to the neutrino losses, triggering the formation of a large convective zone and marking the final location of this burning front.

In the $X_{\mathrm{C}} = 0.17$, model (Fig.~\ref{fig:eps_ratio_burning_fronts}c), while the convective carbon-burning episode is still ongoing, the conditions for neon burning are already reached in the center due to the stronger preceding contraction of the core aided by exceeding $M_{\mathrm{Ch},0}$ (see Sect.~\ref{sec:results:experiment:degeneracy}). While core neon burning is more neutrino-dominated in this model and produces less energy, this simultaneous burning reduces the luminosity of the carbon-burning front, impacting the extend of the convective region above. Once oxygen burning ignites shortly after, it counters the contraction and helps drastically slow down the progression of the now very neutrino-dominated carbon-burning front. In the subsequent phases, neon and oxygen burning dominate the energy generation, followed by core silicon burning. After silicon depletion, the neon-burning front, which moves out rapidly in mass, merges with the carbon-burning front (see Sect.~\ref{sec:results:experiment:shell_mergers}). The slowed C-burning implies a smaller carbon-free core and ultimately, a smaller iron core mass and compactness.

\subsection{Origin of the changing timing of core neon and oxygen ignition}
\label{sec:results:experiment:degeneracy}

\begin{figure}
	\centering
	\includegraphics[width=0.5\textwidth]{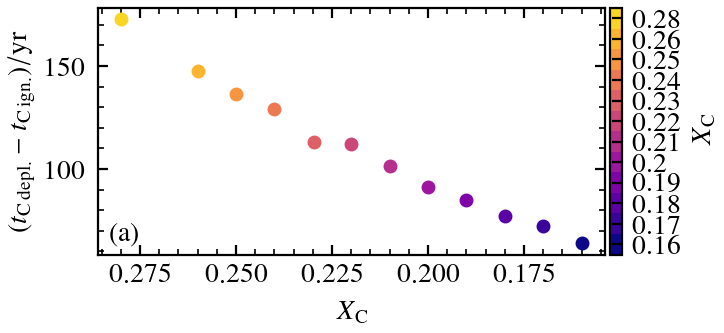}
	\includegraphics[width=0.5\textwidth]{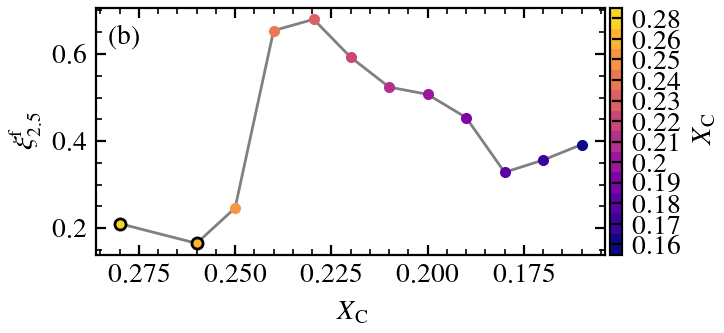}
	\includegraphics[width=0.5\textwidth]{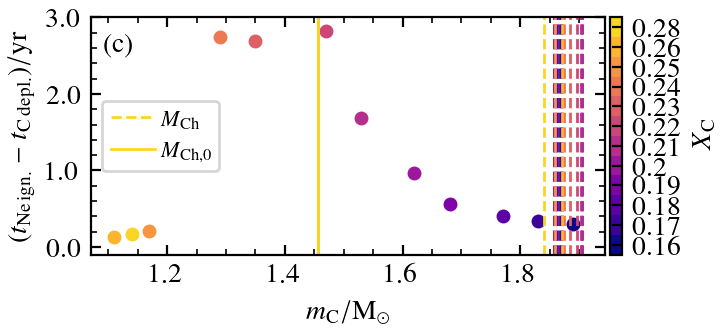}
	\caption{(a) Duration of core carbon burning for models with the same mass and a varying initial core carbon abundance $X_{\mathrm{C}}$, indicated with colors.
		(b) Final compactness as a function of $X_{\mathrm{C}}$. Models that undergo convective core carbon burning are highlighted with black outlines. (c) Duration of the phase between core carbon depletion and core neon ignition as a function of $m_{C}$, the maximum mass coordinate reached by the carbon-burning front before core neon burning. The classical ($M_{\rm{Ch},0}$) and effective ($M_{\rm{Ch}}$) Chandrasekhar mass at core C depletion are shown as vertical dashed and full lines, respectively.}
	\label{fig:mult_xC}
\end{figure}
 Here, we investigate the origin of the systematically earlier ignition of neon and oxygen for lower $X_{\mathrm{C}}$. We find that it is mainly due to a systematic shortening of the duration of core carbon burning for a decreasing amount of fuel ($X_{\mathrm{C}}$). This is shown for a large set of models with the same core mass and a varying core carbon abundance in Fig.~\ref{fig:mult_xC}a. For reference, we show the final compactness of these models in Fig.~\ref{fig:mult_xC}b. The contraction phase that follows core carbon depletion (when $X_{\mathrm{C}} < 10^{-4}$) before core neon burning ignites is significantly shorter than the duration of core carbon burning, on a neutrino-losses accelerated thermal time scale on the order of years (see Fig.~\ref{fig:mult_xC}c). It therefore plays a smaller role than the duration of core carbon burning in determining the timing of core neon ignition. 

\begin{figure*}
	\centering
	\includegraphics[width=\textwidth]{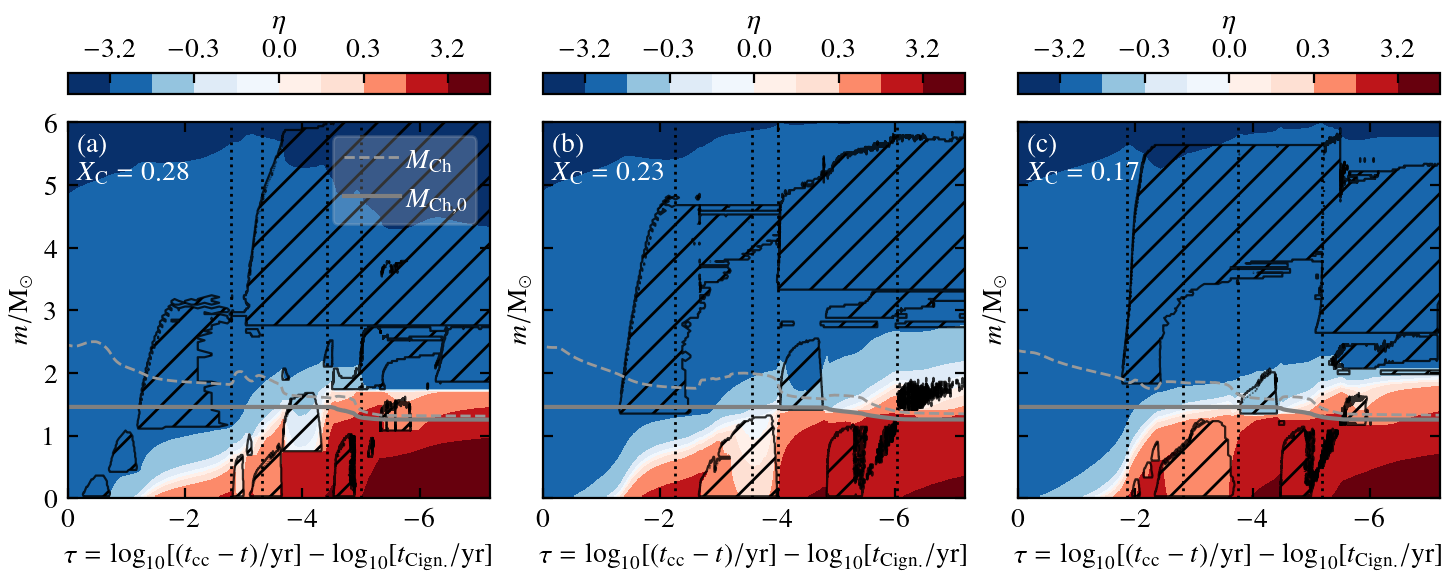}
	\caption{Kippenhahn diagram of the inner 6\Msun core structure of models with the same core mass and different core carbon abundance at core C ignition. Colors indicate the dimensionless electron degeneracy parameter $\eta = \mu/k_{\rm{B}}T$ (electrons are partially degenerate for $\eta \approx 0$ and very degenerate for $\eta \gg 0$). Convective zones are highlighted by the hatched regions and the dotted vertical lines indicate, from left to right, the moments when $^{12}\rm{C}$, $^{20}\rm{Ne}$, $^{16}\rm{O}$, and $^{28}\rm{Si}$ are depleted at the center (central abundance lower than $10^{-4}$). Grey full and dashed line indicates the classical ($M_{\rm{Ch},0}$) and effective ($M_{\rm{Ch}}$) Chandrasekhar mass, respectively. Electron degeneracy during core carbon burning increases for models with lower central carbon abundances.}
	\label{fig:eta_contours}
\end{figure*}

Besides the duration of core carbon burning, a secondary effect related to electron degeneracy also appears to play a role in accelerating the evolution after core carbon burning, as first pointed out by \citet{sukhbold_compactness_2014}. This is shown in Fig.~\ref{fig:eta_contours}, where the evolution of the electron degeneracy parameter $\eta = \mu/k_{\rm{B}}T$ is displayed as colored contours in a Kippenhahn-like plot for each of our example models. $\eta \approx 0$ indicates partial degeneracy, while $\eta \gg 0$ signifies that electrons are strongly degenerate. We also show the evolution of the classical ($M_{\rm{Ch},0}$) and effective Chandrasekhar masses ($M_{\rm{Ch}}$) (see Eq.~\ref{eq:Mch}). For a decreasing $X_{\mathrm{C}}$, the central region where electrons are degenerate during core carbon burning is more extended in mass. This can be understood as a symptom of the overall stronger core contraction experienced by models that are more neutrino-dominated. For the model with $X_{\mathrm{C}}=0.28$ (Fig.~\ref{fig:eta_contours}a), electron degeneracy plays a small role during central carbon burning. The C-burning front (at the base of the convective carbon burning shell) remains significantly below the values of the classical  and effective Chandrasekhar mass. For the compactness peak model ($X_{\mathrm{C}}=0.23$, Fig.~\ref{fig:eta_contours}b), degeneracy pressure helps support the core at the end of core carbon burning, reaching 15\% of the total pressure in the center ($\eta_{c}=0.93$) before it is lifted during core oxygen burning. The carbon burning front remains close to, but below $M_{\rm{Ch},0}$ and $M_{\rm{Ch}}$ during core carbon burning. For the $X_{\mathrm{C}}=0.17$ model (Fig.~\ref{fig:eta_contours}c) the region where electrons are at least partially degenerate extends up $M_{\rm{Ch},0}$. The carbon-burning front exceeds $M_{\rm{Ch},0}$ and even reaches the value of $M_{\rm{Ch}}$.  This appears to help accelerate the core contraction, which is somewhat surprising since the Chandrasekhar mass is known to mainly play a role at full degeneracy, and deserves further study. 

To understand if this effect applies more generally, we investigate the behavior of a larger set of models with varying $X_{\mathrm{C}}$. In Fig.~\ref{fig:mult_xC}c, we show the duration of the phase between core carbon depletion and neon ignition as a function of $m_{C}$, which is defined as the maximum mass coordinate of the carbon-burning front $m_{C}$ during core carbon burning (i.e., the base of the furthest convective carbon-burning shell)\footnote{We note that the computed timescale is somewhat sensitive to the definitions used for core carbon depletion (here, when the central carbon abundance is below $10^{-4}$) and neon ignition (here, when the central neon abundance has decreased by 2\% after core carbon burning). However, the general trends remain, in particular the acceleration of the core contraction for models with $m_{C} > 1.6 \Msun$. }. The changes in this timescale are very close to the compactness pattern shown in Fig.~\ref{fig:mult_xC}b as a function of decreasing core carbon abundance. A similar trend in this timescale was also noted by \citet{chieffi_impact_2021}. At the end of core carbon burning, all models reach the same value of the classical Chandrasekhar $M_{\rm{Ch},0}=1.455\Msun$, while the effective Chandrasekhar mass varies slightly between models due to their different central entropy, with values of $M_{\rm{Ch}}\sim 1.82 - 1.84 \Msun$. Only in the model with the lowest core carbon abundance ($X_{\mathrm{C}}=0.17$, shown in more detail in Fig.~\ref{fig:eta_contours}c), $m_{C}$ exceeds $M_{\rm{Ch}}$.

From the variation of the timescale between core carbon depletion and core neon ignition, we derive the following interpretation. Models with $X_{\mathrm{C}} < 0.25$ ($m_{C} > 1.2 \Msun$) are strongly neutrino-dominated and burn carbon radiatively in the core, experiencing a stronger contraction than those with larger $X_{\mathrm{C}}$ that still burn carbon convectively in their cores. However, degeneracy pressure helps to counter the core contraction towards the end of core carbon burning and to delay the ignition of core neon burning. For models where $m_{C} > 1.5 \Msun$, the entire region below $M_{\rm{Ch},0}$ is at least partially degenerate and degeneracy pressure no longer helps to counter the core contraction, leading to an earlier core neon ignition. 
\citet{sukhbold_compactness_2014,sukhbold_high-resolution_2018}, and \citet{sukhbold_missing_2020} noted a smaller size of the oxygen-burning shell once the base of the carbon-burning shell $m_{C}$ exceeds $M_{\rm{Ch}}$ and linked it to the role of degeneracy. It is noteworthy that we observe the same phenomenology here (see Fig.~\ref{fig:eta_contours}) despite the different assumptions and methods we use. As we discuss in Sect.~\ref{sec:results:stars:regD}, the same mechanism helps explain the second drop in compactness (region D in Fig.\ref{fig:xi_overview}) though in this case it involves the oxygen-burning front and an earlier core Si ignition.

\subsection{Shell mergers}
\label{sec:results:experiment:shell_mergers}

\begin{figure*}
	\centering
	\includegraphics[width=\textwidth]{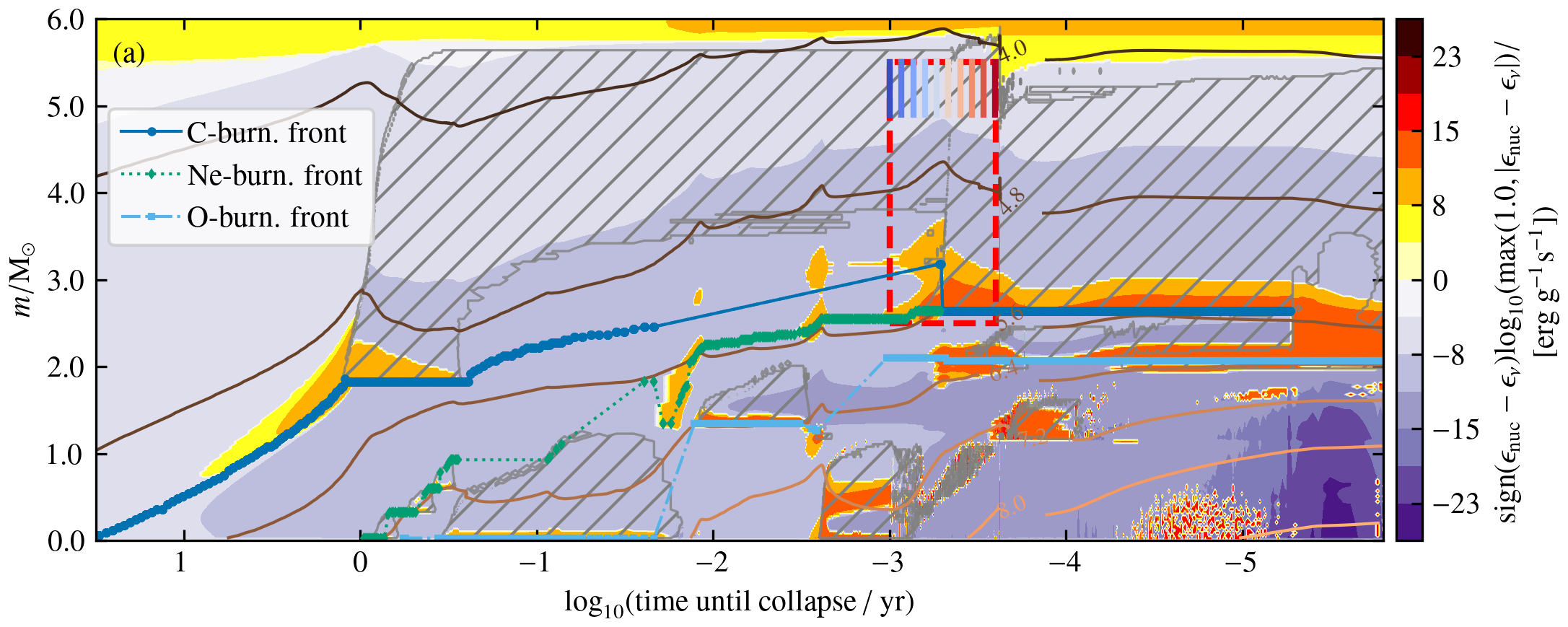}
	\includegraphics[width=\textwidth]{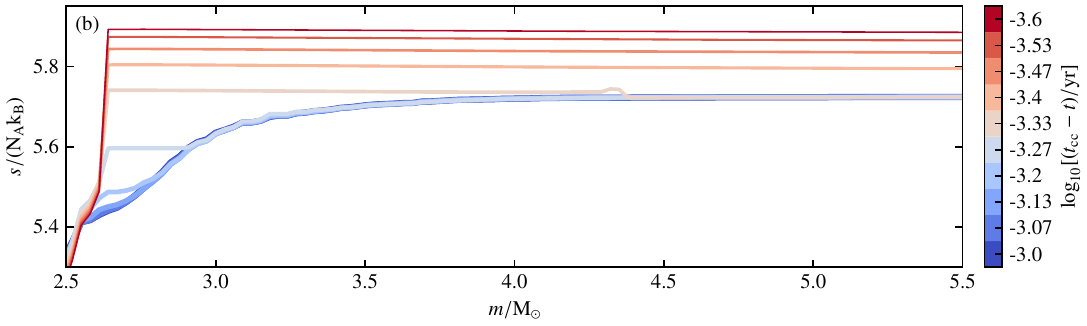}
	\caption{Top: Kippenhahn diagram showing the final evolution of the inner stellar structure of the model with $X_{C} = 0.17$. Colors indicate the regions dominated by nuclear burning or neutrino losses and gray hatched regions indicate convective areas. Brown lines show contours of constant (logarithmic) density. Colored markers trace the burning front for C, Ne, and O burning. The red box highlights the mass and time range shown in the bottom panel. Small colored vertical lines indicate the times at which the entropy profile are shown in the bottom panel. Bottom panel: Entropy profiles in the mass and time range highlighted in the red box in the top panel. The entropy in the neon burning region ($2.5 - 2.9\Msun$) gradually increases until it exceeds the entropy of the C-burning layers above, triggering a merger between these regions.}
	\label{fig:sh_mergers_3cag}
\end{figure*}
In our experiment, the model with the lowest initial carbon abundance $X_{\mathrm{C}} = 0.17$ experiences a merger of the C- and Ne-burning layers shortly after core Si burning. This is highlighted within the red rectangle in the Kippenhahn diagram (Fig.~\ref{fig:sh_mergers_3cag}a). The location of the C-burning and Ne-burning shells are indicated with markers. The origin of this shell merger can be understood by inspecting Fig.~\ref{fig:sh_mergers_3cag}b, which shows the entropy profile in the highlighted region. The entropy in the layers corresponding to the location of the Ne-burning shell (2.5 to 2.9\Msun) rapidly increases. This is because after Si-burning ignites, a rapid contraction of the core below the Ne-burning front leads to enhanced Ne burning, which generates entropy. Eventually, the entropy of these layers exceeds that of the carbon-burning layers above (see the entropy profile at a time of $\log_{10}[(t_{\rm{cc}} - t) / \mathrm{yr}] = -3.33$). This entropy contrast means that these layers are unstable against convection (Schwarzschild criterion). No convective boundary mixing is assumed in these models after core helium burning, which means that the mixing is solely caused by this entropy change. In Appendix~\ref{sec:appendix:res_test}, we perform a resolution test and find that this shell merger is barely affected by numerical uncertainties. The mixing leads to the merging of these shells and to a new configuration of the stellar structure.\\
After the merger, the Ne/C burning shell suddenly drops to a lower mass coordinate and a large convective zone develops above. These layers suddenly expand, leading to a dramatic density drop in these regions (see the density contours in Fig.~\ref{fig:sh_mergers_3cag}a). This particular model additionally experiences a merger of the Ne-burning and O-burning shells (visible at a mass coordinate of 2\Msun at $\log_{10}[(t_{\rm{cc}} - t) / \mathrm{yr}] = -5.2$). These mergers lead to a low final mass coordinate of the C/Ne/and O-burning fronts, eventually limiting the development of the Si-burning front, and thus setting a limit to the maximum mass of the iron-rich core.

\section{Final core structures of actual stellar models}
\label{sec:results:stars}
In the previous section, we have identified key mechanisms that lead to a change in the final core structure through an experiment in which we varied the core carbon abundance for a fixed core mass. We now apply the insights gained to more realistic simulations of massive stars.

The initial conditions for central carbon burning are set by the core mass, which determines the central density and temperature, and by the initial carbon abundance. Hydrostatic equilibrium implies a characteristic trend in the central density and temperature of stars as a function of mass \citep[e.g.,][]{kippenhahn_stellar_2013}. While their central temperatures systematically increase with mass, their central densities decrease. The central carbon mass fraction at core helium depletion decreases systematically as a function of mass. This is a signature of the increased importance of the $^{12}\mathrm{C}(\alpha,\gamma)^{16}\mathrm{O}$ reaction rate with respect to the triple-alpha process at the end of core helium burning for cores of higher mass and lower density, which effectively leads to an increased destruction of carbon \citep[e.g.,][]{arnett_advanced_1972}. For increasing masses, the higher central temperatures, lower central densities, and lower central carbon abundance imply stronger neutrino losses (see Sect.~\ref{sec:results:experiment:c_burn}), with important consequences for the final structure. For reference, a compilation of the changes in interior structure post core helium burning for all models is shown in the Appendix, Figs.~\ref{fig:eps_grav_overview_17_20.5}, \ref{fig:eps_grav_overview_21_25.5}, \ref{fig:eps_grav_overview_26_29}, \ref{fig:eps_grav_overview_30_37}, \ref{fig:eps_grav_overview_38_45}, and \ref{fig:eps_grav_overview_46_50}. 

\subsection{Origin of the increase in final compactness (region A)}
\label{sec:results:stars:regA}
\begin{figure}
	\centering
	\includegraphics[width=0.5\textwidth]{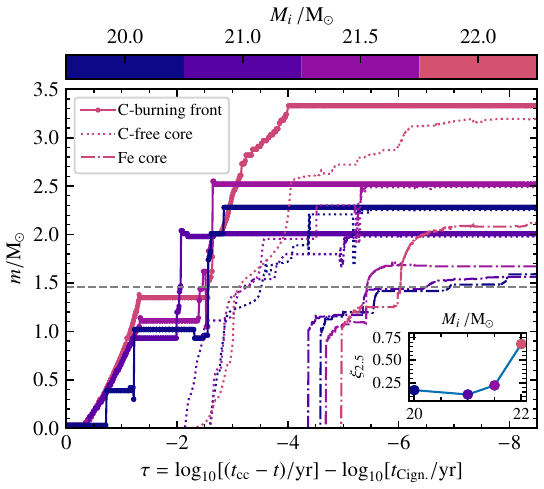}
	\caption{Evolution of the C-burning front, the C-free core (which follows the progression of neon burning), and the iron-rich core as a function of time from carbon burning to core collapse for models before and up to the compactness peak (region A of Fig.~\ref{fig:xi_overview}, see inset). As carbon burning becomes increasingly neutrino-dominated, the C-burning front moves further out in mass, ultimately leading to the formation of a larger C-free and iron-rich core.}
	\label{fig:burning_front_increase}
\end{figure}
\begin{figure}
    \centering
    \includegraphics[width=0.5\textwidth]{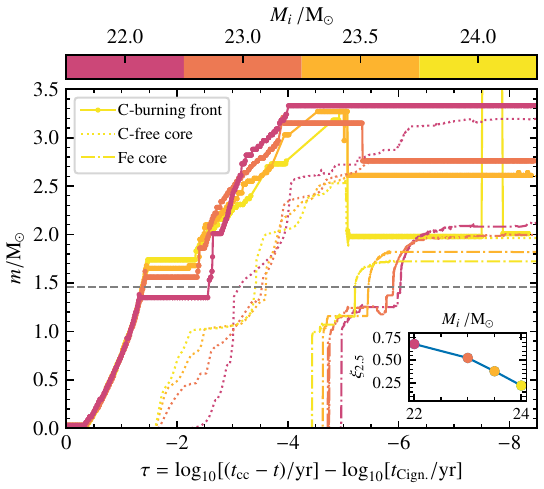}
    \caption{Same as Fig.~\ref{fig:burning_front_increase} for models beyond the compactness peak (region B in Fig.~\ref{fig:xi_overview}, see inset). As a function of initial mass, neon burning, ignites systematically earlier (as traced by the C-free core mass) and slows down the progression of the C-burning front. Shell mergers between the C/Ne/O-burning fronts are responsible for the sudden drops in the C-burning front at $\tau\approx -5$. Ultimately, for increasing masses, these models form systematically smaller iron-rich cores.}
    \label{fig:burning_front_decrease}
\end{figure}

We first focus on stars at the lower-mass end before the compactness peak (region A in Fig.~\ref{fig:xi_overview}). The change in interior structure as a function of mass is summarized in Fig.~\ref{fig:burning_front_increase}, in which we show the progression of the C-burning front, the C-free core mass (which follows the progression of neon burning), and the iron core mass as a function of time.

In the lowest-mass model (20\Msun) core carbon burning is convective initially. The C-burning front remains at the center until the carbon fuel is exhausted (at $\tau \approx -0.75$) and the C-burning front moves further out in mass until the conditions for convective core carbon burning are reached once again. Because of the additional fuel from convective core carbon burning, the progression of the C-burning front proceeds somewhat differently in this 20\Msun model compared to the higher-mass models. The carbon-burning phase lasts longer and as a result, neon burning ignites later while the C-burning front is still moving out in mass (as shown by the growth of the C-free core at $\tau \approx -0.75$, dotted lines in Fig.~\ref{fig:burning_front_increase}). Eventually, the C-burning front reaches a higher value (2.3\Msun) than in the 21\Msun model, which proceeds radiatively. At the end of the evolution, the model forms a relatively low-mass iron core of 1.59\Msun.

The transition from convective to radiative core carbon burning has previously been invoked as the origin of patterns in compactness or iron-core mass \citep{brown_formation_1999,sukhbold_compactness_2014,sukhbold_missing_2020,chieffi_presupernova_2020, takahashi_monotonicity_2023}. However, we can clearly see in Fig.~\ref{fig:burning_front_increase} that the 21\Msun model, which burns carbon radiatively, does not immediately develop a larger iron-core mass (1.56\Msun) than the 20\Msun model (1.59\Msun). This transition in the energy transport mechanism from convection to radiation can be understood as a symptom for neutrino losses becoming increasingly dominant, but it is not the cause of the increase in compactness and iron core mass.

For increasing masses, the models experience larger neutrino losses during core carbon burning because they have higher central temperatures, lower densities, and a lower central carbon abundance (see Sect.~\ref{sec:results:experiment:c_burn}). The neutrino losses imply that the layers above the burning front can contract more significantly than when neutrino losses are not dominant. This causes an overall more significant and earlier core contraction in these stars, which is reflected in the location of the burning fronts. The smaller amount of carbon fuel, together with the increased nuclear energy generation rate in higher-mass stars accelerate the outward progression of the C-burning front. For example, as shown in Fig.~\ref{fig:burning_front_increase}, at the end of the evolution, the C-burning front reaches 2.3\Msun in the 21\Msun model, while it reaches 3.35\Msun in the 22\Msun model (which corresponds to the compactness peak). This progression of the C-burning front halts when the energy generated by carbon burning greatly exceeds the neutrino losses and convection sets in. For the compactness-peak model (22 \Msun) the C-burning front reaches a particularly high final mass coordinate. This is because most of the carbon in the large convective shell that forms during core carbon burning is already exhausted by the time of core neon ignition ($\tau\approx -2.1$), and the carbon-burning front  quickly burns through the remaining fuel. A further progression of the C-burning front is a symptom of a larger core contraction which results in the growth of a systematically more massive C-free core as a function of initial mass (dotted lines in Fig.~\ref{fig:burning_front_increase}). In all models, the C-free core mass grows almost up to the final value of the C-burning front. Consequently, the O- and Si-burning fronts also move further out in mass and eventually, a more massive iron-core grows in the higher-mass models (dash-dotted line in Fig.~\ref{fig:burning_front_increase}).

\subsection{Origin of the decrease in final compactness (region B)}
\label{sec:results:stars:regB}
For models in region B, the early behavior of the carbon-burning front is similar to the neutrino-dominated models in region A (see Fig.~\ref{fig:burning_front_decrease}). For higher-mass models, the C-burning front moves systematically further out in mass (e.g. reaching $m\approx1.35\Msun$ for the 22\Msun model, compared to $m\approx1.6\Msun$ for the 24\Msun model). However, around the time of core carbon depletion ($\tau \approx$ -2 Fig.~\ref{fig:burning_front_decrease}), these models experience a clear reversal of the trends observed in region A. The convective carbon-burning episode at $\tau\approx -2$, where the C-burning front remains at a fixed mass coordinate, is systematically shorter. Afterward, the C-burning front moves out systematically slower in mass for increasing masses and reaches lower final values (e.g., $m\approx3.35\Msun$ for the compactness-peak model with an initial mass of 22\Msun compared to $m\approx2.28\Msun$ for the 25\Msun model). Consequently, the final iron-core mass is smaller in the higher-mass models compared to the lower-mass ones.

\begin{figure*}
	\centering
	\includegraphics[width=\textwidth]{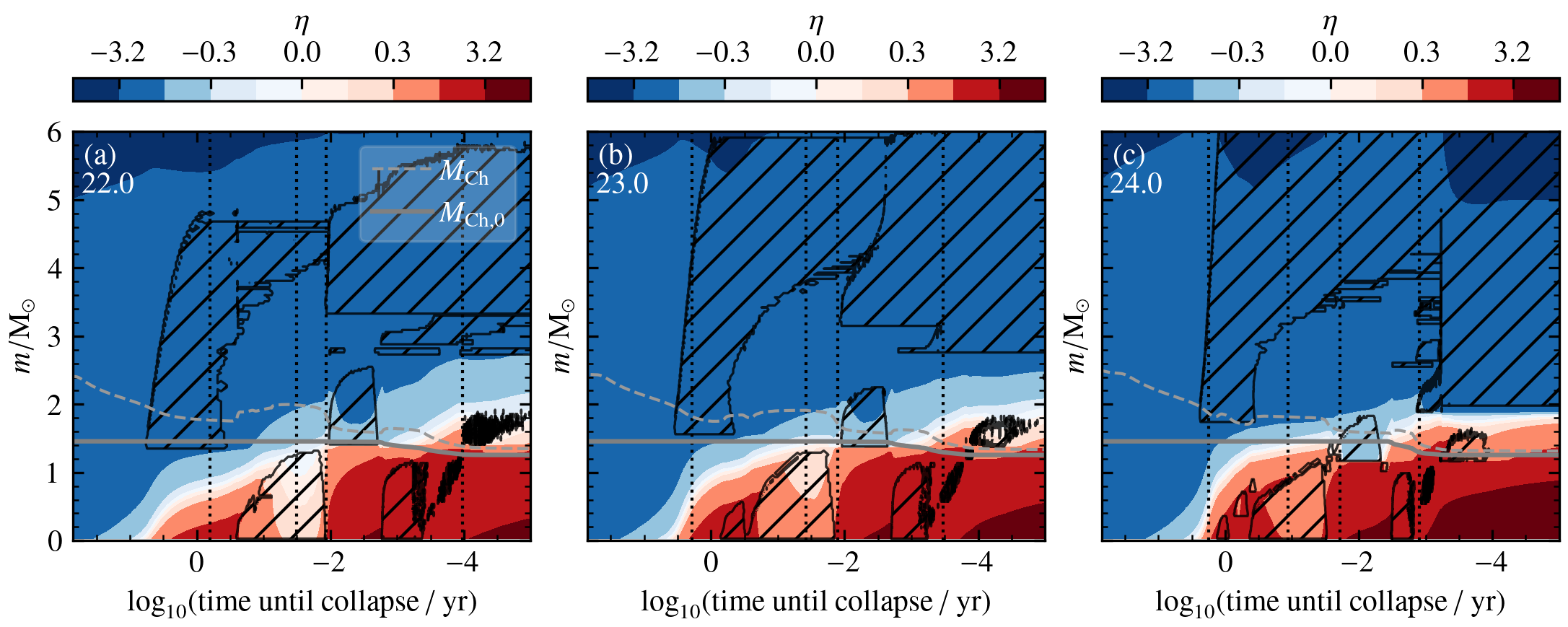}
	\caption{Same as Fig.~\ref{fig:eta_contours} for three stellar models of our grid after the first compactness peak.}
	\label{fig:eta_1st_decrease}
\end{figure*}

The observed change in final compactness and iron core mass can be traced back to the same mechanisms as outlined in Sect.~\ref{sec:results:experiment}. Higher neutrino losses in higher-mass models and the ensuing contraction of the core, together with the low initial core carbon abundance and shorter duration of core carbon burning, mean that the conditions for core neon burning to occur are met systematically earlier than in lower-mass models. This can be observed by the timing of the growth of the C-free core (dotted lines in Fig.~\ref{fig:burning_front_decrease}), which occurs systematically earlier in higher-mass models. Additionally, as shown in Fig.~\ref{fig:eta_1st_decrease}, the extent of the degenerate region within the core increases and helps accelerate the core contraction in models where the C-burning front (at the base of the C-burning shell) exceeds the classical Chandrasekhar mass. 
This earlier central neon ignition increases the core luminosity which slows down the core contraction above the neon-burning front. This causes the observed slower increase in the C-burning front which implies a slower growth of the C-free core.

In addition, central neon burning is more neutrino-dominated for higher mass-models, and leads to a smaller increase in entropy during these burning phases. After core Silicon ignition (as traced by the growth of the iron-rich core, dash-dotted lines in Fig.~\ref{fig:burning_front_decrease}), we observe a phase during which neon and oxygen shell burning become more energetic for higher masses, leading to a higher entropy, which ultimately causes shell mergers just like in our previous experiment (see Sect.~\ref{sec:results:experiment:shell_mergers}). These shell mergers are found for all models in region B beyond the compact peak and can be identified by a sudden drop in the C-burning front shortly after core silicon burning sets in and the iron-core mass starts to increase (e.g. at $\tau\approx 5$ in the 24\Msun model). In most cases, this involves a merger of the neon-burning and carbon-burning shells. In the highest mass models ($M_{i} > 23\Msun$), the oxygen and Ne-burning shell merge additionally. The shell merger implies a reconfiguration of the stellar structure and a large expansion of the layers above the C/O burning shells (see Fig.~\ref{fig:eps_grav_overview_21_25.5}). The lower final location of the C and O-burning fronts after the shell mergers (e.g., $m\approx2.28\Msun$ for the 25\Msun model) limit the progression of the Si-burning front. Consequently, the iron-core mass of these higher-mass models remains small.

\subsection{Origin of the second compactness increase (region C)}
\label{sec:results:stars:regC}
\begin{figure}
	\centering
	\includegraphics[width=0.5\textwidth]{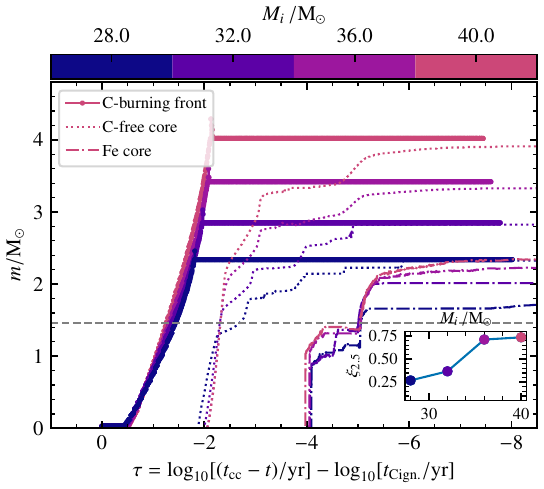}
	\caption{Same as Fig.~\ref{fig:burning_front_increase} for models in the mass range of the second compactness increase (region C in Fig.~\ref{fig:xi_overview}, see inset). In these models, both carbon and neon burning are fully neutrino-dominated.}
	\label{fig:burning_front_2nd_increase}
\end{figure}
\begin{figure}
	\centering
	\includegraphics[width=0.5\textwidth]{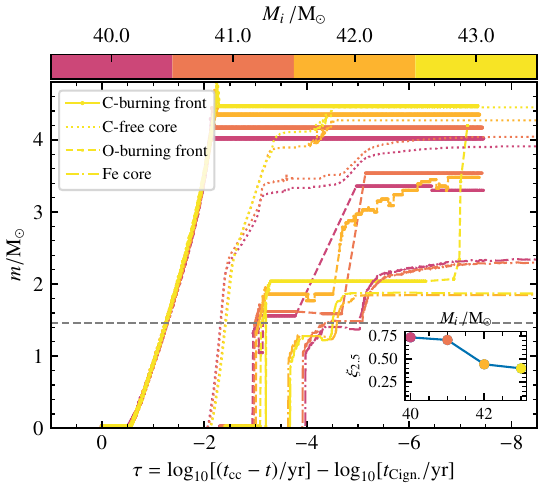}
	\caption{Same as Fig.~\ref{fig:burning_front_increase} for models in the mass range of the second decrease in compactness (region D in Fig.~\ref{fig:xi_overview}, see inset). The dashed line traces the O-burning fronts. For increasing masses, the progression of the O-burning front is slowed by a systematically earlier ignition of core silicon burning (as traced by the iron core mass).}
	\label{fig:burning_front_2nd_decrease}
\end{figure}
For models between 25 and 28\Msun, neutrino losses become increasingly important as the central temperature increases and the initial central carbon abundance decreases further. The core contracts even further during core carbon burning and the carbon-burning front moves further out in mass. Core neon burning ignites even earlier and becomes more neutrino-dominated, which means neon burning proceeds in smaller convective zones (see Fig.~\ref{fig:eps_grav_overview_26_29}). These models are characterized by a large convective oxygen-burning shell that temporarily stops the progression of the carbon-burning front after it reaches its initial location. The final compactness and iron-core mass of these models remains low (see Fig.\ref{fig:xi_overview}).

After an initial mass of about 27\Msun, the final compactness increases again (see Fig.\ref{fig:xi_overview}). This is due to core neon burning becoming fully neutrino dominated. This is best seen in the interior structure evolution of the models with initial masses from 27\Msun to 32\Msun (Fig.~\ref{fig:eps_grav_overview_26_29}). For higher initial masses, neon burning changes from a succession of multiple small convection zones (four in the 27\Msun model), to one or two tiny (< 0.2\Msun) convective zones (27.5--30\Msun) followed by radiative burning, before eventually transitioning to fully radiative, neutrino-dominated neon burning (from models of 31\Msun, where convective core neon burning regions completely disappear, see Fig.~\ref{fig:eps_grav_overview_30_37}). Once again, this transition from convective to radiative burning does not immediately lead to a change in the final compactness. Rather, the compactness gradually increases for fully neutrino-dominated models as the core contracts further to compensate the energy loss due to neutrinos.

In Fig.~\ref{fig:burning_front_2nd_increase} we summarize the changes in the stellar structure for models in region C. All models show a very similar trend as in the neutrino-dominated models in region A. As the neutrino losses increase and the core contracts further, the C-burning front moves further out in mass until it ignites a convective-burning shell. In these models, neon burning ignites even earlier than in region B (as shown by the progression of the C-free core at $\tau \approx -2$, dotted line in Fig.~\ref{fig:burning_front_2nd_increase}). However, for these masses, it becomes neutrino-dominated, which means that the contraction of the layers above the Ne-burning front cannot be slowed significantly by this early ignition of neon. Furthermore, as the initial carbon abundance decreases, so does the neon abundance after core carbon burning, which shortens the duration and impact of core neon burning. Thus core neon ignition barely affects the progression of the C-burning front and neon shell burning no longer leads to a high enough increase in entropy to cause shell mergers with the C-burning layers. The Ne-burning front moves systematically further out in mass for higher-mass models as the core contracts and the C-free core almost reaches the C-burning front (e.g., 2.09\Msun in the 28.0\Msun model, and 3.9\Msun in the 40.0\Msun model, see dotted lines in Fig.~\ref{fig:burning_front_2nd_increase}). Eventually, for increasing masses, the Si-burning front can also move further out in mass and a systematically larger iron-core mass forms (e.g., 1.7\Msun in the 28.0\Msun model compared to 2.35\Msun in the 40\Msun model, see dash-dotted lines in Fig.~\ref{fig:burning_front_2nd_increase}).

\subsection{Origin of the second compactness decrease (region D)}
\label{sec:results:stars:regD}

\begin{figure*}[!ht]
	\centering
	\includegraphics[width=\textwidth]{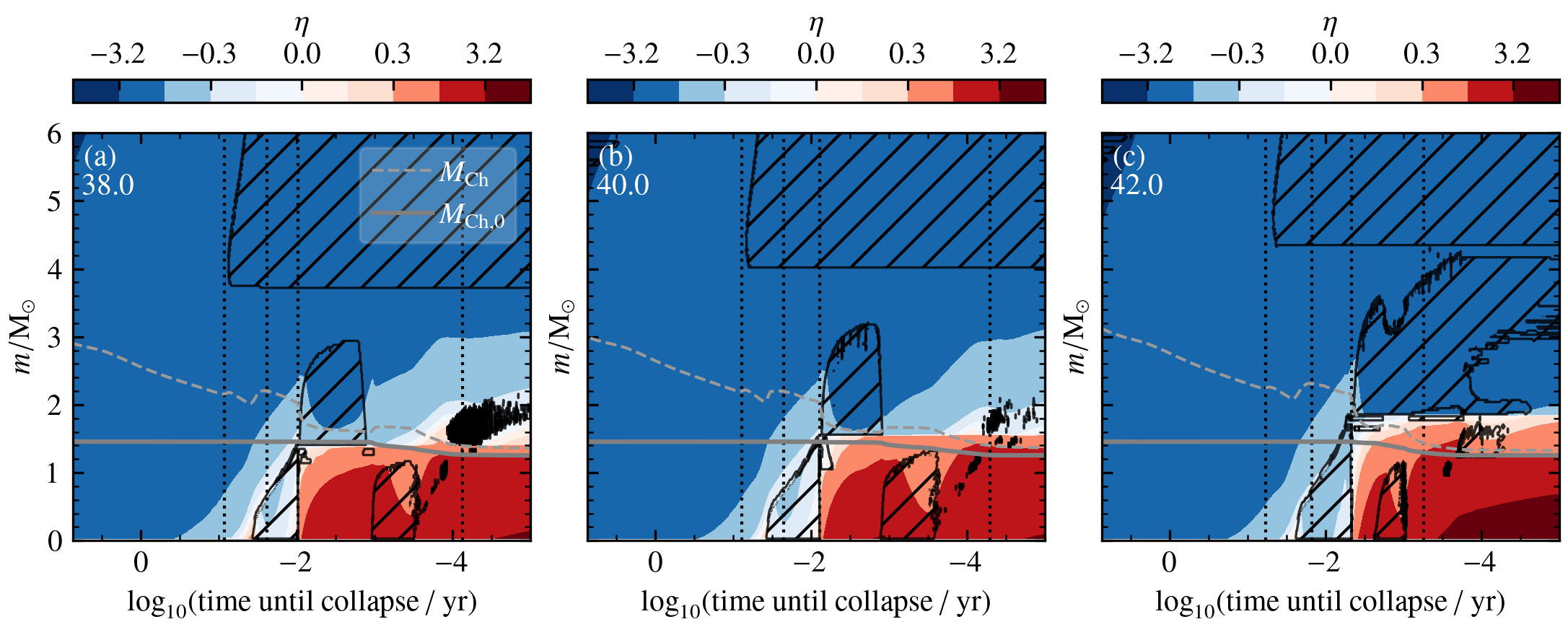}
	\caption{Same as Fig.~\ref{fig:eta_contours} for three stellar models of our grid after the second compactness increase. Here, neon burning is fully neutrino-dominated and radiative, and not visible in the figure. After convective core oxygen burning (hatched region in the center) the degenerate core region reaches up to the oxygen burning front (at the base of the convective oxygen-burning shell) for all models. Panel c: When this front exceeds the (effective) Chandrasekhar mass, the degenerate core contracts and core silicon burning ignites earlier, slowing the contraction of the layers above and affecting the extent of the convective region.}
	\label{fig:eta_2nd_decrease}
\end{figure*}
For models beyond 40\Msun, we once again observe a drop in the final compactness (see Fig.~\ref{fig:xi_overview}). As best observed in Fig.~\ref{fig:eps_grav_overview_38_45}, in these models, both carbon and neon are neutrino-dominated and the respective nuclear burning fronts move out so far in mass from the central regions that they no longer play a large role in setting the final iron core mass (see Fig.~\ref{fig:burning_front_2nd_decrease}). Instead, the core oxygen and silicon-burning phase increase in importance, as shown in Fig.~\ref{fig:eta_2nd_decrease}. Oxygen burning produces a large amount of energy per gram and occurs convectively \citep{woosley_evolution_2002}. In analogy to the effects observed for the C- and Ne-burning front in lower-mass models, after core oxygen burning the oxygen-burning front initially moves further out in mass the more neutrinos dominate in the center (as indicated in Fig.~\ref{fig:burning_front_2nd_decrease})). The timescale between core oxygen and core silicon ignition shortens for increasing masses (as shown by the timing of the growth of the iron-rich core, dash-dotted lines in Fig.~\ref{fig:burning_front_2nd_decrease}). The (at least partially) degenerate core region becomes more extended, as it generally reaches up to the base of the oxygen-burning front (see Fig.~\ref{fig:eta_2nd_decrease}). The core contraction accelerates for models in which the oxygen-burning front reaches a mass coordinate higher than the (classical) Chandrasekhar mass. For example, in the 42\Msun model, the oxygen-burning front (at the base of the oxygen-burning shell) even reaches the effective Chandrasekhar mass (indicated by a dotted line in Fig.~\ref{fig:eta_2nd_decrease}c) with $m\sim 1.85\Msun$ at $\log(t/\rm{yr}) =-2.5$. When silicon burning occurs during the oxygen shell-burning phase, it slows down the contraction of the layers above, and with it the progression of the oxygen-burning front. Eventually, this limits the maximum progression of the Si-burning front, which forms a lower iron-core mass than in models in which core silicon burning occurs later. For example, as shown in Fig.~\ref{fig:eta_2nd_decrease}, in the 43.0\Msun model the oxygen-burning front (dashed lines) remains at a mass coordinate of 2.1\Msun after core silicon burning ignites (as traced by the growth of the iron-rich core, dash-dotted lines) . It only moves further out in mass shortly before core collapse ($\tau = -7$), which does not leave enough time for the iron-core mass to grow further in mass than 1.85\Msun in this model. 

Beyond region D, for models with higher masses the final compactness remains lower than during the second peak (see Fig.~\ref{fig:xi_overview}a) but above a value of 0.4. The 44\Msun model has an unusually high compactness (0.7) compared to the other models with initial masses larger than 40\Msun. We link this back to the oxygen-burning front remaining at a rather low mass coordinate (close to 1.45 \Msun, not exceeding the effective Chandrasekhar mass) after core oxygen burning, triggering a small (0.1\Msun) convective episode (see Fig.~\ref{fig:eps_grav_overview_38_45}). The core contracts less during core oxygen shell burning than for the adjacent models, which means silicon burning ignites later and does not slow down the progression of the oxygen-burning front as effectively, leading to a larger final compactness and iron core mass. Small changes in the energy generation rate or mixing could influence the existence of this small oxygen-burning episode and we therefore consider this model not to be representative of the more general patterns described here. We note that the emergence and disappearance of such small secondary peaks in the final structure landscape can be observed when physical assumptions are varied, such as the efficiency of semiconvection (see Appendix~\ref{sec:appendix:sc}) and deserve further study. Despite the observed variations, it should be noted that for all these models, the binding energy above the inner core is so high (see Fig.~\ref{fig:xi_overview} d) that the formation of a BH is expected even if a successful explosion were triggered in the core.

\section{Global picture}
\label{sec:results:global_picture}

\begin{figure*}[h]
	\centering
	\includegraphics[]{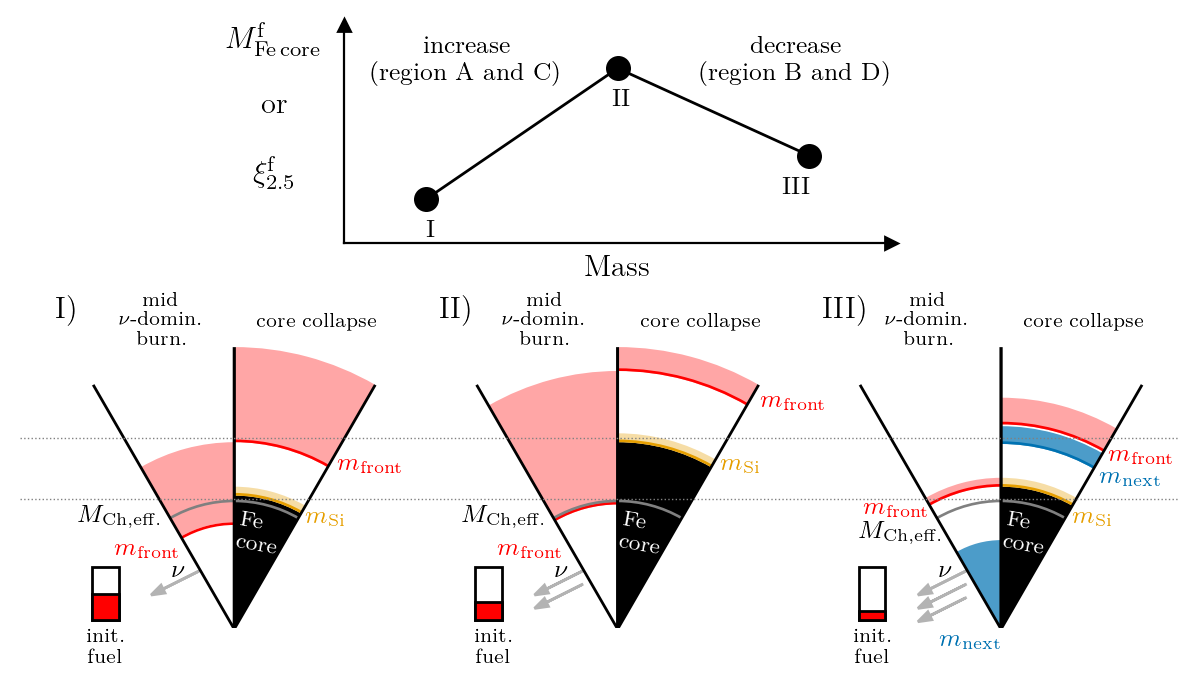}
	\caption{Schematic representation of the mechanisms leading to the emergence and decline of peaks in final iron core mass and compactness. Horizontal gray dotted guidelines are included to aid the comparison. \textit{Star I: Before peak} Schematic stellar structures of a star with a low final iron core mass at two points in the evolution represented by two adjacent wedges. The initial mass fraction of the fuel is still high (see red area in the rectangle) but the main burning stage is neutrino-dominated (arrows). Because neutrinos take away energy, the core contracts and the burning front (red line) moves out in mass as the burning progresses. The large amount of fuel prevents it from moving far out before a convective zone forms above the burning front once the energy generated is high (shaded red area). Ultimately (right wedge), it grows a small fuel-free core and the silicon-burning front (yellow line) cannot move far out in mass, forming a small iron core (black area). \textit{Star II: Peak} With a higher mass and lower initial fuel mass fraction, the burning is even more neutrino dominated and a strong contraction occurs. Hence the burning front moves further out in mass than in I, but stays just below the effective Chandrasekhar mass. Aided by degeneracy pressure support, the star burns through almost all available fuel before the next burning stage ignites, after which it quickly moves out in mass, growing a large fuel-free core and eventually a large iron core. \textit{Star III: Beyond peak:} For an even lower initial fuel abundance and higher mass, the burning is even more neutrino dominated and the core contraction accelerates. The burning front moves further out and exceeds the effective Chandrasekhar mass. This triggers a fast contraction of the partially-degenerate core and the next burning stage (blue area) ignites early. This next stage suppresses nuclear burning at the front above. As a result, the burning front moves out in mass slowly and eventually (right wedge) the star grows a low-mass iron core.}
	\label{fig:cartoon}
\end{figure*}

Based on our findings, we can derive a global picture of the physical processes that determine the final core structures of massive stars. After core helium burning, thermal neutrino emission becomes important and complicates the final evolution by taking away energy from the interior, leading to an acceleration of the evolution. The transition from convective to of radiative carbon and neon burning is a symptom of neutrino losses becoming dominant, but not the cause of the observed changes in the final structure patterns. For higher masses (lower densities, higher temperatures, and lower central carbon abundance), neutrino losses become increasingly important and affect the efficiency and timing of nuclear burning episodes and with it the associated core growth.

We identified physical mechanisms explaining the emergence and decline of the two prominent ``compactness peaks'' (see Fig.~\ref{fig:xi_overview}) found in several independent studies, i.e. mass ranges for which stars develop a high final compactness, iron core mass, central entropy, and binding energy. These are summarized schematically in Fig.~\ref{fig:cartoon}, where we show how neutrino-dominated nuclear burning in the cores of massive stars post core helium depletion leads to different stellar structures. These outcomes are ``written'' in the stellar cores at core helium depletion, when the initial burning conditions of the subsequent burning phases are set.

\subsection{Increase in final iron core mass (Region A and C):}
For increasing stellar masses, the central nuclear burning source (C or Ne) becomes increasingly neutrino-dominated. As observed in multiple studies  \citep[e.g.][]{brown_formation_2001,sukhbold_compactness_2014,sukhbold_high-resolution_2018,sukhbold_missing_2020,chieffi_presupernova_2020,takahashi_monotonicity_2023}, energy transport at the center transitions from convection in a large central region followed by shell burning, to ever smaller and more numerous successive convective burning episodes, until it becomes fully radiative for stars in which nuclear burning is fully neutrino-dominated. In Fig.~\ref{fig:cartoon} I, the structure of a star with a neutrino-dominated, radiative core is shown schematically mid burning (left wedge) and at the onset of core collapse (right wedge). The decreasing amount of fuel together with the neutrino dominance lead to an acceleration of the core contraction and to a faster and further outward progression of the burning front (the point where the maximum nuclear energy generation rate is reached, red lines in Fig.~\ref{fig:cartoon} I) and associated growth of the C/Ne-free core below. When the burning front reaches layers with a high C/Ne abundance, the generated nuclear energy becomes large enough to significantly exceed the neutrino losses, and convection sets in above the front (shaded red area in Fig.~\ref{fig:cartoon} I). The star with the structure I still contains a large amount of fuel, which means (a) that a small core contraction is sufficient for convection to set in and (b) that the newly formed convective region contains enough fuel to remain until core collapse without the need of a large core contraction. This limits the growth of the fuel-free core below as the burning front moves only slowly out in mass. Ultimately (right wedge in Fig.~\ref{fig:cartoon} I), this structure grows a relatively small fuel-free core, which in turn limits the growth of the Si-rich core, which then forms a relatively low-mass iron-rich core.

The final location of the burning front reflects the contraction of the inner core and sets the maximum core growth for the subsequent nuclear-burning episodes. Through this process, for increasing masses, stars in which the central burning source becomes neutrino-dominated grow increasingly massive C/Ne-free cores, which in turn form massive oxygen-free cores and eventually, more massive iron-rich cores. For higher masses, as the cores contract further and their density increases, electron degeneracy also starts to play a role. Degeneracy pressure contributes to supporting the core during the main burning phase (carbon or oxygen, depending on the mass range), delaying the ignition of the next burning phase as long as the burning front does not exceed the effective Chandrasekhar mass.

At the compactness peak (structure II in Fig.~\ref{fig:cartoon}), nuclear burning at the center is strongly neutrino-dominated due to a higher mass (lower initial density) and low initial fuel mass fraction. The strong core contraction leads to an outward progression of the burning front until it nearly reaches the effective Chandrasekhar mass. Support from degeneracy pressure helps slow the contraction during the main burning phase such that nearly all the fuel in the convective region is burned before the next burning stage ignites in the center. Afterward, the burning front quickly burns through the former convective region as the core contracts further. It grows a particularly massive fuel-free core and eventually, a massive iron core.

\subsection{Decrease in final iron core mass (Region B and D)} 
For even higher masses, stars develop structures that are even more neutrino-dominated (star III in Fig.~\ref{fig:cartoon}). Two mechanisms, (a) a lower initial fuel abundance decreasing the duration of the main fuel burning phase (core C-burning or Ne-burning), and (b) electron degeneracy no longer helping support the stellar structure once the main burning front and the fuel-free core it produces exceed the (effective) Chandrasekhar mass, both lead to an acceleration of the core contraction. What follows is an early ignition of the next burning fuel while the main nuclear burning phase is still in progress (e.g. Ne ignition during C burning or Si ignition during Ne/O burning, indicated by the blue central region in Fig.~\ref{fig:cartoon} III). The luminosity of this new nuclear energy source contributes to slowing down the contraction of the layers above and prevents the further growth of the C/O-free core. Consequently, the Si-burning front cannot move far out in mass and grows a lower-mass iron core than in star II.

Additionally, in particular in region B, mergers of the Ne- and C-burning shells, and, in some cases, of the Ne and O-burning shells, can occur due to the high energy generation rate and entropy at these fronts. These lead to a reconfiguration of the stellar structure in which the burning fronts all merge together at the lowest mass coordinate. If these mergers occur before the end of core silicon burning, they limit the progression of the silicon-burning front, and, with it, the growth of the iron-rich core. Eventually, stars that experience such shell mergers form iron cores of even lower mass. Shell mergers before the end of core Si burning are thus an additional mechanism which produces final stellar structures with low iron core masses.

Because of these physical mechanisms, we expect a non-monotonic landscape of final compactness, iron core mass, central entropy, and binding energy as a function of the initial and core masses of massive stars. Based on a neutrino-driven supernova model we have shown that stars with high compactness, iron core masses and binding energy are predicted to form black holes. Hence, the existence of these robust final structure patterns of massive stars observed by multiple independent studies imply the formation of black holes at characteristic mass ranges.

\section{Discussion}
\label{sec:discussion}

\subsection{Uncertainties in stellar physics}
Our stellar evolution models are subject to uncertainties concerning the physics of massive stars that can affect the lanscape of the final structure of stars, and with it, their fate.

\paragraph{Nuclear reaction rates:}
As discussed previously, the final core structures of stars are sensitive to the conditions under which core helium and carbon burning take place. In particular, the $^{12}\mathrm{C}(\alpha,\gamma)^{16}\mathrm{O}$ is a key helium-burning process which is still very uncertain and has a tremendous impact on the final fate of stars \citep[e.g.][]{austin_effective_2014,farmer_constraints_2020}. This nuclear reaction rate sets the amount of carbon left after core helium burning ($X_{C}$) by fusing carbon into oxygen. As shown in Appendix~\ref{sec:appendix:cag}, varying this reaction rate changes the core mass range at which stellar structures reach a high final compactness and iron-core mass and are expected to form black holes. A higher nuclear reaction rate resulting in a lower core carbon abundance means that neutrino-dominated burning occurs at lower CO core masses, systematically shifting the compactness landscape \citep{sukhbold_missing_2020,schneider_bimodal_2023}. For example, a 10\% higher carbon alpha-capture rate leads to a compactness peak occurring at 0.4\Msun lower CO core masses (see Appendix~\ref{sec:appendix:cag}). In turn, this increases the number of stars expected to form black holes, as stars with lower (CO core) masses are more common. It will also affect the final masses of black holes, which is expected to be proportional to the final mass of a star and is generally lower for progenitors with lower masses \citep[see also][]{schneider_bimodal_2023}.

Similarly, the value of the $^{12}\mathrm{C} + ^{12}\mathrm{C}$ cross section also systematically shifts the final core structures of stars \citep{chieffi_impact_2021}. The newly determined, higher rate found by \citet{tumino_increase_2018} due to low-lying resonances shifts the compactness landscape to  slightly lower core and initial masses compared to the classical rate we use in the present work \citep{caughlan_thermonuclear_1988} by allowing core carbon burning to take place at lower temperatures and densities. Carbon thus ignites under conditions where neutrino losses are less severe, which means that the overall core contraction is slightly less significant. This results in a systematic shift in the final compactness to slightly lower values (0.05 in $\xi_{2.5}$, see Fig. 9 in \citealt{chieffi_impact_2021}). 
Finally, the triple-alpha reaction rate also plays an important role, though it is less uncertain \citep{austin_effective_2014}.

\paragraph{Mass loss:}
Wind mass loss (and equivalently, metallicity) has a large impact on the properties of stars and is still very uncertain. Latest observational constraints suggest that wind mass loss at solar metallicity may be less strong than we assume here. When mass loss is strong enough to remove a large fraction of the hydrogen-rich envelope, it can have a large impact on the final structure of stars as it changes the evolution of the helium core, and with it the initial conditions for core carbon burning. In particular, strong wind mass loss can lead to a less efficient hydrogen-burning shell and to a recession of the convective helium-burning core, which results in a large core carbon abundance $X_{C}$ at core helium depletion. This in turn shifts the parameter space for neutrino-dominated carbon burning to higher CO core masses. Stars that experience strong wind mass loss are thus generally more explodable (just like binary-stripped stars, see Sect.~\ref{sec:discussion:binaries}).
In our models with high enough initial masses ($\gtrsim35\Msun$), stars may also enter a Luminous Blue Variable phase in which eruptive mass loss has been observed to occur \citep{smith_missing_2004,davies_luminosities_2018}. The mechanism behind these eruptions has been suggested to be related to helium opacity in the outer layers \citep{jiang_outbursts_2018,grassitelli_wind-envelope_2021}. Unlike line-driven winds, this mechanism would also be important at low metallicity. This highly uncertain mass loss is not taken into account in our models but can lead to the removal of the entire hydrogen-rich envelope, which would significantly affect the final core properties, and lead to a final core structure that explodes more easily.

\paragraph{Mixing:} How mixing proceeds in stellar interiors remains an importance uncertainty. Convective core boundary mixing, computed in our models through step overshooting, influences the size of convective cores. During core hydrogen burning, it helps determine the initial mass to helium core mass relation. Similarly, core overshooting during core helium burning influences the resulting CO core mass and thus has a particularly large influence of their final fate. \citet{temaj_convective-core_2024} showed that core overshooting systematically shifts the compactness landscape of stars as a function of their initial masses, as a larger overshooting allows a star to behave as if it had the core of an initially higher-mass star. This also implies a larger number of black holes formed, as more stars can reach the mass range for black hole formation \citep[see also Fig 6. of ][]{schneider_bimodal_2023}.
As a function of the CO core mass, the compactness landscape is similar to that of single stars, though the compactness peak shifts towards higher CO core masses (1.3\Msun by varying the step overshooting scaling parameter $\alpha_{\rm{OV}}$ from 0.05 to 0.5, \citealt{temaj_convective-core_2024}). This shift is because, as convective overshooting affects the core mass of stars, it also changes their luminosity and therefore mass loss rate. Especially for high initial masses, when the total mass loss is large, $X_{C}$ increases and thus the compactness landscape shifts to higher CO core masses \citep{temaj_convective-core_2024}. Overshooting during core helium burning also influences the ratio of the total mass to the CO core mass, thus the final black hole mass. In stripped helium stars, large overshooting implies a lower black hole mass \citep[see also Fig. 6 of][]{schneider_bimodal_2023}.
 
As we explore in more detail Appendix~\ref{sec:appendix:sc}, semiconvection is another mixing process that has a large effect on the final core structure of stars. It changes the compactness landscape as a function of the initial mass but only slightly affects the compactness landscape as a function of the CO core mass (see Fig.~\ref{fig:entropy_Mi_sc}). Semiconvection affects the stellar structure after core hydrogen exhaustion through the development of intermediate convection zones \citep{sibony_impact_2023}. Furthermore, it can lead to late ingestion of helium during core helium burning depending on the chemical gradient at the edge of the helium core \citep{langer_evolution_1985,langer_evolution_1991} and thus not only shifts the CO core mass but also $X_{C}$ (see Fig.~\ref{fig:XC_MCO_sc}). This can explain the "noise" or "randomness" reported by \citet{sukhbold_high-resolution_2018} in the compactness landscape \citep[see also][]{chieffi_presupernova_2020}. \citet{schootemeijer_constraining_2019} found through a comparison with observations that semi-convection may be an efficient process, i.e. favoring high values of $\alpha_{\rm{sc}}\geq 1$. However, it remains a major uncertainty in our models.

\subsection{Shell mergers}
In our models beyond the compactness peak, we identify several models that experience mergers of their C-, Ne- and in some cases also O-burning shells. As we show in Sect.~\ref{sec:results:experiment:shell_mergers} these are the result of an increasing entropy at a nuclear burning front (e.g. Ne or O) that exceeds the entropy of the layers above, implying that these layers are unstable against convection and must mix. Several studies  have reported the occurrence of such shell mergers in their one-dimensional models for a similar parameter space (i.e models before and just beyond the compactness peak, typically in the CO core mass range from 3 to 7\Msun, \citealt{sukhbold_compactness_2014,sukhbold_high-resolution_2018,sukhbold_missing_2020,collins_properties_2018, laplace_different_2021}). We argue that these shell mergers are not artifacts from our treatment of convective boundary mixing since we do not include overshooting above shells post helium burning. In Appendix~\ref{sec:appendix:res_test}, we verify that these shell mergers are barely affected by our choices of spatial and temporal resolution. The exact predictions for the outcomes of shell mergers will still depend on a more accurate treatment of convective mixing than the mixing-length approximation used here, i.e. informed by multi-dimensional simulations \citep[][]{rizzuti_3d_2023}. Several three-dimensional calculations have also observed the occurrence of shell mergers, strengthening the finding from one-dimensional simulations \citep[e.g.,][]{couch_revival_2013,collins_properties_2018,yoshida_one-_2019,andrassy_3d_2020,yadav_large-scale_2020,mcneill_stochastic_2020}. \citet{collins_properties_2018} shows that shell mergers help develop stellar structures that have extended oxygen shells with high convective Mach numbers. In such stellar structure, perturbations can be induced by this convective oxygen shell burning at the time of core collapse. 3D supernova simulations suggest that such perturbations are crucial for increasing the efficiency of neutrino heating in neutrino-driven supernova explosions and for a successful shock revival \citep[e.g.,][]{couch_revival_2013}. Thus, stellar structures experiencing shell mergers may be particularly prone to successful explosions. In addition, due to the mixing process and ensuing nuclear burning episodes, stars experiencing shell mergers are expected to produce peculiar abundance patterns. Typically, these are enhanced alpha-capture products in the oxygen-shell region, which in turn affect the nucleosynthesis, in particular the Ca/O ratio of core-collapse supernovae \citep[e.g.,][]{ritter_convective-reactive_2018,dessart_radiative-transfer_2020,laplace_different_2021}. In our work, we find that shell mergers are expected in stellar structures with strong neon shell burning. We observe that this systematically occurs in models just beyond the compactness peak ($23 < M_{i}/\Msun < 27$, $6.3 < M_{\rm{CO}}/\Msun < 8.5$), in which carbon burning is neutrino-dominated and core neon burning is becoming neutrino dominated.

\subsection{Final structures of stars that experience binary interactions}
\label{sec:discussion:binaries}
In this work, we have focused on the core structures of single massive stars. However, it is well established that the majority of massive stars live in binary systems and are expected to interact during their lifetime \citep{sana_binary_2012}. As a result of binary interactions,  the core structure of stars can be greatly modified, and lead to a different fate even for stars with the same core mass \citep{laplace_different_2021,schneider_pre-supernova_2021,schneider_bimodal_2023,schneider_pre-supernova_2024}. In addition, the vast majority of black holes are found in binaries observationally. In these systems their properties, in particular their masses, can be measured most accurately \citep{casares_mass_2014}. It is therefore important to discuss the effect of binary interactions on the final structures of stars.

The mechanisms we identified here that set the final structure of stars (i.e. the impact of neutrino-dominated burning) apply generally for any stellar structure, as shown by our experiment in Sect.~\ref{sec:results:experiment} and are mainly influenced by the initial conditions for core carbon burning (i.e. the CO core mass and central carbon abundance, see also \citealt{patton_towards_2020}). These conditions can be influenced by binary interactions or strong wind mass loss, in particular if the structure of a star is significantly affected after core H-burning, which will ultimately change the initial conditions for core carbon burning \citep{brown_formation_1999,brown_formation_2001,woosley_evolution_2019,schneider_pre-supernova_2021,laplace_different_2021}. This implies a shift in the compactness landscape. 

When it comes to black hole formation, binary-stripped stars are particularly important binary products. In isolated massive binary systems, they are the natural progenitors of both black holes that eventually form binary black hole mergers that are now commonly detected as GW sources \citep{schneider_bimodal_2023}. After mass transfer or strong wind mass loss removes the outer envelope of these stars, the conditions under which core helium burning occur can change drastically compared to single stars. A weaker or even absent hydrogen-burning shell, and strong wind mass loss leads to a shrinking of the convective helium-burning core. This results in a higher final core carbon abundance compared to single stars for the same CO core mass \citep{brown_formation_2001,schneider_pre-supernova_2021,laplace_different_2021}. The higher core carbon abundance means that the compactness peak, and the corresponding expected parameter space for BH formation is shifted to higher core masses in these stars ($M_{\rm{CO}}\approx7.5\Msun$, see Fig. 2 in \citealt{schneider_bimodal_2023}). Therefore, binary-stripped stars are expected to be common progenitors of supernovae and to only form black holes for high initial masses (of about 30\Msun), which significantly reduces the parameter space for BH formation \citep{schneider_pre-supernova_2021}. Moreover, the core structures of binary-stripped stars have a weak metallicity dependence, implying the formation of universal BH masses at the compactness peaks, with strong observational signatures \citep{schneider_bimodal_2023}.

The final core structures of accretors and stellar mergers can also be affected by binary interactions. In our recent work \citep{schneider_pre-supernova_2024} we have shown that mass accretion, in particular onto stars that have completed core hydrogen burning, can lead to systematically different final core structures. In particular, the compactness peak in case B and C accretors shifts systematic to lower core masses  \citep{schneider_pre-supernova_2024}. Contrary to binary-stripped stars, this means that accretors form black holes in stars with lower initial masses than in single stars.

\subsection{Compactness peaks and black hole formation}
In this work, we estimate the outcome of core-collapse for our models based on the \citet{ertl_two-parameter_2016} criterion and on the 1D semi-analytical neutrino-driven model of \citet{muller_simple_2016}. We find a correspondence between the formation of black holes and models with high final compactness, which all have a high final binding energy, iron-core mass, and central entropy (see Fig.~\ref{fig:xi_overview}). To first order, compactness (or equivalently, any of these quantities) captures the information that stars with high binding energy are difficult to explode. Thus, independently of the uncertainties regarding the explosion mechanism of core-collapse supernovae, a signature of the characteristic final structure of stars described here is expected in their final remnants and in the properties of their explosions. 

However, while compactness captures the general final structure of the star, it is not a good predictor of the detailed explosion outcome (see Sect.~\ref{sec:results:final_structure}). Explodability criteria and the remnants expected after the core collapse are the subject of active discussion in the supernova modeling community, including within groups studying the neutrino-driven explosion mechanism. For example, some 3D neutrino-driven simulations \citep[e.g.,][]{chan_black_2018,kuroda_full_2018,ott_progenitor_2018,burrows_black-hole_2023} have recently found successful shock revival in models with high compactness. However, the revival of the supernova shock does not necessarily imply a full supernova explosion with complete ejection of the stellar layers, or that these will all form neutron stars. Due to the high accretion rate typical for models with high compactness, these are still likely to form black holes through fallback accretion, especially in stars with high core masses \citep{heger_black_2023}. For example, \citet{chan_black_2018} found that a 40\Msun progenitor star with high compactness forms a black hole early on through fallback accretion. In their simulation, the supernova shock succeeds in unbinding parts of the envelope despite the explosion energy being lower than the binding energy of the envelope. More recently, \citet{burrows_black-hole_2023} found that a 40\Msun model with high compactness results in a successful explosion, though the remnant eventually forms a black hole due to the high accretion rate onto the proto neutron star. Black holes formed through fallback accretion have systematically lower masses than those formed through direct collapse because parts of the star can be expelled. They therefore have distinct properties, which may be constrained through observations \citep{chan_impact_2020}. However, consistently finding explosion of models with high masses and high final compactness would be at odds with observational constraints, which seem to point to a lack of core-collapse supernovae from stars with high core masses \citep[e.g.,][]{smartt_progenitors_2009, davies_red_2020}. 

\subsection{Observational constraints on the pre-supernova structure of stars}
The robustness of the patterns found for the final core structure of massive stars, i.e. the ``compactness peaks'' based on the neutrino-dominated burning mechanism identified here, suggests that signatures of this pattern could be identifiable in observations.

\citet{sukhbold_missing_2020} showed that the luminosity of immediate BH progenitors could be used to test the existence of characteristic core structures corresponding to the compactness peaks observationally. The observation of the disappearing red-supergiant N6946-BH1 could be a piece of evidence \citep{sukhbold_missing_2020}, though more observations are needed to understand the latest constraints on this object \citep{beasor_jwst_2024,kochanek_search_2024}.
Finding more disappearing stars or additional direct observational link between black holes and supernovae, and their progenitors, will provide crucial constraints. For example, \citet{temaj_convective-core_2024} showed that the luminosity of supernova progenitors in hydrostatic equilibrium can be directly linked to their core mass, independently of uncertainties in interior mixing or wind mass loss.\\

If there is a direct link between massive star structures in the "compactness peak" and BH formation, i.e. if it is not fully obscured by supernova physics and fallback dynamics, then black hole masses should also encode information about the final structure of stars in the compactness peak. We expect that such a link should be readily identified in binary-stripped stars, which have characteristic final structures with a weak metallicity dependence (see also Sec.~\ref{sec:discussion:binaries}). We predicted that this would lead to the formation of BHs of universal masses \citep{schneider_bimodal_2023}. Alternatively, if supernova and fallback physics ``wash out'' these signals, then identifying broad features in the BH mass distribution and significant numbers of low-mass black holes may provide further constraints into these processes and their relative importance. The complete absence of features in the BH mass distribution of binary-stripped stars (e.g., a flat black hole mass distribution) would also give particularly strong constraints on stellar, binary, and supernova physics, as this would require narrow ranges of physical ingredients. 

Using GW observations to find features in the chirp-mass distribution of binary BH mergers could provide evidence for the existence of the characteristic final stellar structures of binary-stripped stars \citep{schneider_bimodal_2023, disberg_failed_2023}. Considering a population of merging binary BHs resulting from isolated binary evolution, we predicted peaks in the chirp-mass distribution of binary BH mergers at $\sim8\Msun$ and $\sim14\Msun$, and a dearth between $\sim10 -12\Msun$ \citep{schneider_bimodal_2023}. Based on the published data from LIGO-Virgo-KAGRA (LVK) observations, a peak at a chirp mass of $\sim 8\Msun$ is found to be statistically significant, while the existence of a feature at $\sim 14\Msun$ is less robust \citep{talbot_measuring_2018,tiwari_emergence_2021,tiwari_exploring_2022,abbott_population_2023,edelman_cover_2023, farah_things_2023}. Tentative evidence for the predicted dearth in the distribution between 10 and $12\Msun$ has recently been reported, though more data, even beyond the upcoming O4a data release by the LVK collaboration, is required to further test its existence \citep{adamcewicz_no_2024,galaudage_compactness_2024}.\\
Signatures of the characteristic core structures of massive stars that form black holes may also be identified in the mass distribution of BHs from electromagnetic observations (Nambena et al. in prep.). Studies of black holes in X-ray binaries point to the existence of a characteristic peak in black hole masses of about $\sim8\Msun$ \citep[e.g., ][]{kreidberg_mass_2012,casares_x-ray_2017}, which is similar to the mass we predict for black holes in binary-stripped stars originating from the compactness peak \citep{schneider_bimodal_2023}. The recent discovery of new X-ray quiet black holes in wide binaries \citep{shenar_x-ray-quiet_2022,el-badry_sun-like_2023,el-badry_red_2023,gaia_collaboration_discovery_2024} provides a promising avenue to discover more black holes and better understand their properties, and ultimately, their formation and progenitors.

\section{Conclusions}
\label{sec:conclusion}
Understanding the fate of massive stars, i.e. whether stars form black holes at the end of their lives or successfully explode in a supernova is a major goal of astrophysics. In this work, we have explored how stellar physics determines the pre-supernova core structure and the role it plays in the fate of massive stars by studying the final evolution of single star in the range from 17 to 50\Msun at solar metallicity. Our findings can be summarized as follows:
\begin{itemize}
    \item We confirm that the final structure landscape of massive stars found in multiple independent one-dimensional stellar evolution studies \citep{timmes_neutron_1996,brown_formation_1999,sukhbold_compactness_2014,sukhbold_high-resolution_2018,muller_simple_2016,chieffi_presupernova_2020,chieffi_impact_2021,schneider_pre-supernova_2021,schneider_bimodal_2023,schneider_pre-supernova_2024,takahashi_monotonicity_2023, temaj_convective-core_2024}
    is robust. The prominent features in the final compactness and iron core mass can be traced back to the conditions under which late nuclear burning occurs in the cores of stars at this mass range.
    \item Summarizing quantities such as the final compactness, central entropy, iron core mass, and binding energy are all equivalent and intrinsically linked. Stars with high final compactness all have a high binding energy, and we verify that the formation of a black hole is likely in these stars based on a neutrino-driven supernova model. Independently from uncertainties in the supernova mechanism and explodability of stars, stellar physics thus plays a large role in determining the final remnants of stars, in particular the formation of black holes. We confirm that the final structure of massive stars is already mostly ``written'' in their cores at the end of core helium burning.
    \item We find that the final compactness increases when carbon burning becomes neutrino-dominated (region A in Fig.~\ref{fig:xi_overview}). The transition from convective to radiative core carbon burning and the change in the extent and number of carbon-burning shells identified in previous studies is a symptom of this effect, but not its cause. Strong neutrino losses force a larger core contraction and the development of a larger carbon-free core. Eventually, this enables the formation of a larger iron core. The same mechanism leads to the formation of a second compactness increase once neon burning becomes neutrino dominated (region C in Fig.~\ref{fig:xi_overview}).
    \item We trace back the drop in compactness after the compactness peaks (regions B and D in Fig.~\ref{fig:xi_overview}) to the effect of an earlier ignition of the next nuclear fuel (neon or silicon) in these stars, which slows down the core contraction and eventually leads to the formation of smaller iron core masses. We find that the next burning phase ignites earlier in these stars because of two main mechanisms (1) a shorter duration of the main burning stage due to a decrease in available fuel and increase in temperature and (2) the role of electron degeneracy, which further accelerates the contraction of the core when the fuel-free core exceeds the effective Chandrasekhar mass.
    \item Shell mergers between the C and Ne- burning shells, and in some cases also between the Ne and O-burning shells, occur in our models after the first compactness peak, just like other studies have observed \citep{sukhbold_compactness_2014,sukhbold_high-resolution_2018, collins_properties_2018}. These contribute to the drop in compactness after the compactness peak, lead to smaller iron-core masses, and make these stars more explodable. We find that shell mergers are not numerical artifacts but take place because of energetic neon and oxygen burning which generates enough entropy to mix with the layers above, independently of convective boundary mixing. Shell mergers not only affect the pre-supernova structure but also change the pre-supernova composition and may have observable signatures which call for further investigation.
    \item The exact final core structure landscape of stars is subject to uncertainties in stellar evolution, such as semi-convective mixing, mass loss, and uncertain nuclear reactions, in particular the $^{12}\mathrm{C}(\alpha,\gamma)^{16}\mathrm{O}$ reaction rate. The origin of small variations and secondary peaks in compactness and central entropy observed when nuclear burning conditions change remains to be explored further. However, the robustness of the main features identified in the final structures of massive stars as a function of their CO core masses provides a clear theoretical prediction and an opportunity to constrain stellar physics from the observation of black holes and their immediate progenitors.
\end{itemize}

The final structure landscape discussed here applies for any massive star, including those that interact in binary systems. However, binary interactions shift the compactness landscape to different core masses, which affects the parameter space for expected black hole formation and successful supernovae explosions \citep{schneider_pre-supernova_2021,schneider_bimodal_2023,schneider_pre-supernova_2024}. For accurate predictions of supernova and black hole formation, in particular for understanding the population of gravitational-wave sources, it is crucial to take the different final core structures of single and binary stars into account.

\begin{acknowledgements}
	We thank the referee for constructive comments that helped improve the manuscript.
	This work was funded by the European Research Council (ERC) under the European Union’s Horizon 2020 research and innovation program (Grant agreement No. 945806) and supported by the Klaus Tschira Stiftung. It was also supported by the Deutsche Forschungsgemeinschaft (DFG, German Research Foundation) under Germany’s Excellence Strategy EXC 2181/1-390900948 (the Heidelberg STRUCTURES Excellence Cluster).
\end{acknowledgements}

\bibliographystyle{aa} 
\bibliography{compactness_peak_final.bib}
\begin{appendix}

\section{The effect of the $^{12}\mathrm{C}(\alpha,\gamma)^{16}\mathrm{O}$ reaction rate}\label{sec:appendix:cag}
\begin{figure}[hb!]
	\centering
	\includegraphics[width=0.5\textwidth]{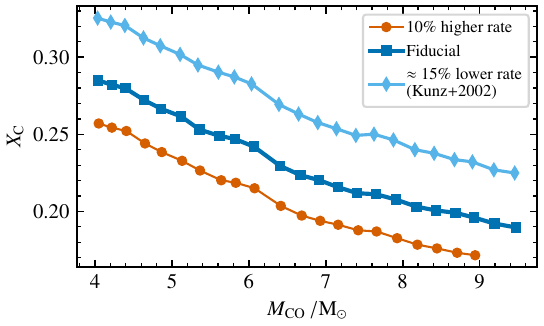}
	\caption{Central carbon mass fraction at core helium depletion as a function of the CO core mass for varying $^{12}\mathrm{C}(\alpha,\gamma)^{16}\mathrm{O}$ reaction rate.}
	\label{fig:XC_MCO_cag}
\end{figure}
\begin{figure*}[hb!]
	\centering
	\includegraphics[width=0.5\textwidth]{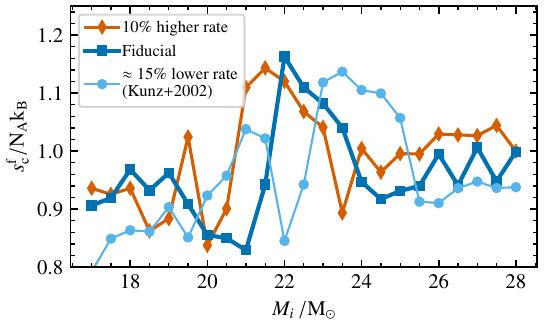}%
	\includegraphics[width=0.5\textwidth]{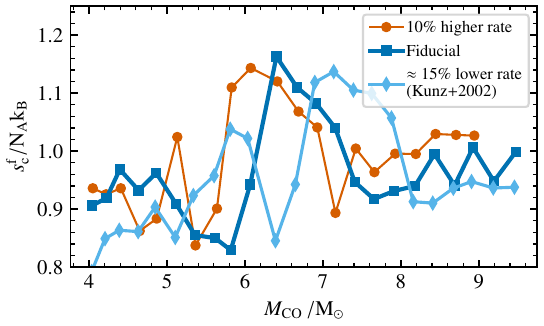}
	\includegraphics[width=0.5\textwidth]{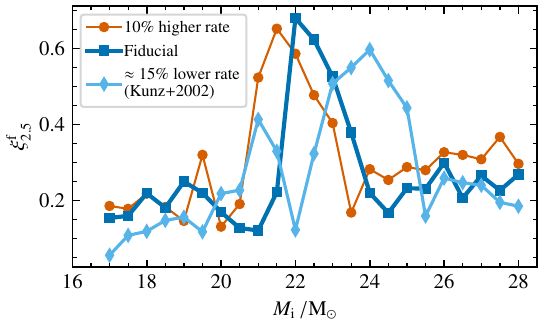}%
		\includegraphics[width=0.5\textwidth]{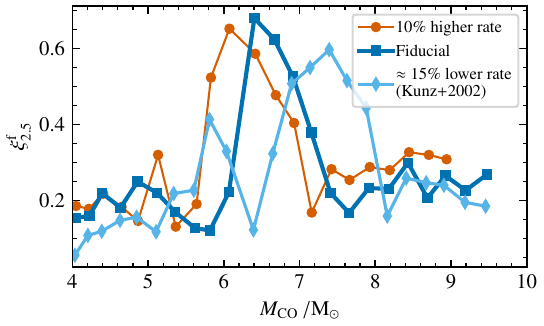}
	\caption{Final central entropy (top) and compactness (bottom) as a function of the core mass (left) and initial mass (right) for a varying $^{12}\mathrm{C}(\alpha,\gamma)^{16}\mathrm{O}$ rate.}
	\label{fig:entropy_xi_cag}
\end{figure*}

Among all nuclear reactions taking place in the interiors of stars, the $^{12}\mathrm{C}(\alpha,\gamma)^{16}\mathrm{O}$ reaction rate is particularly important for determining the fate of stars \citep[e.g.,][]{weaver_nucleosynthesis_1993}. However, it remains notoriously uncertain \citep{farmer_constraints_2020}. Towards the end of core helium burning, this reaction becomes the main helium-burning process. It determines the fraction of remaining carbon and oxygen in the core after core helium burning, as shown in Fig.~\ref{fig:XC_MCO_cag}. For a higher reaction rate, the core carbon mass fraction $X_{\rm{C}}$ reaches systematically lower values. As this sets the initial conditions for the subsequent core carbon-burning stage, this reaction thus plays a crucial role in determining the final structure and fate of stars. In Fig.~\ref{fig:entropy_xi_cag}, we show the effect of varying this reaction rate, from a $\sim15\%$ lower \citep{kunz_astrophysical_2002} to a 10\% higher rate than our fiducial assumptions on the final central entropy and compactness. Both as a function of initial and CO core mass, varying this reaction rate systematically shifts the location of the compactness peak to lower masses for a higher reaction rate. This is because a higher rate leads to a lower core carbon abundance and thus to carbon burning becoming neutrino-dominated at lower masses. This is thus a similar effect as described in our experiment in which we vary the core carbon abundances for models with the same core mass (Sect.~\ref{sec:results:experiment}). A difference comes from the role of the $^{12}\mathrm{C}(\alpha,\gamma)^{16}\mathrm{O}$ reaction rate in setting the mass of the CO core (see the small shift in CO core mass in Fig.~\ref{fig:XC_MCO_cag}). Secondary compactness and central entropy peaks appear at CO core masses between 5 and 6\Msun and 7 to 8\Msun for a varying $^{12}\mathrm{C}(\alpha,\gamma)^{16}\mathrm{O}$ rate. These are discussed in more detail in Appendix~\ref{sec:appendix:sc}. Finally, shifts in the maximum and minimum compactness values can be observed. These can be understood as a varying degree of maximum or minimum contraction in the stellar structure, induced by the varying effect of neutrino losses for different nuclear burning conditions during core carbon burning.

\section{The impact of semi-convective mixing on the final structure}\label{sec:appendix:sc}

\begin{figure}[ht!]
	\centering
	\includegraphics[width=0.5\textwidth]{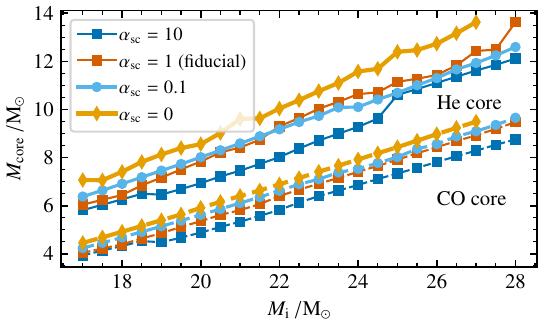}
	\caption{He (full lines) and CO (dashed lines) core mass at core helium depletion for models with a varying semiconvection efficiency. A higher efficiency generally leads to a smaller core mass.}
	\label{fig:core_mass_sc}
\end{figure}

\begin{figure}[hb!]
	\centering
	\includegraphics[width=0.5\textwidth]{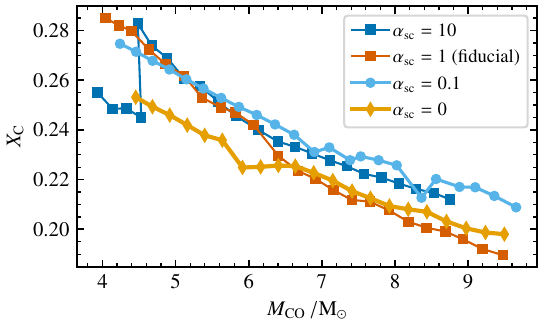}
	\caption{Central carbon mass fraction at core helium depletion as a function of the CO core mass for varying semi-convection efficiency.}
	\label{fig:XC_MCO_sc}
\end{figure}

Semi-convection is a particularly uncertain mixing process that occurs in stellar layers that are unstable against convection because of their superadiabaticity (Schwarzschild criterion for convection) but supported by a stabilizing molecular weight gradient (Ledoux criterion). It occurs for example at the end of core hydrogen burning when a H/He composition gradient is created by the receding convective core. This can form a transient intermediate convection zone (ICZ) and hence change the structure of the core outside the inner convective region \citep{sibony_impact_2023}.  Semi-convection remains difficult to constrain observationally, though a recent study suggest that it may be an efficient process, with an efficiency parameter of $\alpha_{\rm{sc}}\geq 1$ \citep{schootemeijer_constraining_2019}. It has a large influence on the final core structure of stars because it affects the development of the convective helium-burning core \citep{langer_evolution_1985,langer_evolution_1991}, generally leading to smaller core masses for a higher semi-convection efficiency (see Fig.~\ref{fig:core_mass_sc}). In addition, even small semi-convective layers above a convective burning core can drastically change the initial conditions for the next burning phase. For example, late ingestion of helium due to semiconvection during core helium burning can strongly affect the core carbon abundance at core helium depletion \citep{langer_evolution_1991}. This is shown in Fig.~\ref{fig:XC_MCO_sc}, where the core carbon abundance $X_{\mathrm{C}}$ at the end of core helium depletion is plotted as a function of the CO core mass for varying degrees of semi-convective mixing. Between the models with no semi-convection mixing and the $\alpha_{\rm{sc}}=0.1$ and $\alpha_{\rm{sc}}=10$ models, there appears to be a systematic shift to a higher $X_{\mathrm{C}}$. This can be attributed to a systematically smaller helium core mass after core hydrogen burning with respect to the initial mass, which creates a systematically smaller CO core and generates more carbon overall (see also Fig.~\ref{fig:core_mass_sc}). Generally, for lower masses and higher densities, the triple alpha reaction (which depends on the density cubed) is favored during core helium burning and alpha captures onto carbon (which linearly depends on density) play less of a role \citep{brown_formation_2001}, which means less carbon is destroyed at lower masses, leading to a higher $X_{\mathrm{C}}$. However, this trends is no longer reproduced for the $\alpha_{\rm{sc}}=1$ models, where $X_{\mathrm{C}}$ decreases for CO core masses larger than 5\Msun compared to the $\alpha_{\rm{sc}}=0.1$ models. This can be understood as the effect of a late ingestion of helium due to semi-convection during core helium burning in these models, which boost alpha captures onto carbon, which reduce $X_{\mathrm{C}}$. \\

For the $\alpha_{\rm{sc}}=10$ models, we observe a significant jump of $\sim0.03$ in $X_{\mathrm{C}}$ for models with CO core masses larger than 4.5\Msun in Fig.~\ref{fig:XC_MCO_sc}. This is because the lowest-mass models ($M_{i} < 19\Msun$, $M_{\rm{CO}} < 4.5\Msun$) have a different structure compared to higher- models, which all
burn helium as red supergiants. Due to the efficient semiconvection, these low-mass models develop a large ICZ after core hydrogen burning and experience a blue loop, igniting helium as blue supergiants (BSGs). The characteristic strong hydrogen-burning shell of the BSG grows a more massive helium core compared to their initial mass than is the case in RSG structures (Fig.~\ref{fig:core_mass_sc}). For example, the 18.5\Msun model, which develops a BSG structure, has the same He core mass of 6.2\Msun as the 19\Msun model which has a RSG structure. This larger helium core mass and the ensuing different burning conditions during core helium burning favor carbon destruction through alpha captures and result in a systematically lower $X_{\mathrm{C}}$ in these models.

\begin{figure*}[hb!]
	\centering
	\includegraphics[width=0.5\textwidth]{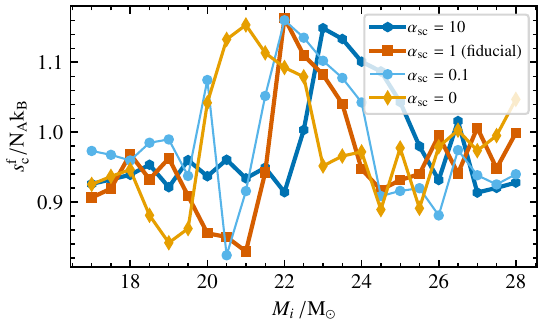}%
	\includegraphics[width=0.5\textwidth]{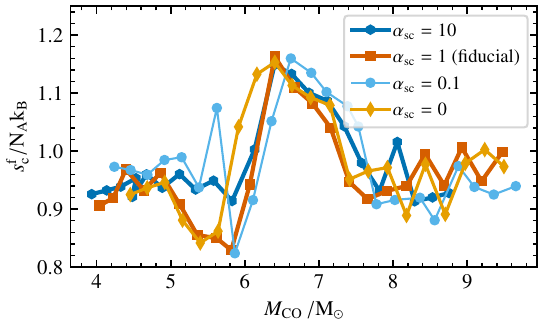}
	\includegraphics[width=0.5\textwidth]{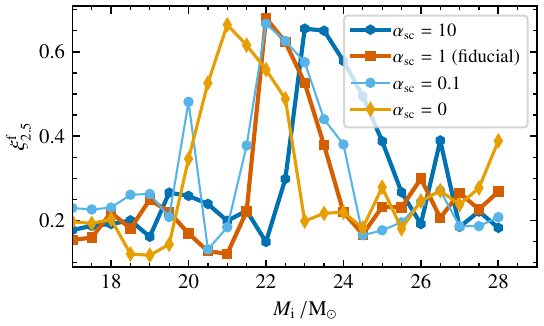}%
	\includegraphics[width=0.5\textwidth]{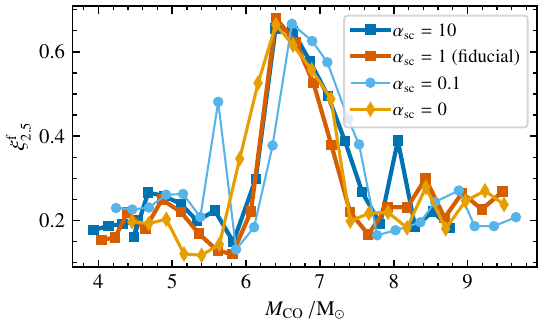}
	\caption{Final central entropy and compactness as a function of the initial mass for varying semi-convection efficiency.}
	\label{fig:entropy_Mi_sc}
\end{figure*}

Changes in semi-convection lead to systematic variations in the final stellar structure, as shown in Fig.~\ref{fig:entropy_Mi_sc}. The shift of the compactness and entropy peaks to higher initial masses (2\Msun for a shift of $\alpha_{\rm{sc}}$ from 0 to 10) reflects the change in the initial mass to helium core mass relation (see Fig.~\ref{fig:core_mass_sc}). However, the CO core mass at which the compactness peak occurs is remarkably similar and varies by only 0.2\Msun between models with different semi-convective mixing. This is because the onset of neutrino-dominated burning happens at a similar core mass ($\sim 6.5$\Msun) and similar $X_{\mathrm{C}}\sim 0.23$. Small variations in the compactness landscape, i.e. the existence of small secondary peaks next to the main compactness peak are observed for different values of semi-convection in Fig.~\ref{fig:entropy_Mi_sc}. For example, the small peak in central entropy and compactness at an initial (CO core mass) of 20 (5.6) \Msun has been identified out in multiple other studies \citep{sukhbold_compactness_2014,sukhbold_high-resolution_2018,chieffi_presupernova_2020,chieffi_impact_2021}. This appears to be a recurrent feature in this narrow CO core mass and $X_{C}$ range. Models with initial (CO core) masses between 24 and 28\Msun (8 and 10\Msun) also show variations in compactness and final entropy for a changing semi-convection efficiency. This is the region at the transition when neon burning becomes neutrino-dominated and develops through several convective shells. We observed that for a varying semi-convection efficiency, the development of this neon-burning front is affected, which leads to small variations in compactness. However, the exact mechanisms behind these secondary peaks remains to be understood. \\
Uncertainties in semi-convective mixing can thus lead to variations in the final stellar structure by influencing the initial conditions for core helium and carbon burning. This creates in part the``noise'' observed by \citealt{sukhbold_high-resolution_2018} in their study of the compactness of massive stars. However, it is noteworthy that semi-convection does not affect the existence of the main features observed in the compactness landscape and does not significantly change the CO core mass range at which the main compactness peak occurs.

\section{Resolution test for the occurrence of shell mergers}\label{sec:appendix:res_test}
\begin{figure*}[hb!]
	\centering
	\includegraphics{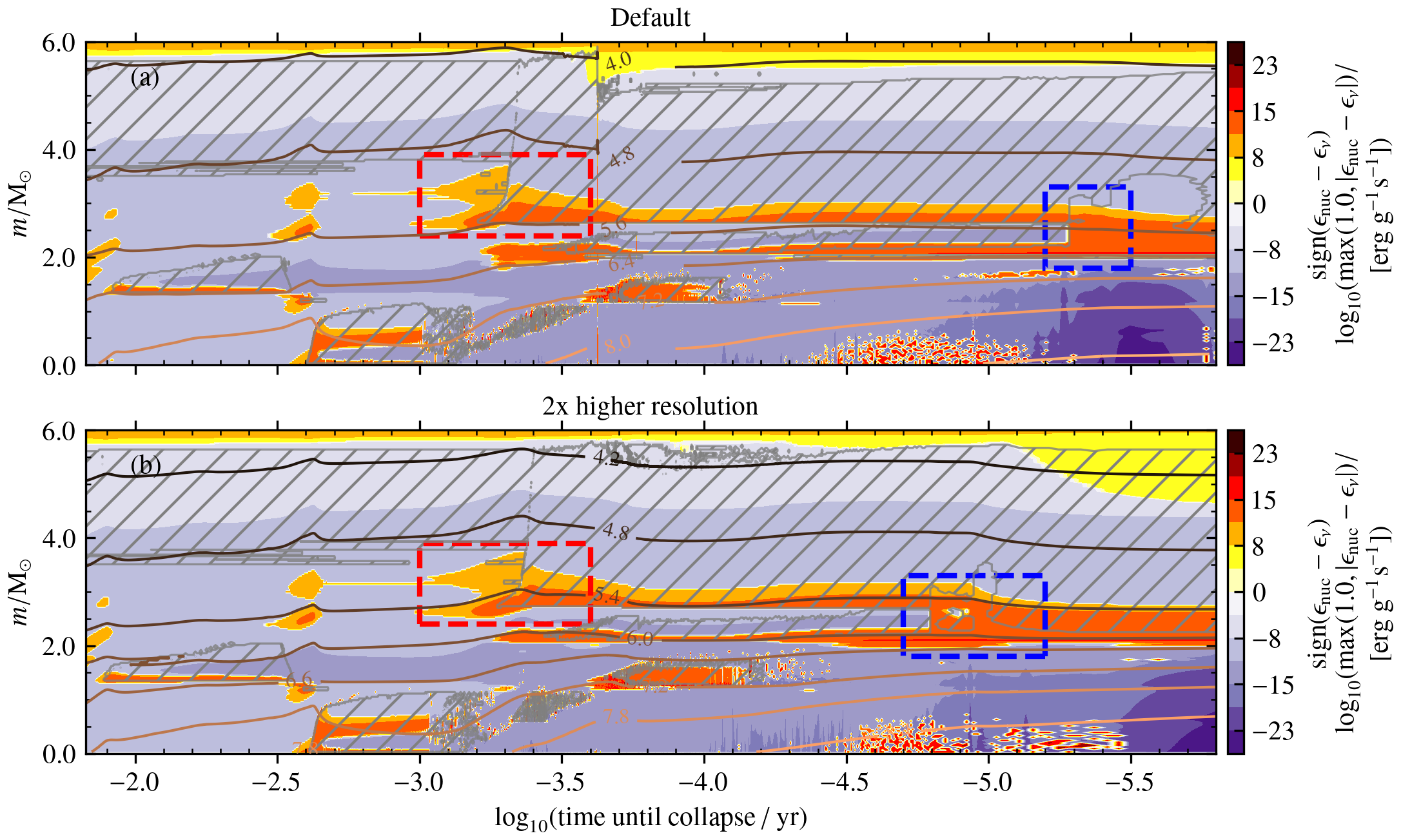}
	\caption{(a) Same as Fig.~\ref{fig:sh_mergers_3cag} but from core oxygen depletion to core collapse. (b) The same model in which we double the spatial and temporal resolution. The red and blue boxes highlight the shell merger between the C and Ne-burning shells, and the C/Ne and O-burning shells, respectively.}
	\label{fig:kip_sh_merger_restest}
\end{figure*}

To test how the occurrence of shell mergers is affected by numerical uncertainties, we perform a resolution test for the model presented in Sect.~\ref{sec:results:experiment:shell_mergers}, i.e. our default $22\Msun$ model with a modified central carbon abundance $X_{\rm{C}}=0.17$, starting from core oxygen exhaustion. We double the spatial and temporal resolution, and in pariticular, double the number of grid points in zones containing composition gradients and nuclear burning regions of C, O, and Ne (MESA settings \texttt{mesh\_logX\_species} and \texttt{mesh\_dlog\_burn\_c\_dlogP\_extra} 
and equivalent). In Fig.~\ref{fig:kip_sh_merger_restest} we show the comparison of the Kippenhahn diagrams of the default model and the model with double the resolution. We find no difference in the occurrence of the shell merger between the C and Ne shell (highlighted by the red box). The timing and extent is nearly identical, and can be traced back to the same mechanism, i.e. the increase in entropy in the Ne-burning shell. We therefore verify that these shell mergers between the C and Ne shell do not originate from poor resolution.
The second shell merger between the C/Ne and the O-burning shells (highlighted by the blue box in Fig.~\ref{fig:kip_sh_merger_restest}) occurs 5 min earlier for the model with higher resolution. Since this shell merger occurs towards the end of Si shell-burning, after most of the iron-rich core has already formed, it barely affects the final core structure. Numerical uncertainties mainly influence the timing of this shell merger event.

\begin{figure}[hb!]
	\centering
	\includegraphics[width=0.5\textwidth]{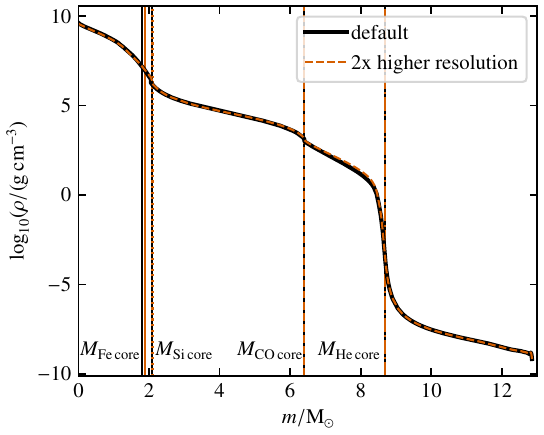}
	\caption{Final density profile of the models in Fig.~\ref{fig:kip_sh_merger_restest} with our default assumptions compared to a model with double the resolution at the onset of core collapse. From left to right, vertical lines indicate the Fe, Si, CO, and He core masses in black and orange for the default and higher resolution model, respectively.}
	\label{fig:rho_comp_sh_merger}
\end{figure}

In Fig.~\ref{fig:rho_comp_sh_merger} we compare the final density profiles at the onset of core collapse. The density profiles are almost identical, except for small variations of order $\rho\sim10 \rm{g}\,\rm{cm}^{-3}$ in the He-rich layer. The final iron core mass is a little larger for the higher-resolution model ($M_{\rm{Fe\,core}} = 1.80$ compared to $M_{\rm{Fe\,core}} = 1.89$). Similarly, the final compactness values is a little larger ($\xi_{2.5} = 0.357$ for the default model and $\xi_{2.5} = 0.364$ for the models with double the resolution). We conclude that the shell mergers identified in this work are barely affected by numerics. The C and Ne shell mergers are not numerical artifacts, but instead originate from the high entropy contrast between the Ne-burning shell and the C shell in this model (see Sec.~\ref{sec:results:experiment:shell_mergers}).

\section{Interior structure evolution diagrams}\label{sec:appendix:kipp}

\begin{figure*}[hb!]
	\centering
	\includegraphics[width=0.5\textwidth]{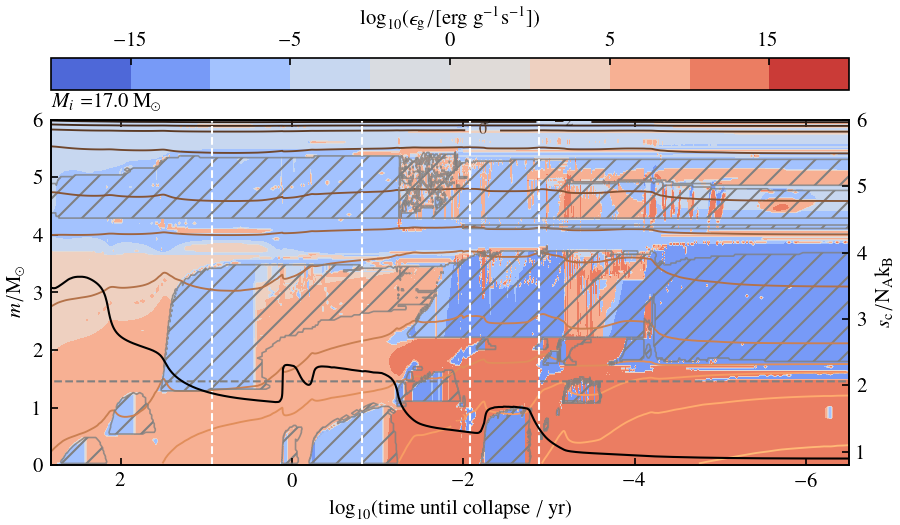}%
	\includegraphics[width=0.5\textwidth]{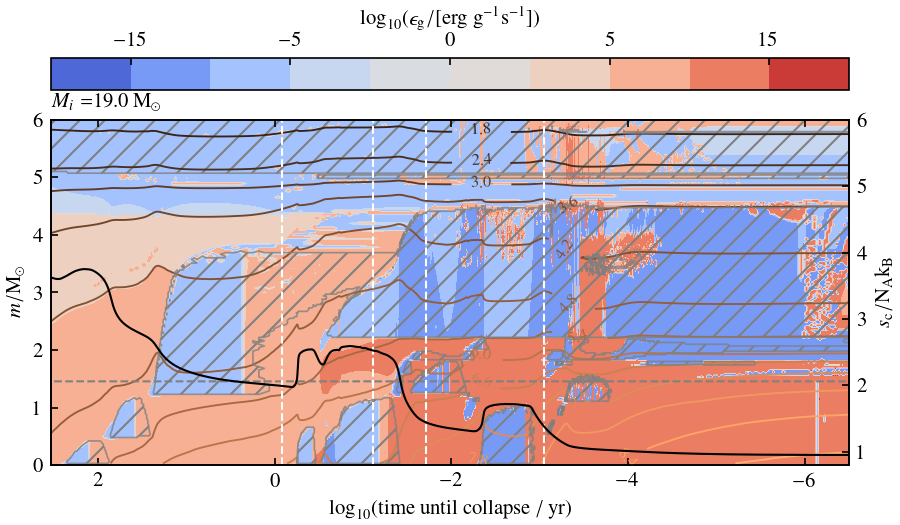}
	
	\includegraphics[width=0.5\textwidth]{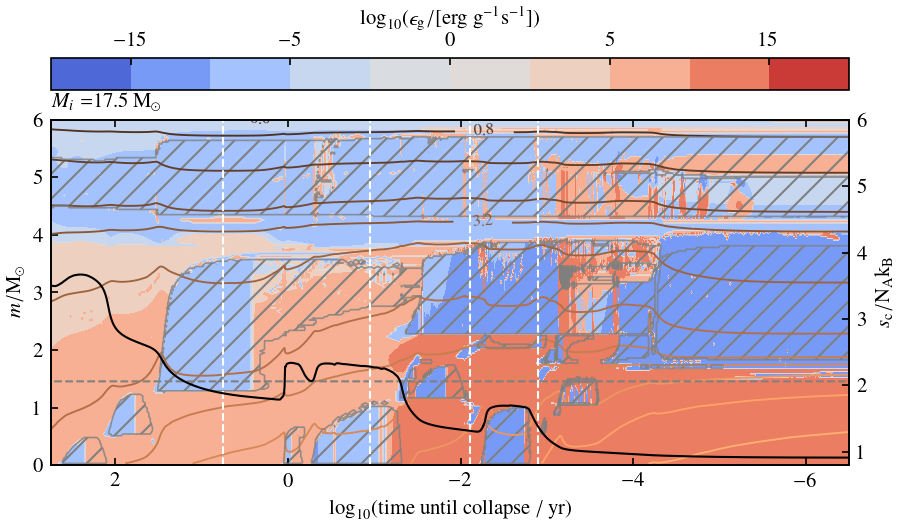}%
	\includegraphics[width=0.5\textwidth]{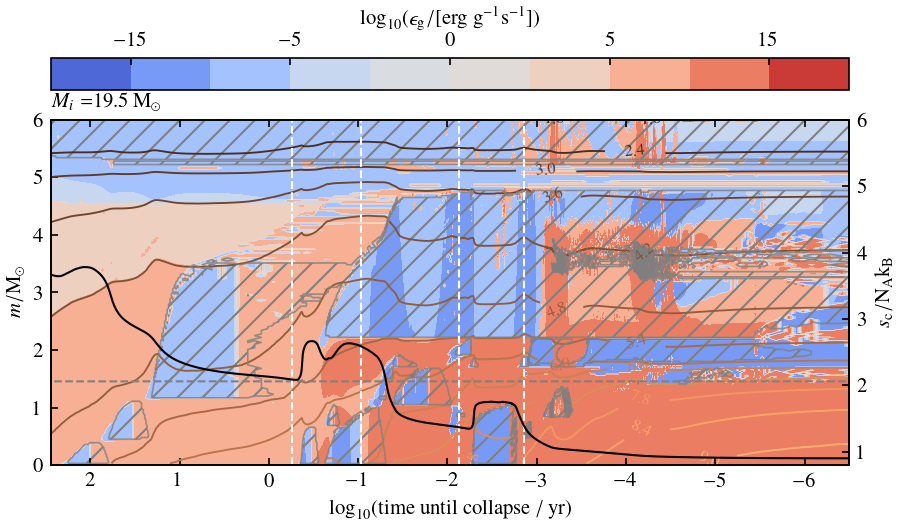}
	
	\includegraphics[width=0.5\textwidth]{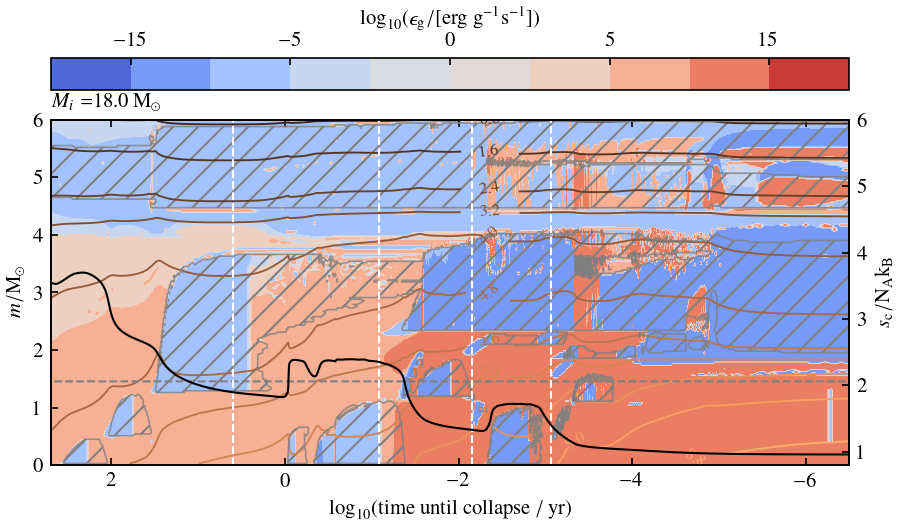}%
	\includegraphics[width=0.5\textwidth]{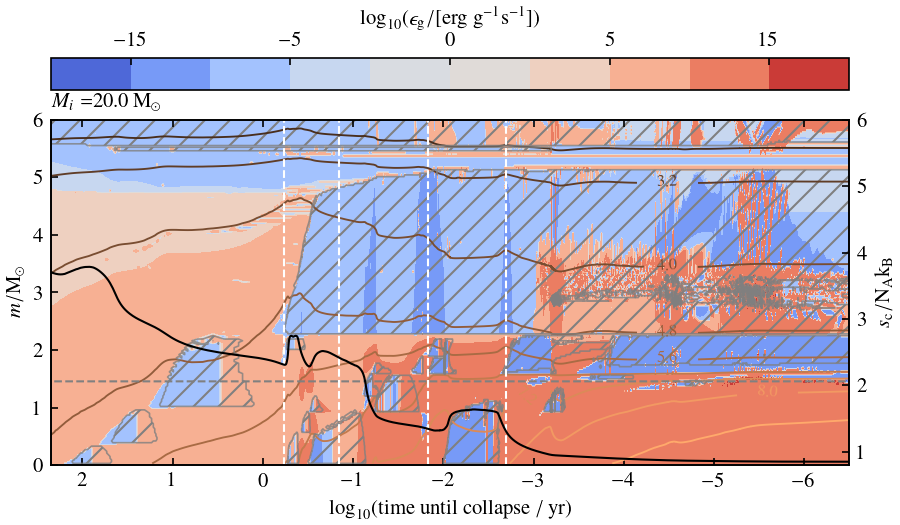}
	
	\includegraphics[width=0.5\textwidth]{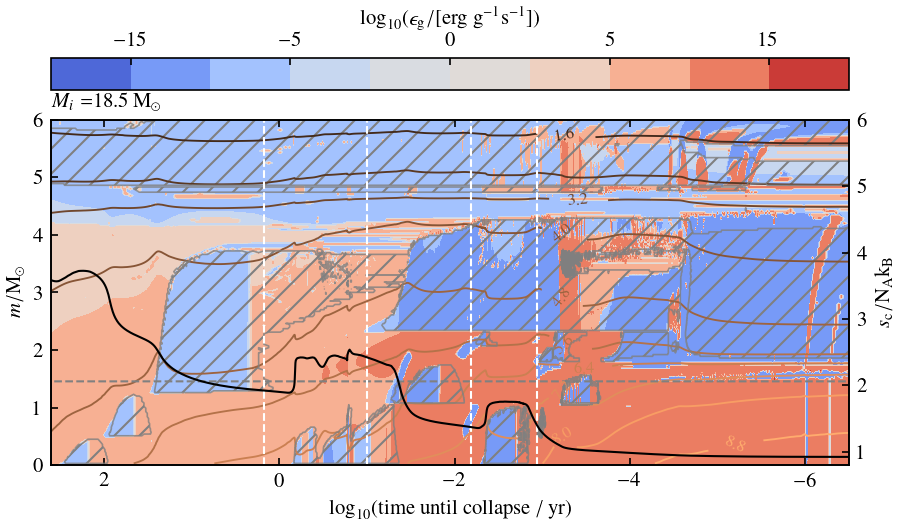}%
	\includegraphics[width=0.5\textwidth]{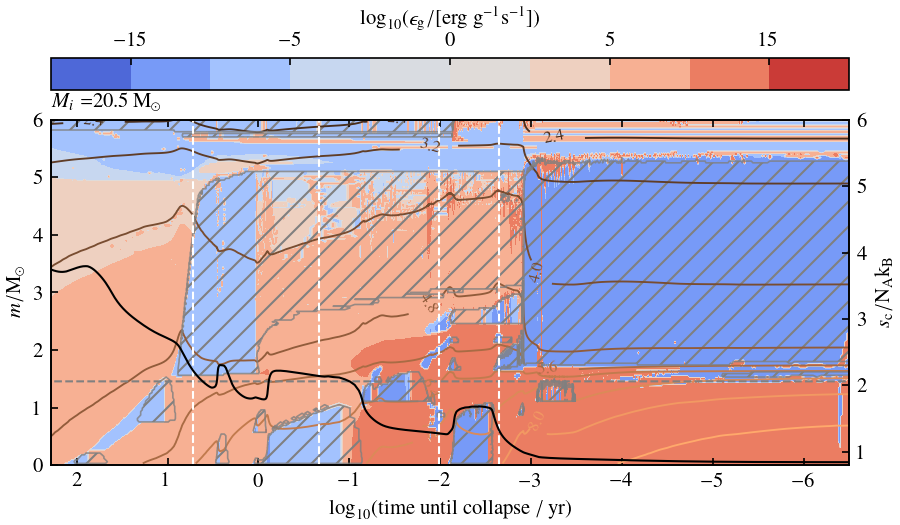}
	
	\caption{Final interior structure evolution of massive stars from core C ignition to the onset of core collapse. Colors indicate the specific gravothermal energy released or used by contraction (red) and expansion (blue). Brown contours show lines of constant (logarithmic) density. Convective regions are shown by hatches and the evolution of the central specific entropy is shown with a black line. From left to right, vertical white dashed lines mark the moments of core C, Ne, O, and Si depletion, respectively. The dashed horizontal line represents the value of the classical Chandrasekhar mass before the end of core oxygen burning. Here we show models from 17\Msun to 20.5\Msun, where central carbon burning is slowly becoming neutrino dominated and the core structures are expected to lead to a successful supernova.}
	\label{fig:eps_grav_overview_17_20.5}
\end{figure*}

\begin{figure*}[hb!]
	\centering
	\includegraphics[width=0.5\textwidth]{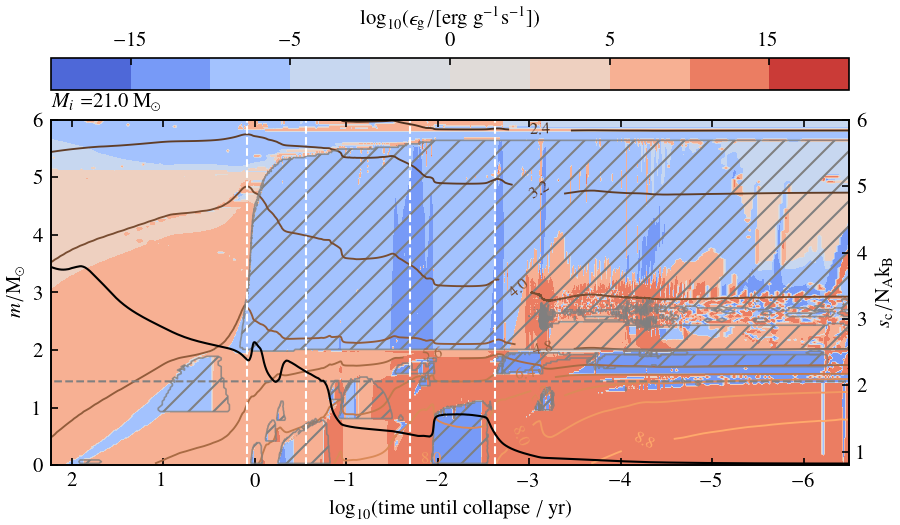}%
	\includegraphics[width=0.5\textwidth]{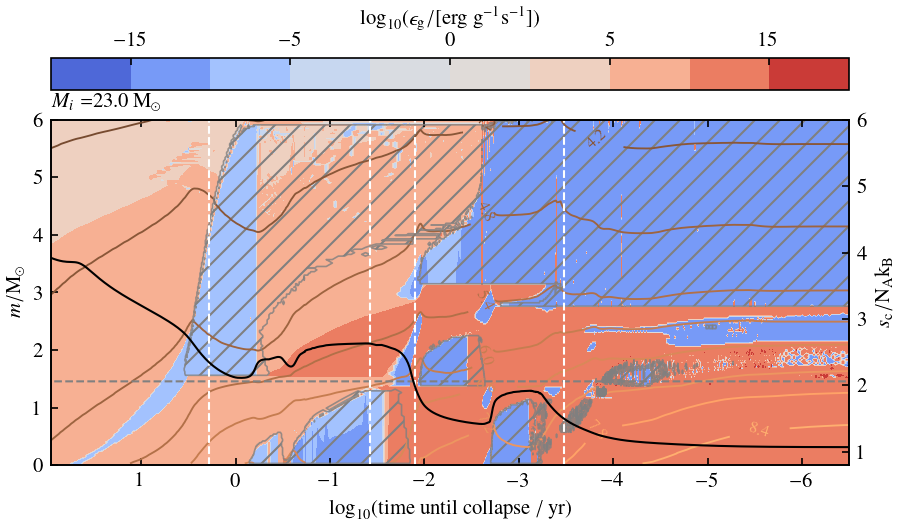}
	
	\includegraphics[width=0.5\textwidth]{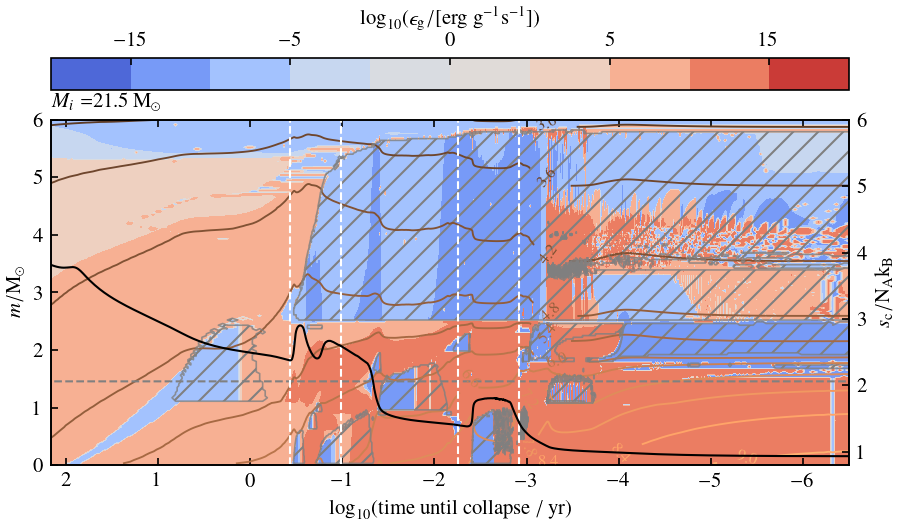}%
	\includegraphics[width=0.5\textwidth]{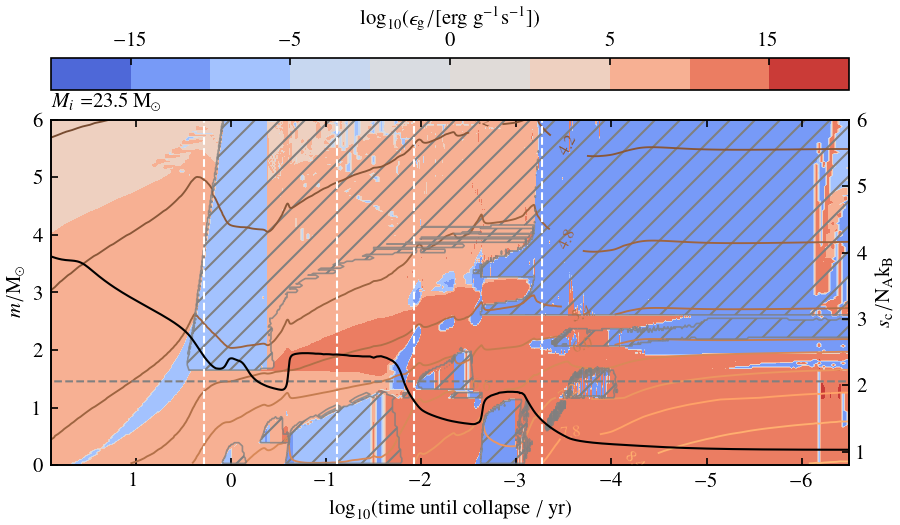}
	
	\includegraphics[width=0.5\textwidth]{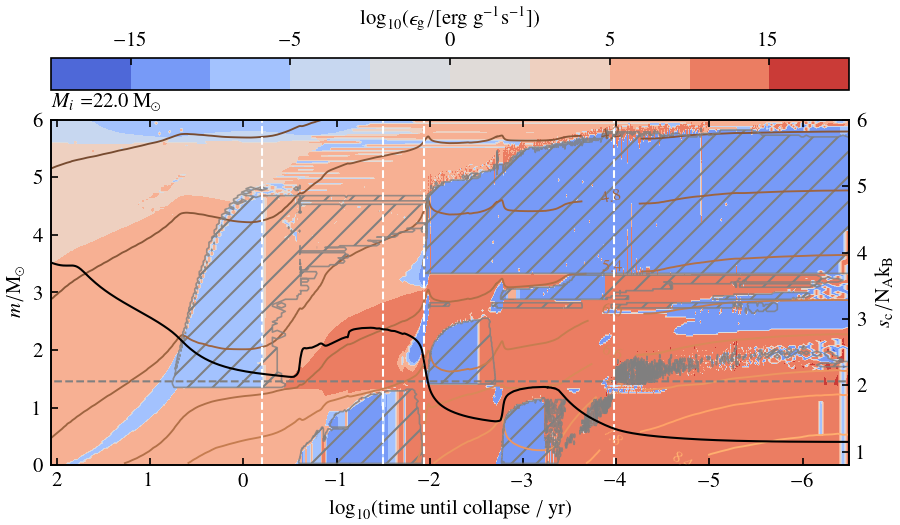}%
	\includegraphics[width=0.5\textwidth]{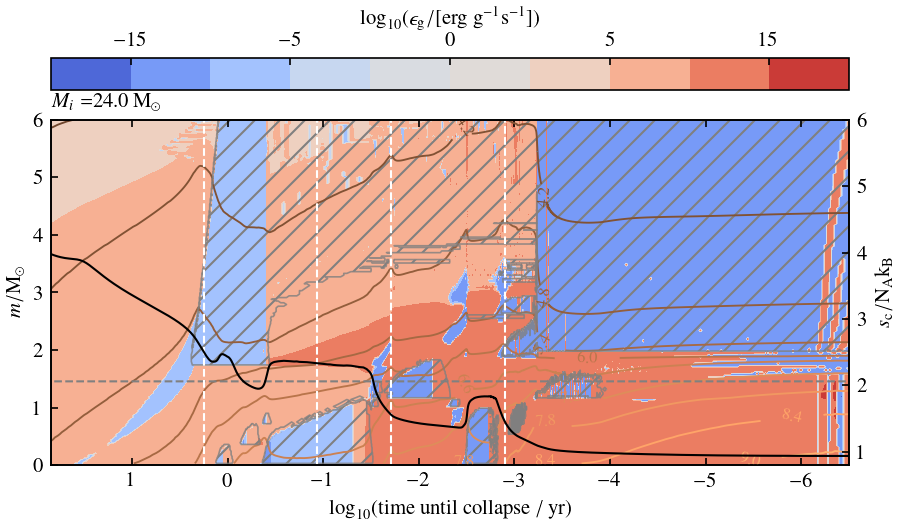}
	
	\includegraphics[width=0.5\textwidth]{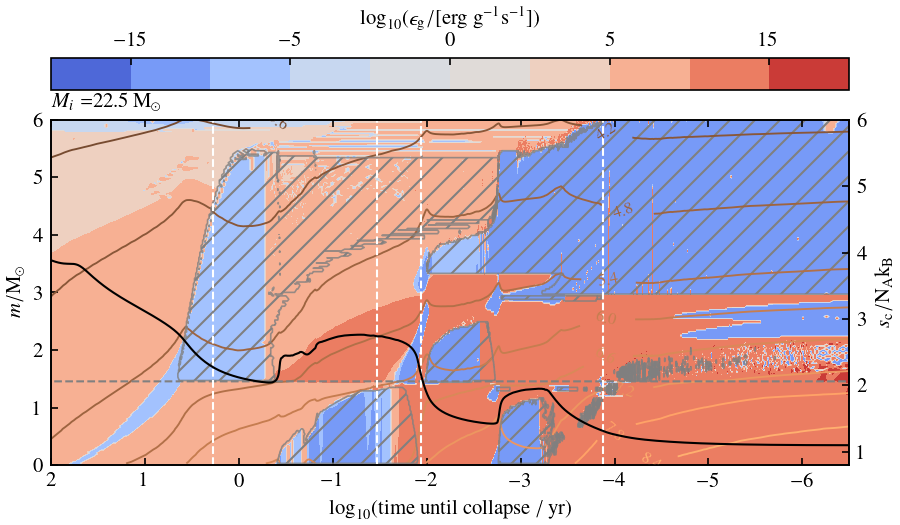}%
	\includegraphics[width=0.5\textwidth]{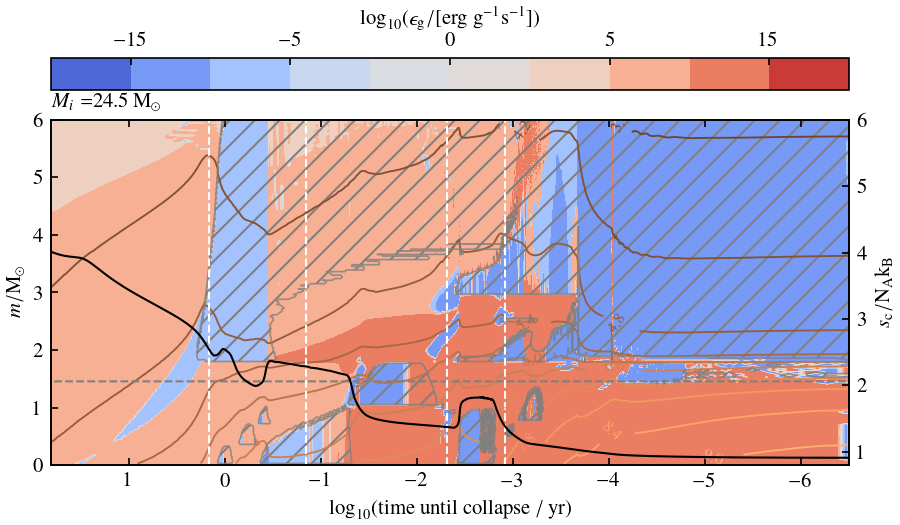}
	
	\caption{Same as Fig.~\ref{fig:eps_grav_overview_17_20.5} with initial masses between 21.0\Msun and 24.5\Msun (region A and B), where central carbon burning is neutrino dominated. The final core structure favors black hole formation for the 22\Msun-- 23\Msun models and successful supernova explosions in the other cases.}
	\label{fig:eps_grav_overview_21_25.5}
\end{figure*}

\begin{figure*}[hb!]
	\centering
	\includegraphics[width=0.5\textwidth]{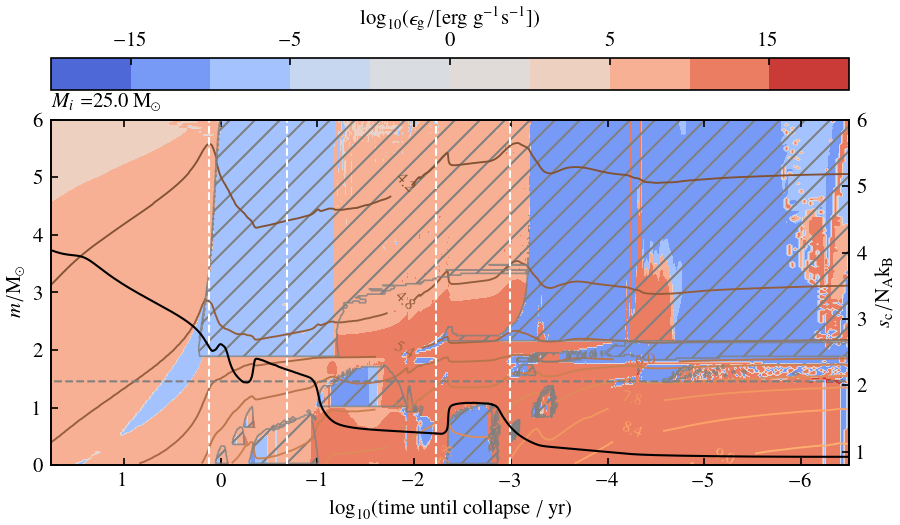}%
	\includegraphics[width=0.5\textwidth]{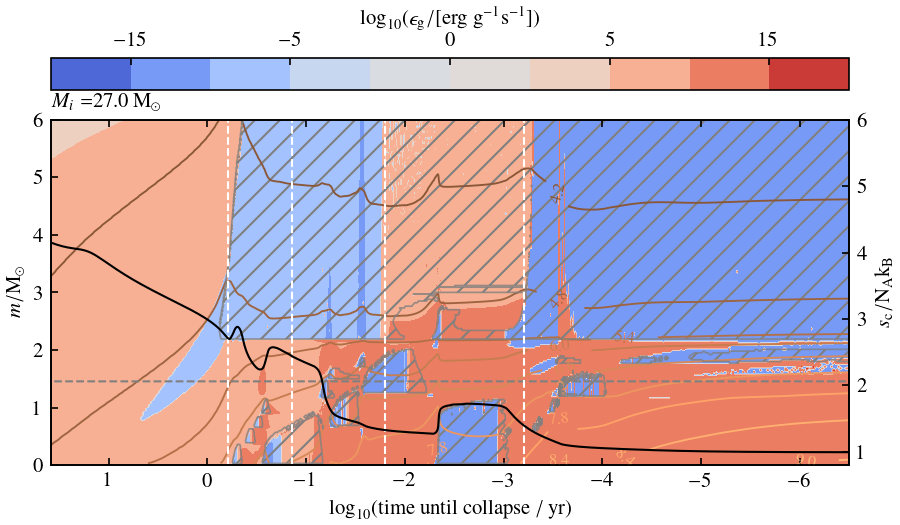}
	
	\includegraphics[width=0.5\textwidth]{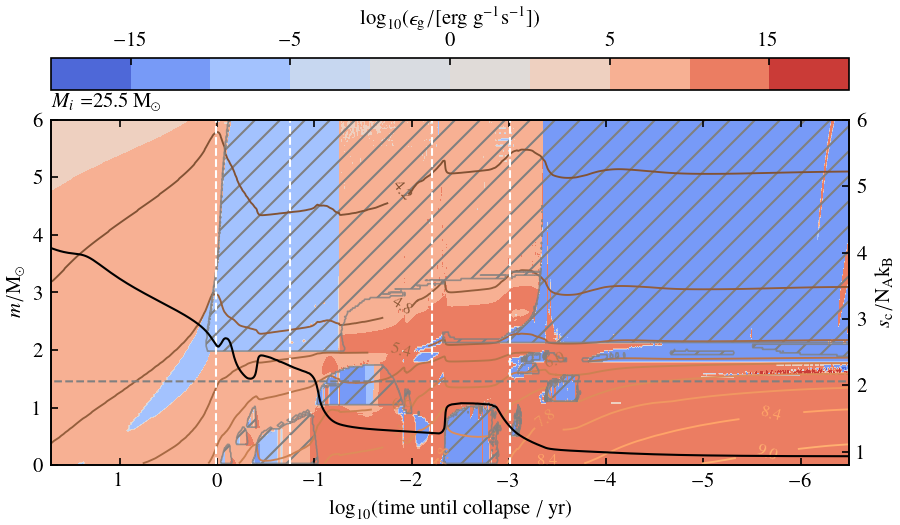}%
	\includegraphics[width=0.5\textwidth]{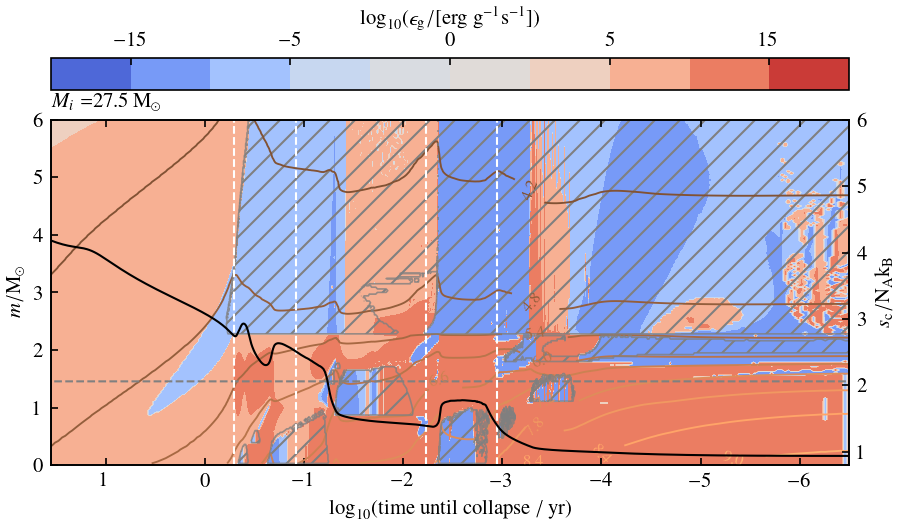}
	
	\includegraphics[width=0.5\textwidth]{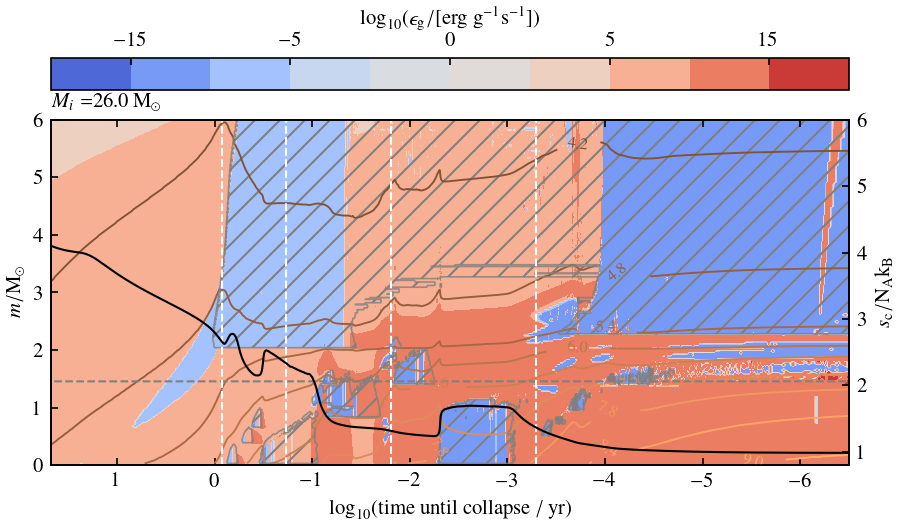}%
	\includegraphics[width=0.5\textwidth]{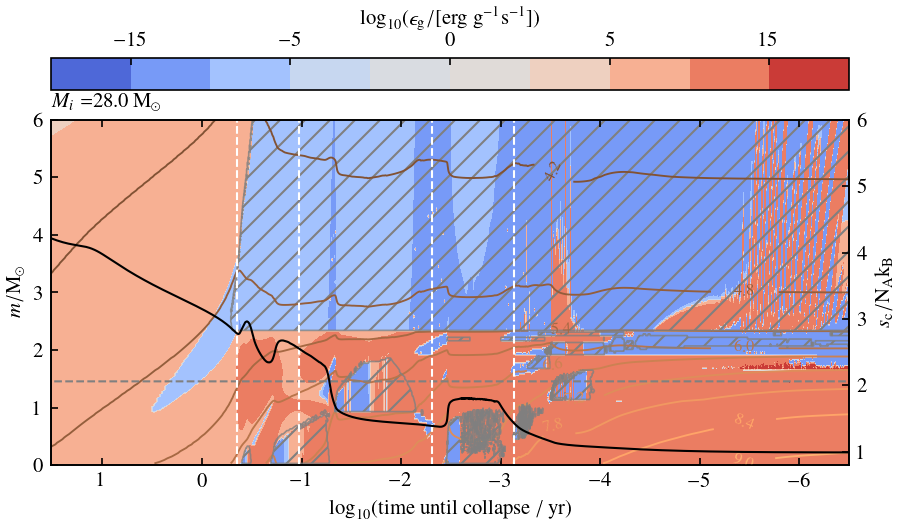}
	
	\includegraphics[width=0.5\textwidth]{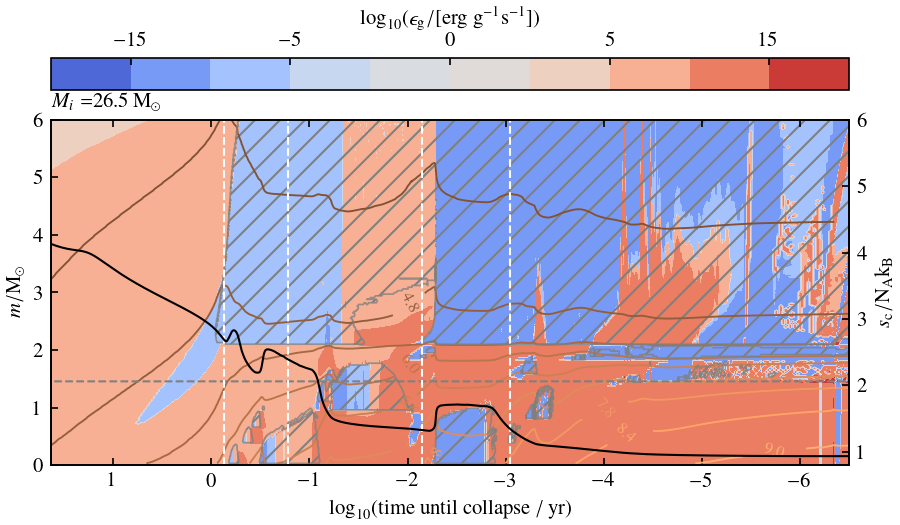}%
	\includegraphics[width=0.5\textwidth]{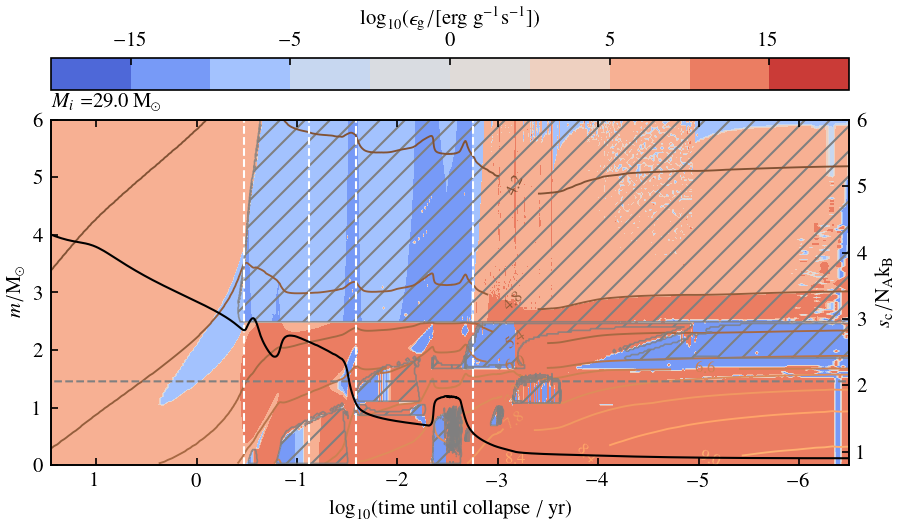}
	
	\caption{Same as Fig.~\ref{fig:eps_grav_overview_17_20.5} for models with initial masses between 25\Msun and 29\Msun, where central carbon burning is neutrino dominated and central neon burning becomes increasingly neutrino-dominated. All models are expected to lead to a successful supernova explosion.}
	\label{fig:eps_grav_overview_26_29}
\end{figure*}

\begin{figure*}[hb!]
	\centering
	\includegraphics[width=0.5\textwidth]{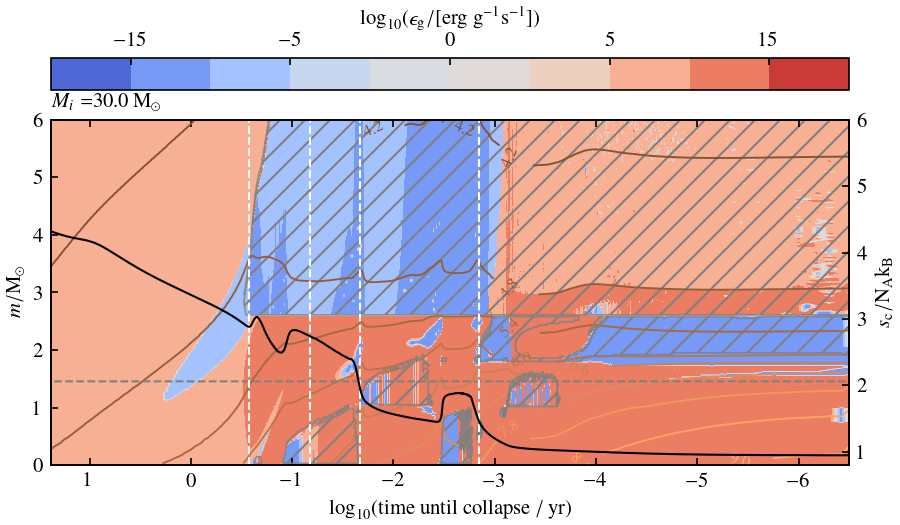}%
	\includegraphics[width=0.5\textwidth]{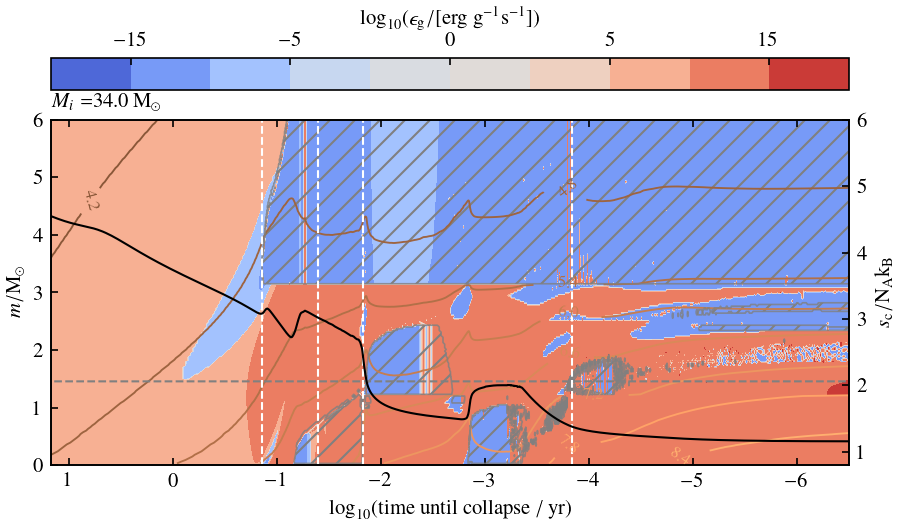}
	
	\includegraphics[width=0.5\textwidth]{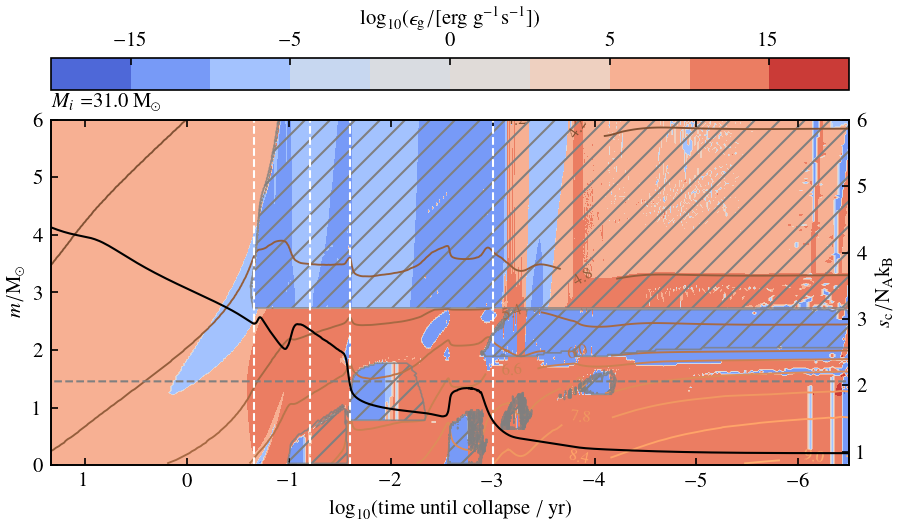}%
	\includegraphics[width=0.5\textwidth]{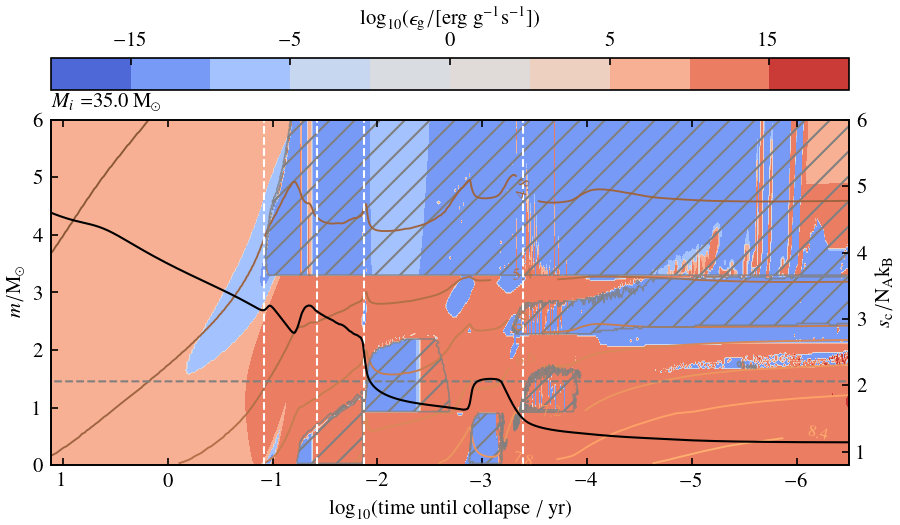}
	
	\includegraphics[width=0.5\textwidth]{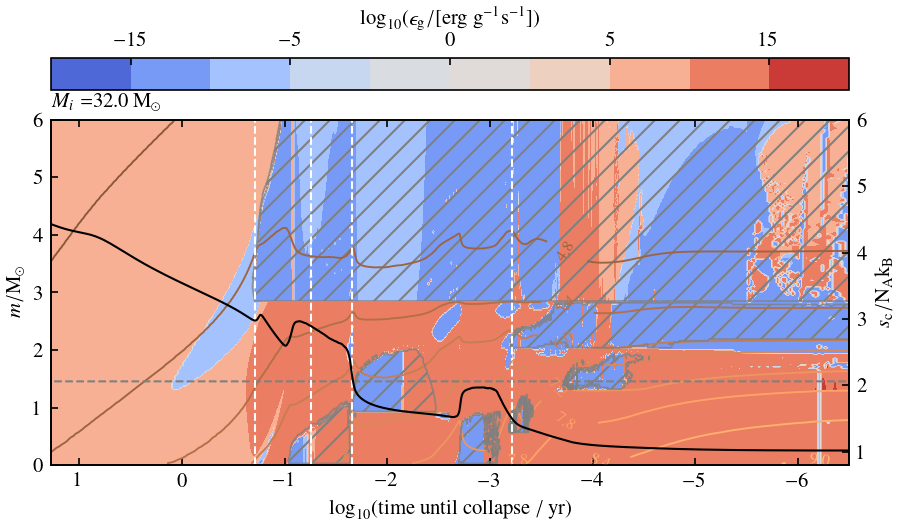}%
	\includegraphics[width=0.5\textwidth]{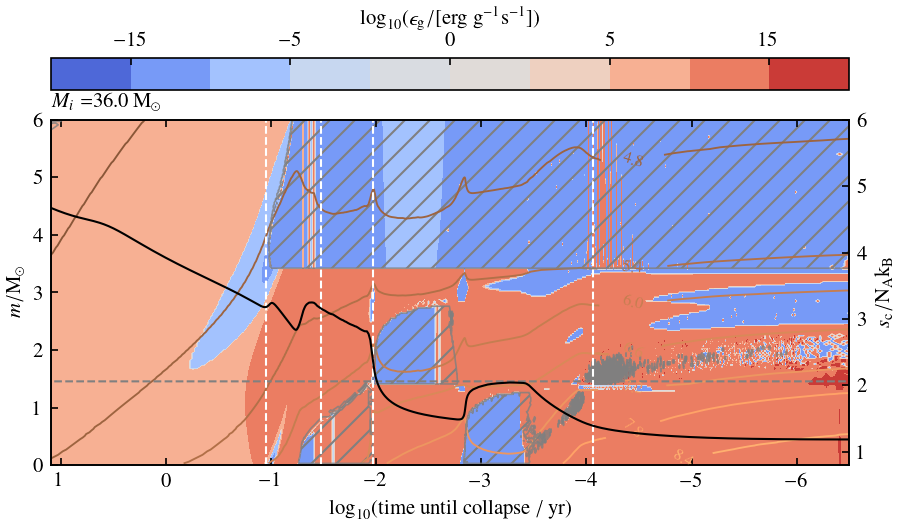}
	
	\includegraphics[width=0.5\textwidth]{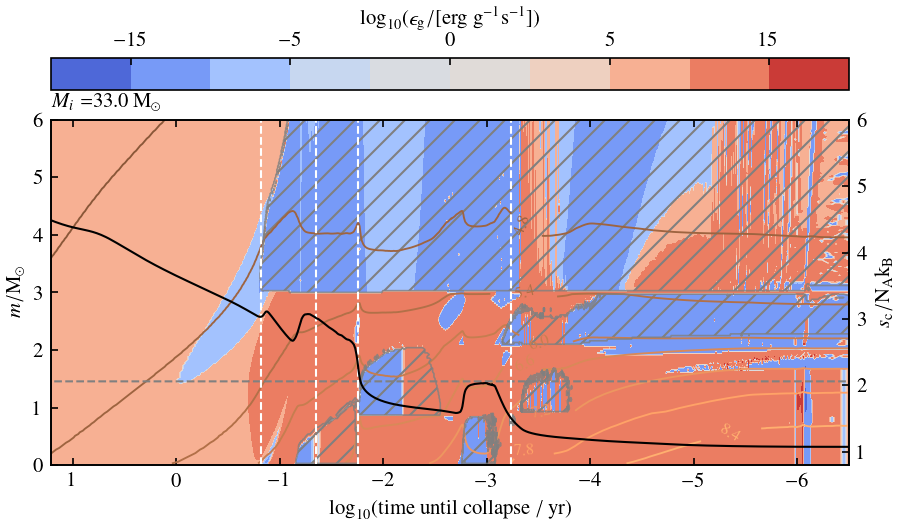}%
	\includegraphics[width=0.5\textwidth]{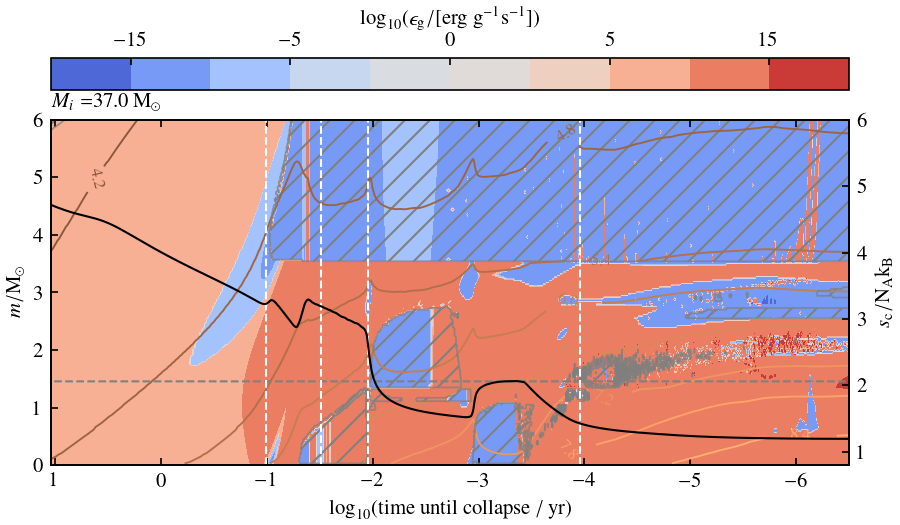}
	
	\caption{Same as Fig.~\ref{fig:eps_grav_overview_17_20.5} for models with initial masses between 30.0\Msun and 37.0\Msun (region C), where central carbon burning and central neon burning are both neutrino-dominated. Above an initial mass of 34.0\Msun, Because of the high final binding energy of the layers above the central core, these models are expected to lead to black hole formation.}
	\label{fig:eps_grav_overview_30_37}
\end{figure*}

\begin{figure*}[hb!]
	\centering
	\includegraphics[width=0.5\textwidth]{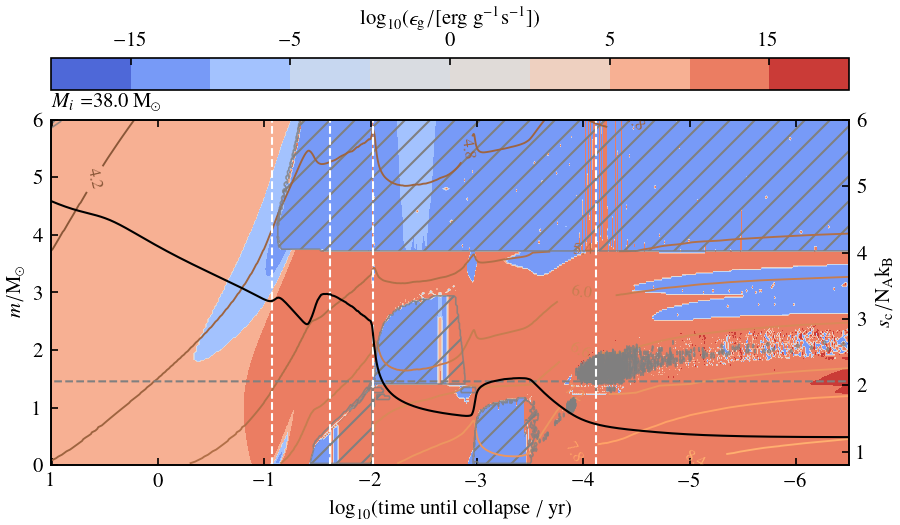}%
	\includegraphics[width=0.5\textwidth]{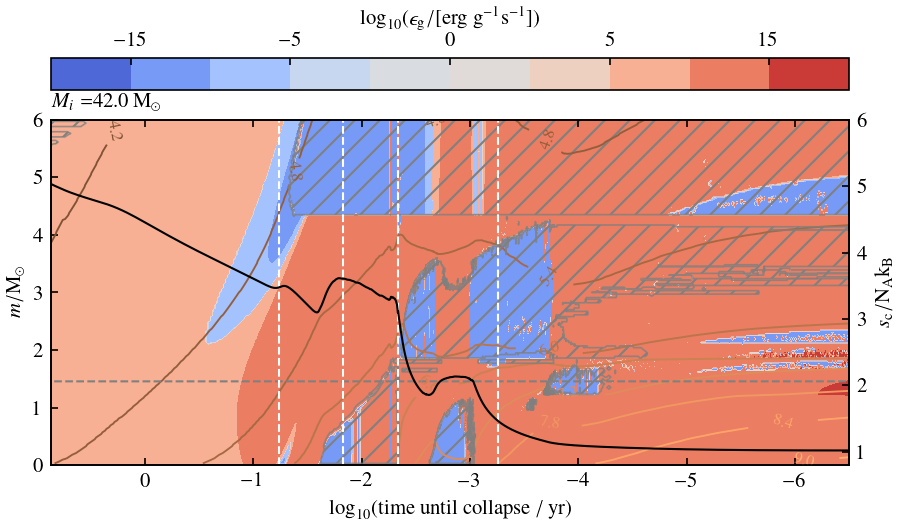}
	
	\includegraphics[width=0.5\textwidth]{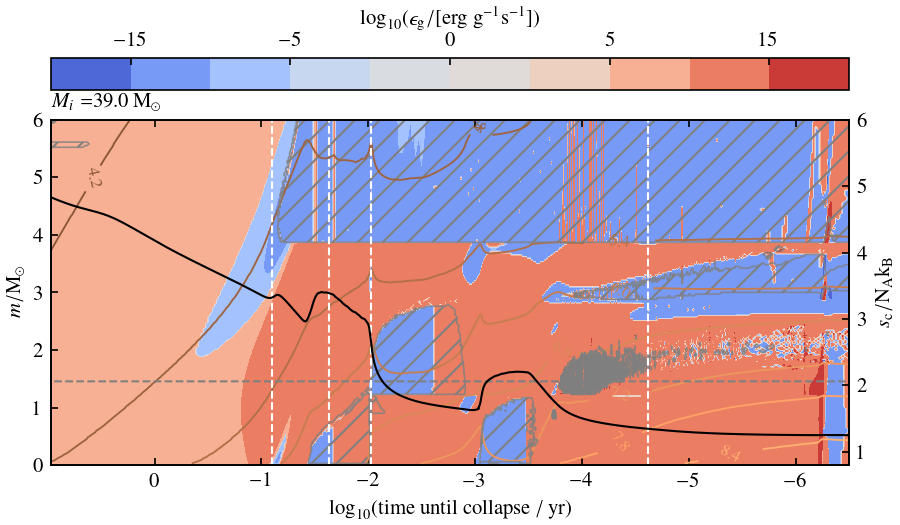}%
	\includegraphics[width=0.5\textwidth]{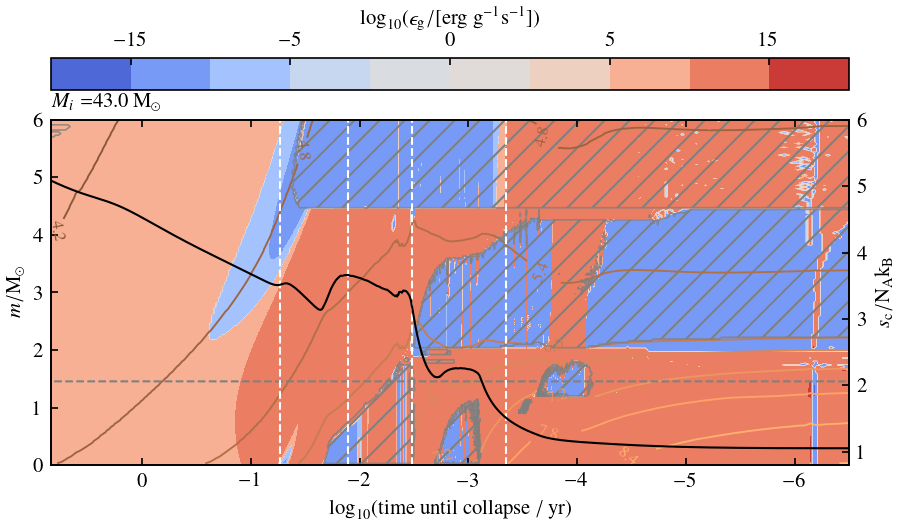}
	
	\includegraphics[width=0.5\textwidth]{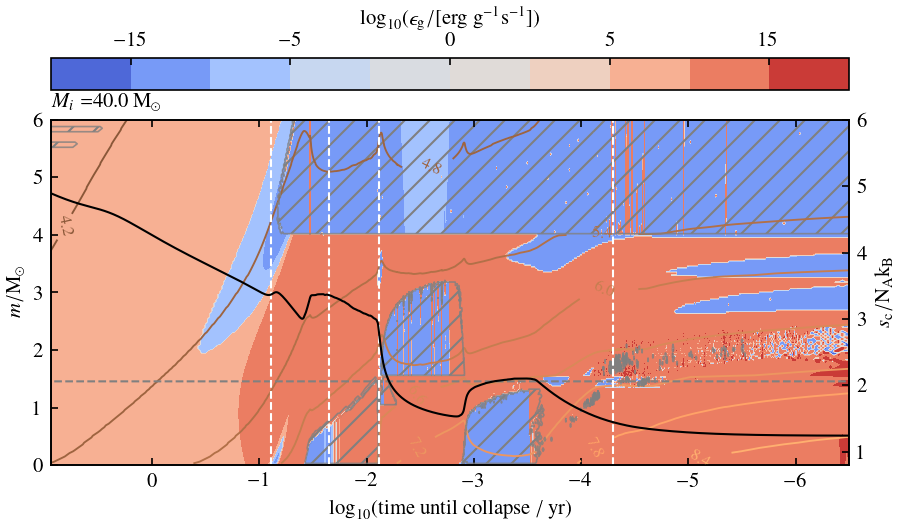}%
	\includegraphics[width=0.5\textwidth]{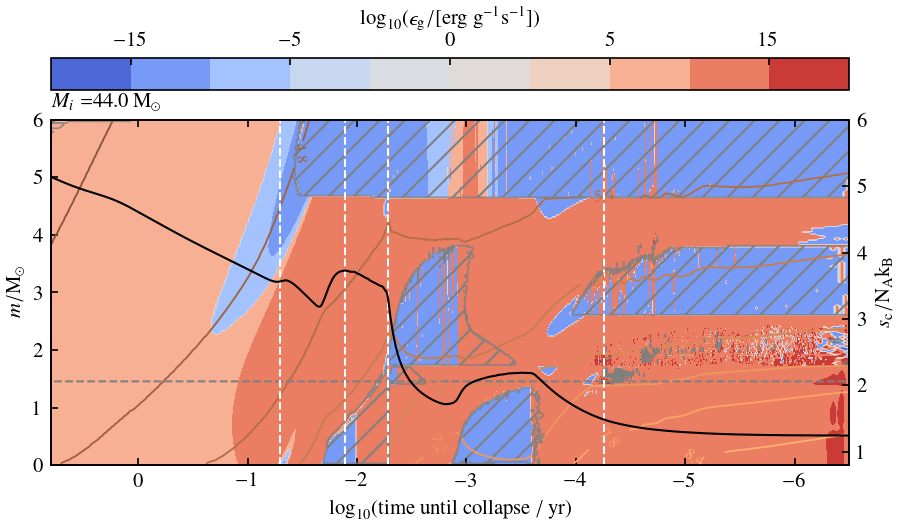}
	
	\includegraphics[width=0.5\textwidth]{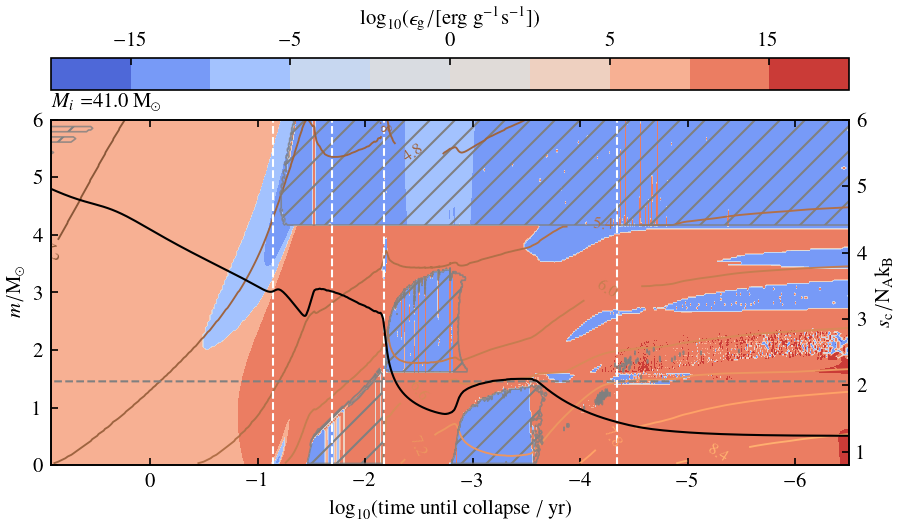}%
	\includegraphics[width=0.5\textwidth]{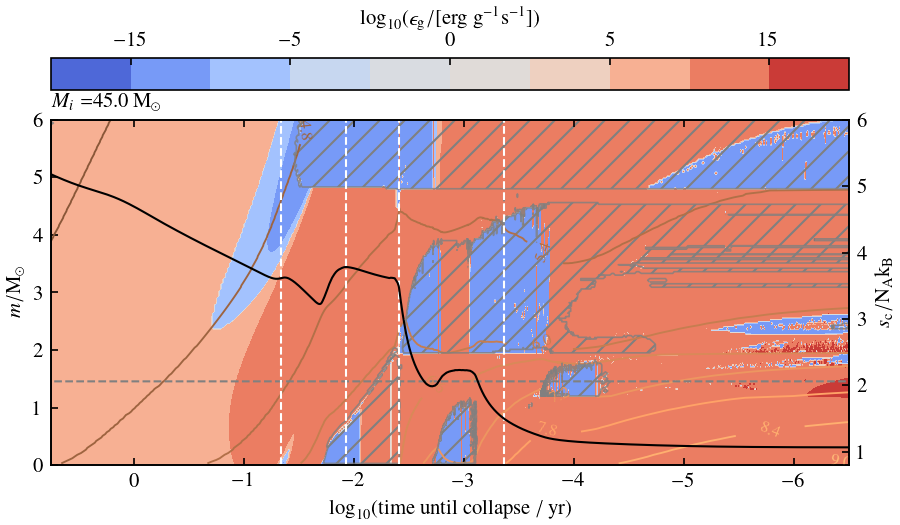}
	
	\caption{Same as Fig.~\ref{fig:eps_grav_overview_17_20.5} for models with initial masses between 38.0\Msun and 45.0\Msun (region D and beyond), where central carbon burning and central neon burning are both neutrino-dominated, and core oxygen burning is increasingly neutrino dominated. Because of the high final binding energy of the layers above the central core, these models are expected to lead to black hole formation.}
	\label{fig:eps_grav_overview_38_45}
\end{figure*}

\begin{figure*}[hb!]
	\centering
	\includegraphics[width=0.5\textwidth]{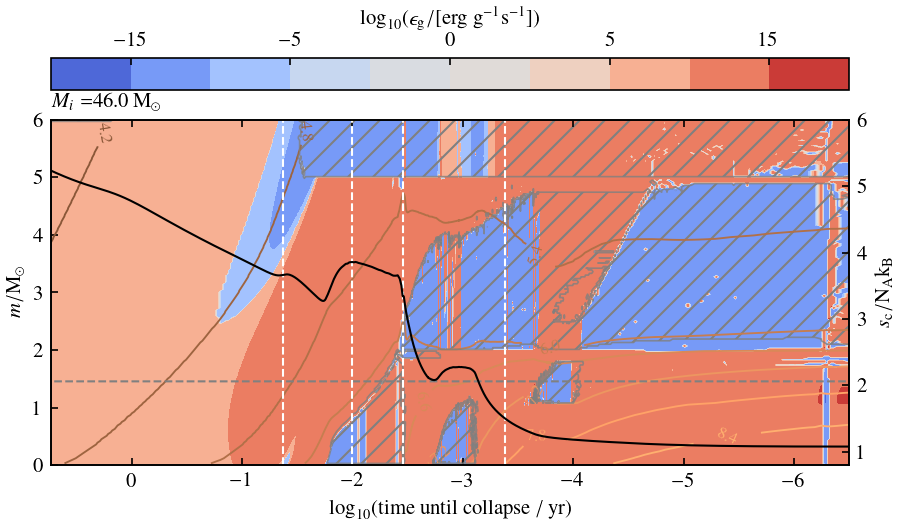}%
	\includegraphics[width=0.5\textwidth]{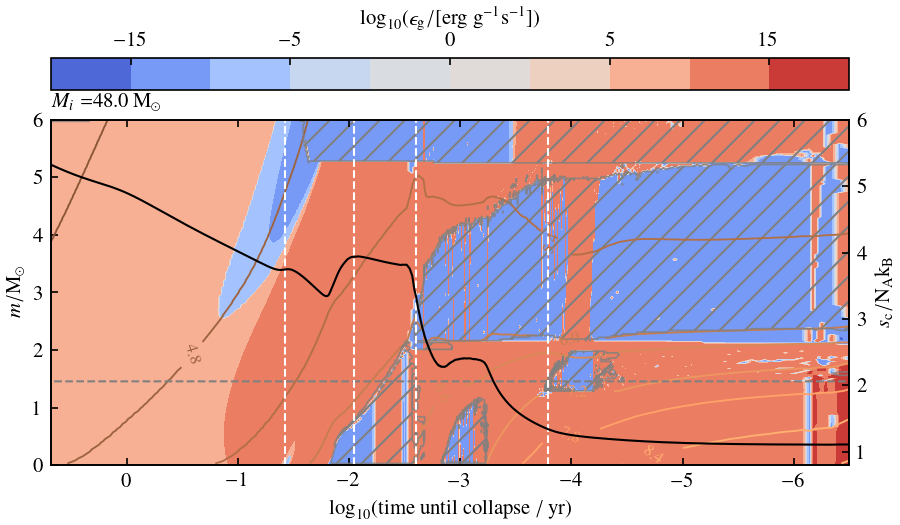}
	
	\includegraphics[width=0.5\textwidth]{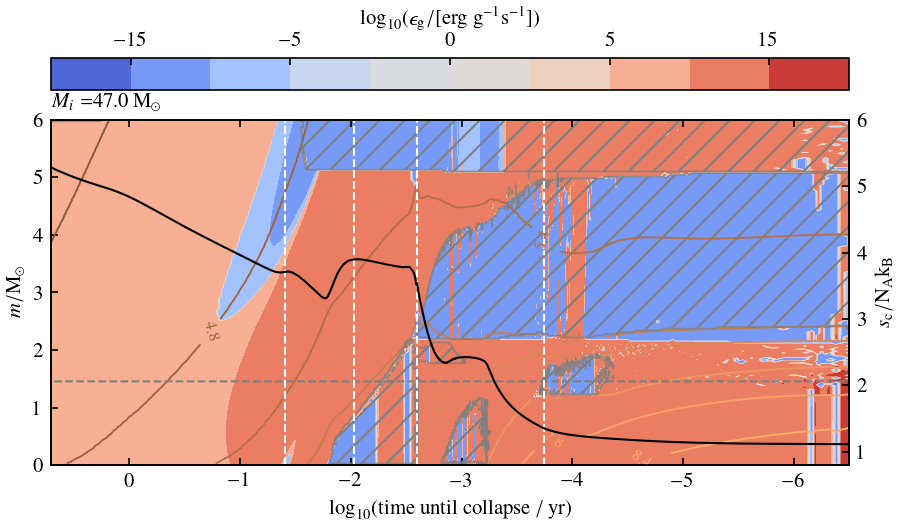}%
	\includegraphics[width=0.5\textwidth]{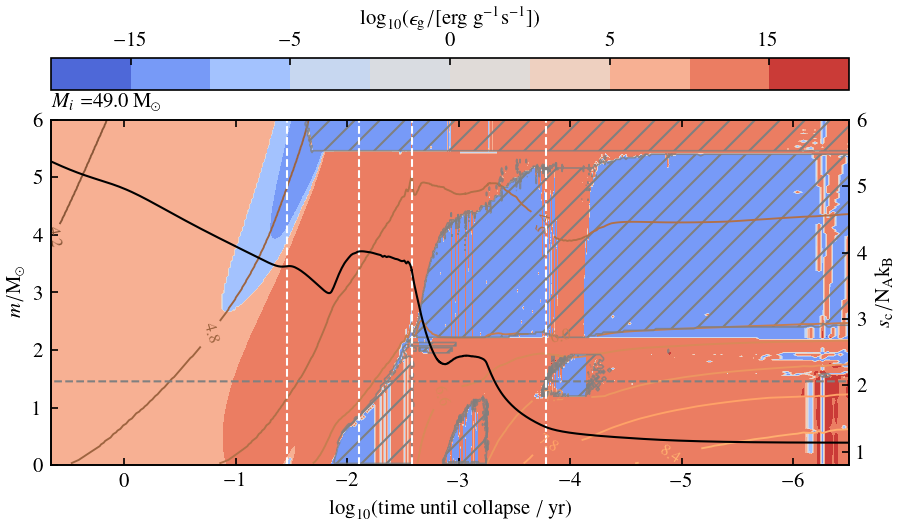}
	
	\includegraphics[width=0.5\textwidth]{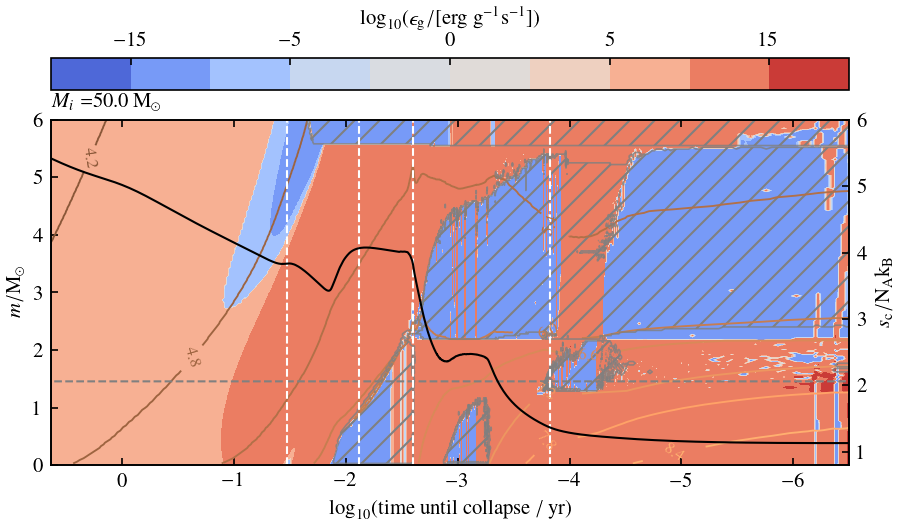}
	\caption{Same as Fig.~\ref{fig:eps_grav_overview_17_20.5} for models with initial masses between 46.0\Msun and 50.0\Msun. Because of the high final binding energy of the layers above the central core, these models are expected to lead to black hole formation.}
	\label{fig:eps_grav_overview_46_50}
\end{figure*}

We provide an overview of the evolution of the interior 6\Msun structure of all our models from core carbon ignition to core collapse, shown in Fig.~\ref{fig:eps_grav_overview_17_20.5}, \ref{fig:eps_grav_overview_21_25.5}, \ref{fig:eps_grav_overview_26_29}, \ref{fig:eps_grav_overview_30_37}, \ref{fig:eps_grav_overview_38_45}, and \ref{fig:eps_grav_overview_46_50}.

\end{appendix}

\end{document}